\documentclass[twoside,12pt]{article}
\usepackage{epsfig}
\usepackage{bbm}
\usepackage{slashed}
\begin{document}

\title{ \vspace{1cm} Chiral Perturbation Theory: Introduction and Recent Results
in the One-Nucleon Sector}
\author{S.\ Scherer\\ \\
Institut f\"ur Kernphysik, Johannes Gutenberg-Universit\"at Mainz,\\
J.~J.~Becher-Weg 45, D-55099 Mainz, Germany}
\date{April 10, 2009}
\maketitle

\begin{abstract}
   We provide an introduction to the basic concepts of chiral perturbation
theory and discuss some recent developments in the manifestly Lorentz-invariant
formulation of the one-nucleon sector.
\end{abstract}
%\eject
\tableofcontents
\section{Introduction}
   Effective field theory (EFT) is a powerful tool in the description of the
strong interactions at low energies.
   The central idea is due to Weinberg \cite{Weinberg:1978kz}:
   \begin{quote}
"...  if one writes down the most general possible Lagrangian,
including all terms consistent with assumed symmetry principles, and
then calculates matrix elements with this Lagrangian to any given
order of perturbation theory, the result will simply be the most
general possible S--matrix consistent with analyticity, perturbative
unitarity, cluster decomposition and the assumed symmetry
principles."
\end{quote}
   In general, an EFT is an approximation to a (more) fundamental theory, designed
to be valid in a certain kinematical domain.
   Instead of solving the underlying theory, the processes under investigation
are described in terms of a suitable set of effective degrees of freedom, dominating
the phenomena in the particular energy region.
   In the context of the strong interactions, the underlying theory is
quantum chromodynamics
(QCD)---a gauge theory with color SU(3) as the gauge group.
   Under normal conditions, the fundamental degrees of freedom of QCD,
namely, quarks and gluons, do not show up as free particles.
   One assumes that any asymptotically observed hadron must be in a
color-singlet state, i.e., a physically observable state is invariant under
SU(3) color transformations.
   The strong increase of the running coupling for large distances possibly
provides a mechanism for the color confinement.

   For the low-energy properties of the strong interactions and the setting up of
a corresponding EFT description, another phenomenon is of vital importance.
   The masses of the up and down quarks and, to a lesser extent, also of the
strange quark are sufficiently small that the dynamics of QCD in the chiral
limit, i.e., for massless quarks, is believed to resemble that of the ``real'' world.
   Although a rigorous mathematical proof is not yet available, there are good reasons
to assume that a dynamical spontaneous symmetry breaking emerges from
the chiral limit.
   Examples of indications for this to happen are the comparatively small masses
of the pseudoscalar octet, the absence of a parity doubling in the low-energy spectrum
of hadrons, and a non-vanishing scalar singlet quark condensate.

   According to the Goldstone theorem, a breakdown of the chiral
$\mbox{SU(3)}_L\times\mbox{SU(3)}_R$ symmetry at the Lagrangian level to
the $\mbox{SU(3)}_V$ symmetry in the ground state implies the existence of
eight massless pseudoscalar Goldstone bosons.
   The finite masses of the pseudoscalar octet of the real world are attributed to
the explicit chiral symmetry breaking due to the quark masses in the QCD Lagrangian.
   Due to the vanishing of the Goldstone boson masses in the chiral limit
in combination with their vanishing interactions in the zero-energy limit, a
derivative and quark-mass expansion is the natural scenario for an EFT.
   The corresponding method is called (mesonic) chiral perturbation
theory (ChPT) \cite{Gasser:1983yg}, with the Goldstone bosons as the relevant effective
degrees of freedom (see Table \ref{1:table_QCD_EFT}).
\begin{table}[t]
\begin{center}
\renewcommand{\arraystretch}{1.5}
\begin{tabular}{c|c|c}
 &Fundamental theory & Effective field theory\\\hline
 Theoretical framework&QCD&ChPT\\\hline
 Degrees of freedom & Quarks and gluons & Goldstone bosons (+ other hadrons)\\
 \hline
 Parameters & $g_3$ + quark masses & Low-energy coupling constants + quark masses
\end{tabular}
\label{1:table_QCD_EFT}
\caption{Comparison of QCD and ChPT.}
\end{center}
\end{table}

   Using these effective degrees of freedom, physical quantities are calculated in
terms of an expansion in $q/\Lambda_\chi$, where $q$ stands for
momenta or masses of the pseudoscalar octet that are smaller than the energy/mass scale
$\Lambda_\chi={\cal O}(1\,\mbox{GeV})$ associated with spontaneous symmetry
breaking.
   Since an EFT is based on the most general Lagrangian, which includes
all terms that are compatible with the symmetries of the underlying
theory, the corresponding Lagrangian contains an infinite number of
terms, where each term is accompanied by a low-energy coupling constant
(LEC).
   The method that allows one to decide which terms
contribute in a calculation up to a certain accuracy is called
Weinberg's power counting.
   In the mesonic sector, the combination of dimensional regularization
with the modified minimal subtraction scheme of ChPT leads to
a straightforward correspondence between the loop expansion and the
chiral expansion in terms of momenta and quark masses at a fixed ratio.
   In actual calculations only a finite number of terms in the expansion in
$q/\Lambda_\chi$ has to be considered and thus one has predictive power.
   What distinguishes the EFT approach from purely phenomenological approaches
is the possibility of a systematic improvement.
   Mesonic ChPT has been tremendously successful and may be considered as
a full-grown and mature area of low-energy particle physics.

   The situation gets more complicated once other hadronic degrees of freedom beyond
the Goldstone bosons are considered.
   Together with such hadrons, another scale of the order of the chiral symmetry
breaking scale $\Lambda_\chi$ enters the problem and the methods of the
pure Goldstone-boson sector cannot be transferred one to one.
   For example, in the extension to the one-nucleon sector %\cite{Gasser:1988rb}
the correspondence between the loop expansion and the chiral expansion,
at first sight, seems to be lost: higher-loop
diagrams can contribute to terms as low as ${\cal O}(q^2)$ \cite{Gasser:1987rb}.
   For a long time this was interpreted as the absence of a systematic power
counting in the relativistic formulation of ChPT.
   However, over the last decade new developments in devising a suitable renormalization
scheme have led to a simple and consistent power counting for the renormalized
diagrams of a manifestly Lorentz-invariant approach.

   The purpose of this article is to first provide a pedagogical introduction
to the basic concepts of ChPT and to then present the
more recent developments of a manifestly Lorentz-invariant approach to the
one-nucleon sector.
   It is definitely {\em not} intended to give a survey of the vast literature
on ChPT and its various extensions in terms of chiral
effective field theories.
   For further information the interested reader is referred to review articles
and lecture notes addressing different topics with various priorities
\cite{Bijnens:1993xi},
\cite{Georgi:1994qn},
\cite{Ecker:1994gg},
\cite{Pich:1995bw},
\cite{Bernard:1995dp},
\cite{Hemmert:1997ye},
\cite{Burgess:1998ku},
\cite{Scherer:2002tk},
\cite{Scherer:2005ri},
\cite{Epelbaum:2005pn},
\cite{Bijnens:2006zp},
\cite{Bernard:2006gx},
\cite{Bernard:2007zu}.

   The article is organized as follows.
   Section \ref{section_chiral symmetry} contains a discussion of the chiral symmetry
of QCD, spontaneous symmetry breaking, and the Goldstone theorem.
   In Sec.\ \ref{section_mchpt}, the basic concepts of mesonic ChPT
are developed.
   Section \ref{section_bchpt} is devoted to baryonic ChPT.
   The power-counting problem is illustrated and solutions in terms of suitable
renormalization conditions are presented.
   Section \ref{section_applications} contains a few selected applications of
the manifestly Lorentz-invariant approach to nucleon properties.

\section{Chiral symmetry and spontaneous symmetry breaking}
\label{section_chiral symmetry}
   The essential ingredients to setting up chiral perturbation theory
as the effective field theory of the strong interactions are the
chiral $\mbox{SU(3)}_L\times\mbox{SU(3)}_R$ symmetry of QCD for
massless $u$, $d$, and $s$ quarks and the emergence of a spontaneous
breakdown to the vectorial subgroup SU(3)$_V$.

\subsection{Quantum chromodynamics and chiral symmetry}
   QCD is the gauge theory of the strong interactions
\cite{Gross:1973id:2}, \cite{Weinberg:un:2},\linebreak \cite{Fritzsch:pi:2} with color SU(3)
as the underlying gauge group.
   Historically, the color
degree of freedom was introduced into the quark model to account for
the Pauli principle in the description of baryons as three-quark
states.
   The matter fields of QCD are the so-called quarks which are
spin-1/2 fermions, with six different flavors $(u, d, s, c, b, t)$ in addition to their
three possible colors (see Table \ref{2:2:table:quarks}).
   Since quarks have not been observed as asymptotically free
states, the meaning of quark masses and their numerical values are
tightly connected with the method by which they are extracted from
hadronic properties (see Ref.\ \cite{Manohar_PDG} for a thorough
discussion).

\begin{table}
\begin{center}
{\renewcommand{\baselinestretch}{1.5}\small\normalsize
\begin{tabular}{|l|c|c|c|}
\hline
Flavor&$u$&$d$&$s$\\
\hline
Charge [e] &$2/3$&$-1/3$&$-1/3$\\
\hline
Mass [MeV]&$1.5 - 3.3$ & $3.5 - 6.0$ & $70 - 130$\\
\hline \hline
Flavor&$c$&$b$&$t$\\
\hline
Charge [e] &$2/3$&$-1/3$&$2/3$\\
\hline
Mass [GeV] & $1.27^{+0.07}_{-0.11}$ & $4.20^{+0.17}_{-0.07}$ &$171.2\pm 2.1$\\
\hline
\end{tabular}
}
\caption{\label{2:2:table:quarks} Quark flavors and their charges
and masses. See \cite{Manohar_PDG} for details.}
\end{center}
\end{table}

\subsubsection{The QCD Lagrangian} \label{sec_qcdl}
   The QCD Lagrangian can be obtained from the
Lagrangian for free quarks by applying the gauge principle with
respect to the group SU(3) of all unitary, unimodular, $3\times 3$ matrices.
   Denoting the quark field components by
$q_{f,A,\alpha}$,
where $f=1,\cdots, 6$ refers to the flavor index,
$A=1,2,3$ to the color index, and $\alpha=1,\cdots,4$ to the Dirac spinor
index, respectively, the ``free'' quark Lagrangian without interaction
may be regarded as the sum of $6\times 3=18$ free fermion Lagrangians:
\begin{equation}
\label{2:1:1:lfq}
{\cal L}_{\rm free\, quarks}
=\sum_{f=1}^6 \sum_{A=1}^3\sum_{\alpha,\alpha'=1}^4
\bar{q}_{f,A,\alpha}(\gamma^\mu_{\alpha\alpha'}i\partial_\mu
-m_f\delta_{\alpha\alpha'})q_{f,A,\alpha'}.
\end{equation}
   Suppressing the Dirac spinor index and introducing for each quark flavor $f$ a color triplet
\begin{equation}
\label{2:1:1:qtriplet}
 q_f=\left(\begin{array}{c}q_{f,1}
\\q_{f,2}\\q_{f,3}\end{array}\right),
\end{equation}
the gauge principle is applied with respect to the group SU(3), i.e., all
$q_f$ are subject to the same local SU(3) transformation:
\begin{equation}
\label{2:1:1:qtraf}
q_f\mapsto q_f'=\exp\left(-i\sum_{a=1}^8 \Theta_a
\frac{\lambda_a^c}{2}\right)q_f=U q_f,
\end{equation}
where the eight $\lambda_a^c$ denote Gell-Mann matrices acting in color space and
the $\Theta_a$ are smooth, real functions in Minkowski space.
   Whenever convenient, we will make use of the summation convention implying a summation
over repeated indices.
   Introducing eight gauge potentials ${\cal A}_{a\mu}$, transforming as
\begin{equation}
\label{2:1:1:atraf}
{\cal A}_\mu\equiv{\cal A}_{a\mu}\frac{\lambda_a^c}{2}\mapsto
{\cal A}_\mu'=
U{\cal A}_\mu U^\dagger +\frac{i}{g_3}\partial_\mu U U^\dagger,
\end{equation}
the covariant derivative of the quark field, by construction, transforms as the quark field:
\begin{equation}
\label{2:1:1:cdq}
D_\mu q_f \equiv (\partial_\mu
 +ig_3{\cal A}_\mu)q_f\mapsto
(D_\mu q_f)'=D'_\mu
q'_f=U D_\mu q_f.
\end{equation}
   In Eq.\ (\ref{2:1:1:cdq}), $g_3$ denotes the strong coupling constant.
   In order to treat the gauge potentials as dynamical degrees of freedom,
one defines a generalization of the field strength tensor to the non-Abelian case
as
\begin{equation}
\label{2:1:1:gmunu}
{\cal G}_{a\mu\nu}=\partial_\mu {\cal A}_{a\nu}-\partial_\nu {\cal A}_{a\mu}
-g_3 f_{abc}{\cal A}_{b\mu} {\cal A}_{c\nu},
\end{equation}
where, suppressing the superscript $c$ in the Gell-Mann matrices,
the standard totally antisymmetric SU(3) structure constants are given by
(see Table \ref{table:2:1:1:su3structurconstants})
\begin{equation}
\label{2:1:1:fabc}
f_{abc}=\frac{1}{4i}\mbox{Tr}([\lambda_a,\lambda_b]\lambda_c).
\end{equation}
\begin{table}[t]
\begin{center}
{\renewcommand{\baselinestretch}{1.5}\small\normalsize
\begin{tabular}{|r|r|r|r|r|r|r|r|r|r|}
\hline
$abc$&123&147&156&246&257&345&367&458&678\\
\hline
$f_{abc}$&1&$\frac{1}{2}$&$-\frac{1}{2}$&$\frac{1}{2}$&
$\frac{1}{2}$&$\frac{1}{2}$&$-\frac{1}{2}$&$\frac{1}{2}\sqrt{3}$&
$\frac{1}{2}\sqrt{3}$\\
\hline
\end{tabular}
}
\caption{\label{table:2:1:1:su3structurconstants}
Totally antisymmetric non-vanishing structure constants of SU(3):
$[\frac{\lambda_a}{2},\frac{\lambda_b}{2}]=if_{abc}\frac{\lambda_c}{2}$.
}
\end{center}
\end{table}
   Given Eq.\ (\ref{2:1:1:atraf}), the field strength tensor transforms under
SU(3) as
\begin{equation}
\label{2:1:1:gtrafo} {\cal G}_{\mu\nu}\equiv
{\cal G}_{a\mu\nu} \frac{\lambda_a^c}{2}\mapsto U{\cal G}_{\mu\nu}
U^\dagger.
\end{equation}
   The QCD Lagrangian obtained by applying the gauge principle to the
free Lagrangian of Eq.~(\ref{2:1:1:lfq}), finally, reads
\begin{equation}
\label{2:1:1:lqcd} {\cal L}_{\rm QCD}=\sum_{f={u,d,s, \atop c,b,t}}
\bar{q}_f(i D\hspace{-.6em}/ -m_f)q_f -\frac{1}{4}{\cal
G}_{a\mu\nu}{\cal G}^{\mu\nu}_a.
\end{equation}

   From the point of view of gauge invariance the strong-interaction
Lagrangian could also involve a term of the type
\begin{equation}
\label{2:1:1:ltheta} {\cal
L}_\theta=\frac{g^2_3\bar{\theta}}{64\pi^2}\epsilon_{\mu\nu\rho\sigma}
{\cal G}_a^{\mu\nu}{\cal G}_a^{\rho\sigma},\quad \epsilon_{0123}=1,
\end{equation}
where $\epsilon_{\mu\nu\rho\sigma}$ denotes the totally
antisymmetric Levi-Civita tensor.
   The so-called $\theta$ term of Eq.\ (\ref{2:1:1:ltheta}) implies an explicit
$P$ and $CP$ violation of the strong interactions which, for
example, would give rise to an electric dipole moment of the
neutron.
   The present empirical information indicates that the $\theta$ term is
small and, in the following, we will omit Eq.\ (\ref{2:1:1:ltheta})
from our discussion.

\subsubsection{Chiral limit}
   The terminology chiral limit refers to massless quarks, resulting in
an important additional global symmetry of the QCD Lagrangian which will
be discussed in the following.
    We introduce the chirality matrix
$\gamma_5=\gamma^5=i\gamma^0\gamma^1\gamma^2\gamma^3=\gamma_5^\dagger$,
$\{\gamma^\mu,\gamma_5\}=0$, $\gamma_5^2=\mathbbm 1$, and define the projection operators
\begin{equation}
\label{2:1:2:prpl}
P_L=\frac{1}{2}({\mathbbm 1}-\gamma_5)=P_L^\dagger,\quad
P_R=\frac{1}{2}({\mathbbm 1}+\gamma_5)=P_R^\dagger.
\end{equation}
These operators satisfy the completeness relation $P_L+P_R=\mathbbm 1$,
are idempotent, $P_L^2=P_L$, $P_R^2=P_R$, and respect the orthogonality relations
$P_L P_R=P_R P_L=0$.
   When applied to the solutions of the free massless Dirac equation,
the operators $P_R$ and $P_L$ project to the
positive and negative helicity eigenstates, hence the subscripts
$R$ and $L$ for right-handed and left-handed, respectively.

   Omitting color and flavor indices, we introduce left- and right-handed quark fields as
\begin{equation}
\label{2:1:2:qlr}
q_L=P_L q\quad\mbox{and}\quad q_R=P_R q.
\end{equation}
   A quadratic form containing any of the 16 independent $4\times 4$ matrices
$\{{\mathbbm 1},\gamma^\mu,\gamma_5,\gamma^\mu\gamma_5,\sigma^{\mu\nu}\}$
can be decomposed as
\begin{equation}
\label{2:1:2:qgq}
\bar{q}\,\Gamma_i q=\left \{\begin{array}{lcl}
\bar{q}_L\Gamma_1 q_L+\bar{q}_R\Gamma_1 q_R&\mbox{for}&
\Gamma_1\in\{\gamma^\mu,\gamma^\mu\gamma_5\}\\
\bar{q}_R\Gamma_2 q_L +\bar{q}_L\Gamma_2 q_R&\mbox{for}& \Gamma_2
\in\{{\mathbbm 1},\gamma_5,\sigma^{\mu\nu}\}
\end{array}
\right.,
\end{equation}
where
$$ \bar{q}_R=\bar{q}P_L\quad\mbox{and}\quad \bar{q}_L=\bar{q}P_R.
$$
  The validity of Eq.\ (\ref{2:1:2:qgq}) is general
and does not refer to ``massless'' quark fields.

   From a phenomenological point of view the $u$ and $d$ quarks and to a lesser
extent also the $s$ quark have relatively small masses in comparison to a
typical hadronic scale of the order of 1 GeV.
   On the other hand, we will neglect the three heavy quarks $c$, $b$, and $t$,
because we will restrict ourselves to energies well below the production threshold
of particles containing a heavy (anti-) quark.
   In the following, we will approximate the full QCD Lagrangian by its
light-flavor version, and will consider the chiral limit for the three light
quarks $u$, $d$, and $s$.
   To that end, we apply Eq.\ (\ref{2:1:2:qgq}) to the term containing the contraction
of the covariant derivative with $\gamma^\mu$.
   This quadratic quark form decouples into the sum of two
terms which connect only left-handed with left-handed and right-handed
with right-handed quark fields.
   The QCD Lagrangian in the chiral limit can then be written as
\begin{equation}
\label{2:1:2:lqcd0lr}
{\cal L}^0_{\rm QCD}=\sum_{l=u,d,s}
(\bar{q}_{R,l}iD\hspace{-.6em}/\hspace{.3em}q_{R,l}+\bar{q}_{L,l}iD
\hspace{-.6em}/\hspace{.3em}
q_{L,l})-\frac{1}{4}{\cal G}_{a\mu\nu} {\cal G}^{\mu\nu}_a.
\end{equation}
   Note that because of Eq.\ (\ref{2:1:2:qgq}) the quark-mass term generates a
coupling between left- and right-handed quark fields.

\subsubsection{Global symmetry currents of the light quark sector}
\label{subsubsec_agsl}
    Due to the flavor independence of the covariant derivative,
${\cal L}^0_{\rm QCD}$ is invariant under the infinitesimal global transformations
of the left- and right-handed quark fields,
\begin{eqnarray}
\label{2:1:3:u3lu3r}
q_L\equiv
\left(\begin{array}{c}u_L\\d_L\\s_L\end{array}\right)&\mapsto&
\left({\mathbbm 1}-i\sum_{a=1}^8
\epsilon^L_a\frac{\lambda_a}{2}-i\epsilon^L\right)q_L,\nonumber\\
q_R\equiv\left(\begin{array}{c}u_R\\d_R\\s_R\end{array}\right) &\mapsto& \left({\mathbbm 1}-i\sum_{a=1}^8
\epsilon^R_a\frac{\lambda_a}{2}-i\epsilon^R\right)q_R.
\end{eqnarray}
   Note that the Gell-Mann matrices act in flavor space.
   ${\cal L}^0_{\rm QCD}$ is said to have a classical
{\em global} $\mbox{U(3)}_L\times\mbox{U(3)}_R$ symmetry.
   Applying Noether's theorem \cite{Noether:1918}, \cite{Hill:1951}, \cite{Gell-Mann:1960np},
from such an invariance one would expect a total of $2\times(8+1)=18$ conserved currents:
\begin{equation}
\label{2:1:3:str} L^{\mu}_a=\bar{q}_L\gamma^\mu \frac{\lambda_a}{2}q_L,\quad
L^{\mu}=\bar{q}_L\gamma^\mu q_L,\quad
R^{\mu}_a=\bar{q}_R\gamma^\mu \frac{\lambda_a}{2}q_R,\quad
R^{\mu}=\bar{q}_R\gamma^\mu q_R.
\end{equation}
Making use of
\begin{displaymath}
P_L\gamma^\mu P_R\pm P_R\gamma^\mu P_L
=\left\{\begin{array}{l}\gamma^\mu\\\gamma^\mu\gamma_5\end{array}\right.,
\end{displaymath}
we introduce the linear combinations
\begin{eqnarray}
\label{2:1:3:v}
V^{\mu}_a&=& R^{\mu}_a+L^{\mu}_a=\bar{q}\gamma^\mu\frac{\lambda_a}{2}q,\\
\label{2:1:3:a}
A^{\mu}_a&=&R^{\mu}_a-L^{\mu}_a=\bar{q}\gamma^\mu\gamma_5
\frac{\lambda_a}{2}q,
\end{eqnarray}
which under a parity transformation of the quark fields,
$q(t,\vec x)\mapsto \gamma_0 q(t,-\vec x)$,
transform as vector and axial-vector current densities,
respectively,
\begin{eqnarray}
\label{2:1:3:pv}
P:V^{\mu}_a(t,\vec{x})\mapsto V_{a\mu}(t,-\vec{x}),\\
\label{2:1:3pa} P: A^{\mu}_a(t,\vec{x})\mapsto -A_{a\mu}(t,-\vec{x}).
\end{eqnarray}
   The conserved singlet vector current results from a transformation of all
left-handed and right-handed quark fields by the {\em same} phase,
\begin{equation}
\label{2:1:3:sv}
V^\mu=R^{\mu}+L^{\mu}=\bar{q}\gamma^\mu q.
\end{equation}
   The singlet axial-vector current originates from a transformation of all left-handed
quark fields with one phase and all right-handed with the {\em
opposite} phase,
\begin{equation}
\label{2:1:3:sav}
A^\mu= R^{\mu}-L^{\mu}=\bar{q}\gamma^\mu\gamma_5q.
\end{equation}
   Quantum fluctuations destroy the singlet axial-vector current conservation
and there will be extra terms, referred to as anomalies \cite{Bell:1969ts}, \cite{Adler:1969gk},
\cite{Adler:1969er}, resulting in
\begin{displaymath}
\partial_\mu A^\mu=\frac{3 g^2_3}{32\pi^2}\epsilon_{\mu\nu\rho\sigma}
{\cal G}^{\mu\nu}_a {\cal G}^{\rho\sigma}_a,\quad \epsilon_{0123}=1.
\end{displaymath}
   The factor of three originates from the number of flavors.
   In the large $N_c$ (number of colors) limit of Ref.~\cite{'tHooft:1973jz} the singlet axial-vector
current is conserved, because the strong coupling constant behaves
as $g^2_3\sim N_c^{-1}$.

\subsubsection{Chiral algebra}
\label{subsec_ca}
   The invariance of ${\cal L}^0_{\rm QCD}$ under global
$\mbox{SU(3)}_L\times\mbox{SU(3)}_R\times\mbox{U(1)}_V$ transformations
implies that also the QCD Hamilton operator in the chiral limit,
$H^0_{\rm QCD}$, exhibits a global
$\mbox{SU(3)}_L\times\mbox{SU(3)}_R\times\mbox{U(1)}_V$ symmetry.
   As usual, the charge operators are defined as the space integrals
of the charge densities,
\begin{eqnarray}
\label{2:1:4:ql}
Q_{aL}(t)&=&\int \mbox{d}^3x\,q^\dagger_L(t,\vec{x})\frac{\lambda_a}{2}q_L(t,\vec{x}),\\
\label{2:1:4:qr}
Q_{aR}(t)&=&\int \mbox{d}^3x\,q^\dagger_R(t,\vec{x})\frac{\lambda_a}{2}q_R(t,\vec{x}),\\
\label{2:1:4:qv}
Q_V(t)&=&\int \mbox{d}^3x\,\left[q^\dagger_L(t,\vec{x})q_L(t,\vec{x})+
q^\dagger_R(t,\vec{x})q_R(t,\vec{x})\right].
\end{eqnarray}
   For conserved symmetry currents, these operators are time independent,
i.e., they commute with the Hamiltonian,
\begin{equation}
\label{2:1:4:vrhq} [Q_{aL},H^0_{\rm QCD}]=[Q_{aR},H^0_{\rm
QCD}]=[Q_V,H^0_{\rm QCD}]=0.
\end{equation}
The commutation relations among the charge operators reflect the underlying Lie algebra
of $\mbox{SU(3)}_L\times\mbox{SU(3)}_R\times\mbox{U(1)}_V$,
\begin{eqnarray}
\label{2:1:4:crqll}
[Q_{aL},Q_{bL}]&=&if_{abc}Q_{cL},\\
\label{2:1:4:crqrr}
{[Q_{aR},Q_{bR}]}&=&if_{abc}Q_{cR},\\
\label{2:1:4:crqlr}
{[Q_{aL},Q_{bR}]}&=&0,\\
\label{2:1:4:crqlvrv} {[Q_{aL},Q_V]}&=&[Q_{aR},Q_V]=0.
\end{eqnarray}
   Equations (\ref{2:1:4:crqll}) - (\ref{2:1:4:crqlvrv}) are verified by expressing the
commutators in terms of equal-time anti-commutation relations of the
quark fields.

  It should be stressed that, even without being able to explicitly solve
the equation of motion of the quark fields entering the charge
operators of Eqs.\ (\ref{2:1:4:crqll}) - (\ref{2:1:4:crqlvrv}), we
know from the equal-time commutation relations and the symmetry of
the Lagrangian that these charge operators are the generators of
infinitesimal transformations of the Hilbert space associated with
$H^0_{\rm QCD}$.
   Furthermore, their commutation relations with a given operator
specify the transformation behavior of the operator in question
under the group $\mbox{SU(3)}_L\times\mbox{SU(3)}_R\times\mbox{U(1)}_V$.

\subsubsection{Quark masses and chiral symmetry breaking}
\label{subsec_csbdqm}
   So far, we have discussed an idealized world with massless light quarks.
   The finite $u$-, $d$-, and $s$-quark masses explicitly break the chiral symmetry and
generate divergences of the symmetry currents.
   As a consequence, the charge operators are, in general, no longer
time independent.
   However, as first pointed out by Gell-Mann \cite{Gell-Mann:xb:3:5}, the equal-time commutation
relations still play an important role even if the symmetry is
explicitly broken.

   Defining the quark-mass matrix as
\begin{displaymath}
{\cal M}=\mbox{diag}(m_u,m_d,m_s),
\end{displaymath}
   the quark-mass term in the QCD Lagrangian leads to a mixing
of left- and right-handed fields [see Eq.~(\ref{2:1:2:qgq})],
\begin{equation}
\label{2:1:5:lm} {\cal L}_{\cal M}= -\bar{q}{\cal M}q= -(\bar{q}_R
{\cal M} q_L +\bar{q}_L {\cal M} q_R).
\end{equation}
   Inserting the transformations of Eqs.~(\ref{2:1:3:u3lu3r}) into
the quark-mass term of Eq.~(\ref{2:1:5:lm}) results in the variation
$\delta{\cal L}_{\cal M}$, from which one obtains for the divergences
\begin{eqnarray}
\label{2:1:5:dslr}
\partial_\mu L^{\mu}_a&=&\frac{\partial \delta {\cal L}_{\cal M}}{\partial \epsilon^L_a}
=-i\left(\bar{q}_L\frac{\lambda_a}{2}{\cal M} q_R -\bar{q}_R {\cal
M} \frac{\lambda_a}{2}
q_L\right),\nonumber\\
\partial_\mu L^{\mu}&=&\frac{\partial \delta {\cal L}_{\cal M}}{\partial \epsilon^L}
=-i\left(\bar{q}_L {\cal M} q_R -\bar{q}_R {\cal M} q_L\right).
\end{eqnarray}
   The analogous expressions for $\partial_\mu R^{\mu}_a$ and
$\partial_\mu R^{\mu}$ are obtained from Eqs.~(\ref{2:1:5:dslr}) through the substitution
$R\leftrightarrow L$.
   The divergences are proportional to the mass parameters which is
the origin of the expression current-quark mass.
   In terms of the vector and axial-vector currents the divergences read
\begin{eqnarray}
\label{2:1:5:dsva}
\partial_\mu V^{\mu}_a&=&
i\bar{q}[{\cal M},\frac{\lambda_a}{2}]q,\nonumber\\
\partial_\mu A^{\mu}_a&=&
i\bar{q}\{{\cal M},\frac{\lambda_a}{2}\}\gamma_5q,\nonumber\\
\partial_\mu V^\mu&=&0,\nonumber\\
\partial_\mu A^\mu&=&2i\bar{q}{\cal M}\gamma_5 q+
\frac{3 g^2_3}{32\pi^2}\epsilon_{\mu\nu\rho\sigma} {\cal G}^{\mu\nu}_a
{\cal G}^{\rho\sigma}_a,\quad \epsilon_{0123}=1.
\end{eqnarray}

   We are now in the position to summarize the various (approximate)
symmetries of the strong interactions in combination with the corresponding
currents and their divergences.

\begin{itemize}
\item In the limit of massless quarks, the sixteen currents $L^\mu_a$
and $R^\mu_a$ or, alternatively, $V^\mu_a$ and $A^\mu_a$ are
conserved.
   The same is true for the singlet vector current $V^\mu$, whereas the
singlet axial-vector current $A^\mu$ has an anomaly.
\item For any value of quark masses, the individual flavor currents
$\bar{u}\gamma^\mu u$, $\bar{d}\gamma^\mu d$, and
$\bar{s}\gamma^\mu s$ are always conserved in the strong
interactions reflecting the flavor independence of the strong
coupling and the diagonality of the quark-mass matrix.
   Of course, the singlet vector current $V^\mu$, being the sum of
the three flavor currents, is always conserved.
\item In addition to the anomaly, the singlet axial-vector current
has an explicit divergence due to the quark masses.
\item For equal quark masses, $m_u=m_d=m_s$, the eight vector currents
$V^\mu_a$ are conserved, because $[\lambda_a,{\mathbbm 1}]=0$.
   Such a scenario is the origin of the SU(3) symmetry
originally proposed by Gell-Mann and Ne'eman
\cite{EightfoldWay:3:6}.
   The eight axial-vector currents $A^\mu_a$ are not conserved.
   The divergences of the octet axial-vector currents of Eq.\ (\ref{2:1:5:dsva})
are proportional to pseudoscalar quadratic forms.
   This can be interpreted as the microscopic origin of the
PCAC relation (partially conserved axial-vector current)
\cite{Gell-Mann:1964tf:3:6}, \cite{Adler:1968:3:6} which states that the
divergences of the axial-vector currents are proportional to
renormalized field operators representing the lowest-lying
pseudoscalar octet.
\item Taking $m_u=m_d\neq m_s$ reduces SU(3) flavor symmetry to SU(2) isospin symmetry.
\item Taking $m_u\neq m_d$ leads to isospin symmetry breaking.
\item Various symmetry-breaking patterns are discussed in great detail in
Ref.\ \cite{Pagels:1974se}.
\end{itemize}

\subsubsection{Green functions, chiral Ward identities, and generating functional}
\label{subsubsection_gfcwigf}
  For conserved currents, the spatial integrals of the
charge densities are time independent, i.e., in a quantized theory
the corresponding charge operators commute with the Hamilton operator.
   These operators are generators of infinitesimal transformations
on the Hilbert space of the theory.
   The mass eigenstates should organize themselves in degenerate multiplets
with dimensionalities corresponding to irreducible representations of the Lie
group in question.
   For the moment, we assume that the dynamical system described
by the Hamiltonian does not lead to a spontaneous symmetry breakdown.
   We will come back to this point later.
   Which irreducible representations ultimately appear, and what
the actual energy eigenvalues are, is determined by the dynamics of
the Hamiltonian.
   For example, SU(2) isospin symmetry of the strong interactions reflects
itself in degenerate SU(2) multiplets such as the nucleon doublet,
the pion triplet, and so on.
   Ultimately, the actual masses of the nucleon and the pion should
follow from QCD.

   It is also well-known that symmetries imply relations between $S$-matrix
elements. For example, applying the Wigner-Eckart theorem to pion-nucleon
scattering, assuming the strong-interaction Hamiltonian to be an isoscalar,
it is sufficient to consider two isospin amplitudes describing transitions
between states of total isospin $I=1/2$ or $I=3/2$.
   All the dynamical information is contained in these isospin amplitudes and
the results for physical processes can be expressed in terms of
these amplitudes together with geometrical coefficients, namely,
the Clebsch-Gordan coefficients.

   In quantum field theory, the objects of interest are the
Green functions which are vacuum expectation values of time-ordered
products.
   Later on, we will also refer to matrix elements
of time-ordered products between states other than the vacuum as
Green functions.
   The physical scattering amplitudes are obtained from the Green functions
using the reduction formalism \cite{Lehmann:1954rq}.
   Symmetries provide strong constraints not only for scattering amplitudes,
i.e.~their transformation behavior, but, more generally speaking, also for
Green functions and, in particular, {\em among} Green functions.
  Even if a symmetry is broken, i.e., the infinitesimal
generators are time dependent, conditions related to the
symmetry-breaking terms can still be obtained using equal-time commutation
relations.

   The symmetry currents relevant to the global
$\mbox{SU(3)}_L\times\mbox{SU(3)}_R\times\mbox{U(1)}_V$ of QCD are given
in Eqs.~(\ref{2:1:3:v}), (\ref{2:1:3:a}), and (\ref{2:1:3:sv}).
   Moreover, since we also want to discuss explicit symmetry breaking, we
introduce the scalar and pseudoscalar densities
\begin{equation}
\label{2:1:6:quadraticforms}
S_a=\bar{q}\lambda_a q,\quad
P_a=i\bar{q}\gamma_5 \lambda_a  q,\quad
a=0,\cdots, 8,
\end{equation}
where $\lambda_0=\sqrt{2/3}\,\mathbbm 1$.
   For example, linear combinations of $S_a$ and $P_a$ are needed to describe
the divergences of the currents in Eqs.~(\ref{2:1:5:dsva}).
  Whenever it is more convenient, we will also use
\begin{equation}
\label{2:1:6:SP}
S(x)=\bar{q}(x) q(x),\quad
P(x)=i\bar{q}(x)\gamma_5 q(x),
\end{equation}
instead of $S_0$ and $P_0$.

   For example, the following Green functions of the
``vacuum'' sector
\begin{eqnarray*}
\langle 0|T[A^\mu_a(x) P_b(y)]|0\rangle,\\
\langle 0|T[P_a(x)J^\mu(y) P_c(z)]|0\rangle,\\
\langle 0|T[P_a(w)P_b(x)P_c(y)P_d(z)]|0\rangle
\end{eqnarray*}
are related to pion decay, the pion electromagnetic form factor ($J^\mu$
is the electromagnetic current), and
pion-pion scattering, respectively.
  One may also consider similar time-ordered products
evaluated between a single nucleon in the initial and final states
in addition to the vacuum Green functions.
   This allows one to discuss properties of the nucleon as well as
dynamical processes involving a single nucleon, such as
\begin{eqnarray*}
\langle N|J^\mu(x)|N\rangle&\leftrightarrow&\mbox{nucleon electromagnetic form factors},\\
\langle N|A^\mu_a(x)|N\rangle&\leftrightarrow&\mbox{axial form factor + induced pseudoscalar form factor},\\
\langle N|T[J^\mu(x) J^\nu(y)]|N\rangle&\leftrightarrow&\mbox{Compton scattering},\\
\langle N|T[J^\mu(x) P_a(y)]|N\rangle&\leftrightarrow&\mbox{pion electroproduction}.
\end{eqnarray*}

   Generally speaking, a chiral Ward identity relates the divergence
of a Green function containing at least one factor of $V^{\mu}_a$ or
$A^{\mu}_a$ to some linear combination of other Green functions.
   The terminology {\em chiral} refers to the underlying $\mbox{SU(3)}_L\times
\mbox{SU(3)}_R$ group.
   To make this statement more precise, let us consider as a simple example
the two-point Green function involving an axial-vector current and a
pseudoscalar density,
\begin{equation}
\label{2:1:6:gfaav}
G^\mu_{AP,ab}(x,y)=\langle 0| T[A^\mu_a(x) P_b(y)]|0\rangle
=\Theta(x_0-y_0)\langle 0|A^\mu_a(x) P_b(y)|0\rangle
+\Theta(y_0-x_0)\langle 0|P_b(y) A^\mu_a(x)|0\rangle,
\end{equation}
   and evaluate the divergence
\begin{eqnarray*}
\partial_\mu^x G^{\mu}_{AP,ab}(x,y)
&=&
 \partial_\mu^x [\Theta(x_0-y_0)\langle 0| A^\mu_a(x) P_b(y)|0\rangle
+\Theta(y_0-x_0)\langle 0|  P_b(y)A^\mu_a(x)|0\rangle ]\\
&=&\delta(x_0-y_0)\langle 0| A_0^a(x) P_b(y)|0\rangle
-\delta(x_0-y_0)\langle 0|  P_b(y)A_0^a (x)|0\rangle\\
&&
+\Theta(x_0-y_0)\langle 0|\partial_\mu^x A^\mu_a(x) P_b(y)|0\rangle
+\Theta(y_0-x_0)\langle 0| P_b(y)\partial_\mu^x A^\mu_a(x)|0\rangle\\
&=&\delta(x_0-y_0)\langle 0|[A^a_0(x),P_b(y)]|0\rangle
+\langle 0|T[\partial_\mu^x A^\mu_a(x) P_b(y)]|0\rangle,
\end{eqnarray*}
   where we made use of $\partial_\mu^x \Theta(x_0-y_0)=\delta(x_0-y_0)
g_{0\mu}=-\partial_\mu^x \Theta(y_0-x_0)$.
   This simple example already shows the main features of (chiral) Ward
identities.
   From the differentiation of the theta functions
one obtains equal-time commutators between a charge density and the
remaining quadratic forms.
   The results of such commutators are a reflection of the underlying symmetry.
   As a second term, one obtains the divergence of the current operator in
question.
   If the symmetry is perfect, such terms vanish identically.
   If the symmetry is only approximate, an additional term involving the
symmetry breaking appears.
   For a soft breaking such a divergence can be treated as a perturbation.

   The time ordering of $n+1$ points $x,x_1,\cdots,x_n$ gives rise to $(n+1)$! distinct orderings, each
involving products of $n$ theta functions.
   Via induction, the generalization of the above simple example to an
$(n+1)$-point Green function is symbolically of the form
\begin{eqnarray}
\label{2:1:6:gendmug}
\lefteqn{\partial_\mu^x \langle 0|T\{J^\mu(x) A_1(x_1)\cdots A_n(x_n)
\}|0\rangle=}\nonumber\\
&&\langle 0|T\{[\partial_\mu^x J^\mu(x)]
A_1(x_1)\cdots A_n(x_n)\}|0\rangle\nonumber\\
&&+\delta(x^0-x_1^0)\langle 0|T\{[J_0(x),A_1(x_1)] A_2(x_2)\cdots A_n(x_n)\}|
0\rangle\nonumber\\
&&+\delta(x^0-x_2^0)\langle 0|T\{A_1(x_1)[J_0(x),A_2(x_2)]
\cdots A_n(x_n)\}|0\rangle\nonumber\\
&&+\cdots+\delta(x^0-x_n^0)
\langle 0|T\{A_1(x_1)\cdots [J_0(x),A_n(x_n)]\}|0\rangle,
\end{eqnarray}
where $J^\mu$ stands generically for any of the Noether currents.

  The discussion so far assumes that one explicitly works out the
particular chiral Ward identity one is interested in.
   However, there is an elegant way of obtaining {\em all} chiral Ward
identities from a single expression.
   To that end we introduce into the Lagrangian of QCD the couplings of the
nine vector currents, eight axial-vector currents, nine scalar quark densities,
and nine pseudoscalar quark densities
to external c-number fields \cite{Gasser:1984gg}:
\begin{equation}
\label{2:1:6:lQCD+ext}
{\cal L}={\cal L}^0_{\rm QCD}+{\cal L}_{\rm ext},
\end{equation}
where
\begin{eqnarray}
\label{2:1:6:lqcds}
{\cal L}_{\rm ext}&=&
\sum_{a=1}^8v^\mu_a\, \bar q\gamma_\mu\frac{\lambda_a}{2} q
+v^\mu_{(s)}\, \frac{1}{3}\bar q\gamma_\mu q
+ \sum_{a=1}^8a^\mu_a\,\bar q\gamma_\mu\gamma_5\frac{\lambda_a}{2} q
-\sum_{a=0}^8 s_a\, \bar q\lambda_a q
+\sum_{a=0}^8 p_a\, i\bar q\gamma_5\lambda_a q
\nonumber\\
&=&\bar{q}\gamma_\mu (v^\mu
+\frac{1}{3}v^\mu_{(s)} +\gamma_5 a^\mu )q -\bar{q}(s-i\gamma_5 p)q.
\end{eqnarray}
   The 35 real functions $v^\mu_a(x)$, $v^\mu_{(s)}(x)$, $a^\mu_a(x)$,
$s_a(x)$, and $p_a(x)$, will collectively be denoted by $[v,a,s,p]$.
   A precursor of this method was already used by Bell and Jackiw \cite{Bell:1969ts}
in their discussion of the anomalous divergences in the $\pi^0\to\gamma\gamma$ decay.
   The Green functions of the vacuum sector may be combined in the generating functional
\begin{equation}
\label{2:1:6:gfgfvs}
\exp(i Z[v,a,s,p]) =\langle 0|T\exp\left[i\int \mbox{d}^4 x\, {\cal
L}_{\rm ext}(x) \right]|0\rangle_0.
\end{equation}
   Note that both the quark field operators $q$ in ${\cal L}_{\rm ext}$ and the ground
state $|0\rangle$ refer to the chiral limit, indicated by the subscript 0 in Eq.~(\ref{2:1:6:gfgfvs}).
   A particular Green function is then obtained through a functional derivative with
respect to the external fields.
   As an example, suppose we are interested in the scalar $u$-quark condensate in the chiral
limit, $\langle 0| \bar{u}u|0\rangle_0$.
   We express $\bar u u$ as
\begin{displaymath}
\bar u u=\frac{1}{2}\sqrt{\frac{2}{3}}\bar q\lambda_0 q
+\frac{1}{2}\bar q\lambda_3 q +\frac{1}{2}\frac{1}{\sqrt{3}} \bar
q\lambda_8 q
\end{displaymath}
and obtain
\begin{displaymath}
\langle 0|\bar{u}(x) u(x)|0\rangle_0 = \left.\frac{i}{2}\left[\sqrt{\frac{2}{3}}\frac{\delta}{\delta
s_0(x)} +\frac{\delta}{\delta s_3(x)} +\frac{1}{\sqrt{3}}
\frac{\delta}{\delta s_8(x)}\right]
\exp(iZ[v,a,s,p])\right|_{v=a=s=p=0}.
\end{displaymath}

   From the generating functional,  we can even obtain Green functions of the
``real world,'' where the  quark fields and the ground state are those with
finite quark masses.
   For example, the two-point function of two
axial-vector currents of the ``real world,'' i.e., for
$s=\mbox{diag}(m_u,m_d,m_s)$, and the ``true vacuum'' $|0\rangle$,
is given by
\begin{equation}
\label{2:1:6:grw}
\langle 0|T[A^\mu_a(x) A^\nu_b(0)]|0\rangle =\left.
(-i)^2 \frac{\delta^2}{\delta a_{a\,\mu}(x)\delta
a_{b\,\nu}(0)}
\exp(iZ[v,a,s,p])\right|_{v=a=p=0,s=\mbox{diag}(m_u,m_d,m_s)}.
\end{equation}
   Note that the left-hand side involves the quark fields and the ground state of the
``real world,'' whereas the right-hand side is the generating functional defined in
terms of the quark fields and the ground state of the chiral limit.
   The actual value of the generating functional for a given
configuration of external fields $v$, $a$, $s$, and $p$ reflects the
dynamics generated by the QCD Lagrangian.

   The (infinite) set of {\em all} chiral Ward identities resides in an
invariance of the generating functional under a {\em local}
transformation of the external fields \cite{Gasser:1983yg}, \cite{Leutwyler:1993iq}.
   The use of local transformations allows one to also consider divergences
of Green functions.
   We require $\cal L$ of Eq.~(\ref{2:1:6:lQCD+ext}) to be a Hermitian Lorentz scalar, to be
even under $P$, $C$, and $T$, and to be invariant under {\em local} chiral transformations.
  In fact, it is sufficient to consider $P$ and $C$, only, because
$T$ is then automatically incorporated owing to the $CPT$ theorem.

   Under parity, the quark fields transform as
\begin{equation}
\label{2:1:6:qtrafop}
q_f(t,\vec{x})\stackrel{\mbox{$P$}}{\mapsto}\gamma_0 q_f(t,-\vec{x}),
\end{equation}
   and the requirement of parity conservation,
\begin{equation}
\label{2:1:6:parinv}
{\cal L}(t,\vec{x}) \stackrel{\mbox{$P$}}{\mapsto} {\cal L}(t,-\vec{x}),
\end{equation}
leads, using the results of Table \ref{2:1:6:parity}, to the following
constraints for the external fields,
\begin{equation}
\label{2:1:6:eftrafop}
v^\mu\stackrel{\mbox{$P$}}{\mapsto}v_\mu,\quad
v^\mu_{(s)}\stackrel{\mbox{$P$}}{\mapsto}v_\mu^{(s)},\quad
a^\mu\stackrel{\mbox{$P$}}{\mapsto}-a_\mu,\quad
s\stackrel{\mbox{$P$}}{\mapsto}s,\quad
p\stackrel{\mbox{$P$}}{\mapsto}-p.
\end{equation}
   In Eq.\ (\ref{2:1:6:eftrafop}) it is understood that the arguments change
from $(t,\vec{x})$ to $(t,-\vec{x})$.

\begin{table}[t]
\begin{center}
\renewcommand{\arraystretch}{1.5}
\begin{tabular}{|c|c|c|c|c|c|}
\hline $\Gamma$& $\mathbbm 1$& $\gamma^\mu$& $\sigma^{\mu\nu}$& $\gamma_5$&
$\gamma^\mu\gamma_5$\\
\hline $\gamma_0 \Gamma \gamma_0$& $\mathbbm 1$& $\gamma_\mu$&
$\sigma_{\mu\nu}$& $-\gamma_5$& $-\gamma_\mu\gamma_5$
\\
\hline
\end{tabular}
\end{center}
\caption{\label{2:1:6:parity} Transformation properties of the Dirac
matrices $\Gamma$ under parity.}
\end{table}

   Similarly, under charge conjugation the quark fields transform as
\begin{equation}
\label{2:1:6:qtrafc}
q_{f,\alpha}\stackrel{\mbox{$C$}}{\mapsto}C_{\alpha\beta}\bar{q}_{f,\beta},
\quad
\bar{q}_{f,\alpha}\stackrel{\mbox{$C$}}{\mapsto}
-q_{f,\beta}C^{-1}_{\beta\alpha},
\end{equation}
   where the subscripts $\alpha$ and $\beta$ are Dirac
spinor indices,
\begin{displaymath}
C=i\gamma^2\gamma^0
=\left(\begin{array}{cccc}0&0&0&-1\\0&0&1&0\\
0&-1&0&0\\
1&0&0&0
\end{array}\right)
\end{displaymath}
is the usual charge conjugation matrix, and $f$ refers to flavor.
   Taking Fermi statistics into account, one obtains
\begin{displaymath}
\bar{q}\,\Gamma F q=-\bar{q}\, C \Gamma^T C F^T q,
\end{displaymath}
where $F$ denotes a matrix in flavor space.
   In combination with Table
\ref{2:1:6:chargeconjugation} it is straightforward to show that invariance
of ${\cal L}_{\rm ext}$ under charge conjugation requires the transformation
properties
\begin{equation}
\label{2:1:6:eftrafoc}
v_\mu\stackrel{C}{\rightarrow}-v_\mu^T,\quad
v_\mu^{(s)}\stackrel{C}{\rightarrow}-v_\mu^{(s)T},\quad
a_\mu\stackrel{C}{\rightarrow}a_\mu^T,\quad
s,p\stackrel{C}{\rightarrow}s^T,p^T,
\end{equation}
where the transposition refers to the flavor space.

\begin{table}[t]
\begin{center}
\renewcommand{\arraystretch}{1.5}
\begin{tabular}{|c|c|c|c|c|c|}
\hline
$\Gamma$&$\mathbbm 1$&$\gamma^\mu$&$\sigma^{\mu\nu}$&$\gamma_5$&$\gamma^\mu\gamma_5$\\
\hline
$-C\Gamma^TC
$&$\mathbbm 1$&$-\gamma^\mu$&$-\sigma^{\mu\nu}$&$\gamma_5$&$\gamma^\mu\gamma_5$
\\
\hline
\end{tabular}
\end{center}
\caption{\label{2:1:6:chargeconjugation}
Transformation properties of the Dirac matrices $\Gamma$
under charge conjugation.}
\end{table}

   Finally, we need to discuss the requirements to be met by the external
fields under local $\mbox{SU(3)}_L\times\mbox{SU(3)}_R\times\mbox{U}(1)_V$
transformations.
   In a first step, we write Eq.\ (\ref{2:1:6:lqcds}) in terms of the
left- and right-handed quark fields.
   Using the projection operators of Eq.~(\ref{2:1:2:prpl}) the Lagrangian
of Eq.~(\ref{2:1:6:lqcds}) reads
\begin{equation}
\label{2:1:6:lqcdsn}
{\cal L}={\cal L}_{\rm QCD}^0
+\bar{q}_L\gamma^\mu\left(l_\mu+\frac{1}{3}v^{(s)}_\mu\right)q_L
+\bar{q}_R\gamma^\mu\left(r_\mu+\frac{1}{3}v^{(s)}_\mu\right)q_R-\bar{q}_R(s+ip)q_L-\bar{q}_L(s-ip)q_R.
\end{equation}
   Equation (\ref{2:1:6:lqcdsn}) remains invariant under {\em local}
transformations
\begin{eqnarray}
\label{2:1:6:qrl}
q_R&\mapsto&\exp\left(-i\frac{\Theta(x)}{3}\right) V_R(x) q_R,\nonumber\\
q_L&\mapsto&\exp\left(-i\frac{\Theta(x)}{3}\right) V_L(x) q_L,
\end{eqnarray}
where $V_R(x)$ and $V_L(x)$ are independent space-time-dependent SU(3)
matrices, provided the external fields are subject
to the transformations
\begin{eqnarray}
\label{2:1:6:sg}
r_\mu&\mapsto& V_R r_\mu V_R^{\dagger}
-i \partial_\mu V_R V_R^{\dagger},\nonumber\\
l_\mu&\mapsto& V_L l_\mu V_L^{\dagger}
-i\partial_\mu V_L V_L^{\dagger},
\nonumber\\
v_\mu^{(s)}&\mapsto&v_\mu^{(s)}-\partial_\mu\Theta,\nonumber\\
s+ip&\mapsto& V_R(s+ip)V_L^{\dagger},\nonumber\\
s-ip&\mapsto& V_L(s-ip)V_R^{\dagger}.
\end{eqnarray}
   The derivative terms in Eq.\ (\ref{2:1:6:sg}) serve the same purpose as
in the construction of gauge theories, i.e., they cancel analogous
terms originating from the kinetic part of the quark Lagrangian.
   Note that the external currents are coupled with an ``opposite'' sign in comparison
with our convention for gauge theories.

   There is another, yet, more practical aspect of the local invariance,
namely: such a procedure allows one to also discuss a coupling to external
gauge fields in the transition to the effective theory to be discussed later.
   For example, a coupling of the electromagnetic field to point-like
fundamental particles results from
gauging a U(1) symmetry.
  Here, the corresponding U(1) group is to be understood
as a subgroup of a local $\mbox{SU(3)}_L\times\mbox{SU(3)}_R$.
   Another example deals with the interaction of the light quarks
with the charged and neutral gauge bosons of the weak interactions.

   Let us consider both examples explicitly. The coupling of quarks
to an external electromagnetic field ${\cal A}_\mu$ is given by
\begin{equation}
\label{2:1:6:rla}
r_\mu=l_\mu=-e Q {\cal A}_\mu,
\end{equation}
where $Q=\mbox{diag}(2/3,-1/3,-1/3)$ is the quark charge matrix and $e>0$ the
elementary charge:
\begin{displaymath}
{\cal L}_{\rm ext}=-e {\cal A}_\mu(\bar{q}_L Q\gamma^\mu q_L
+\bar{q}_R Q \gamma^\mu q_R)
=-e {\cal A}_\mu \bar{q}Q\gamma^\mu q
=-e {\cal A}_\mu\left(\frac{2}{3}\bar{u}\gamma^\mu u
-\frac{1}{3} \bar{d}\gamma^\mu d -\frac{1}{3}\bar{s}\gamma^\mu s\right).
\end{displaymath}
   On the other hand, if one considers only the two-flavor version of QCD one
has to insert for the external fields
\begin{equation}
\label{2:1:6:rlasu2}
r_\mu=l_\mu=-e\frac{\tau_3}{2}{\cal A}_\mu,\quad
v_\mu^{(s)}=-\frac{e}{2}{\cal A}_\mu.
\end{equation}

   In the description of semi-leptonic interactions such as
$\pi^-\to \mu^-\bar{\nu}_\mu$,  $\pi^-\to\pi^0e^-\bar{\nu}_e$, or
neutron decay $n\to p e^-\bar{\nu}_e$ one needs the interaction of quarks with
the massive charged weak bosons
${\cal W}^\pm_\mu=({\cal W}_{1\mu}\mp i {\cal W}_{2\mu})/\sqrt{2}$,
\begin{equation}
\label{2:1:6:rlw} r_\mu=0,\quad l_\mu=-\frac{g}{\sqrt{2}} ({\cal W}^+_\mu T_+ +
\mbox{H.c.}),
\end{equation}
where H.c.~refers to the Hermitian conjugate and
$$
T_+=\left(\begin{array}{rrr}0&V_{ud}&V_{us}\\0&0&0\\0&0&0\end{array}\right).
$$
   Here, $V_{ij}$ denote the elements of the
Cabibbo-Kobayashi-Maskawa quark-mixing matrix describing the
transformation between the mass eigenstates of QCD and the weak
eigenstates \cite{PDG_2008},
$$|V_{ud}|=0.97418\pm 0.00027,\quad
|V_{us}|=0.2255\pm 0.0019.
$$
   At lowest order in perturbation theory, the Fermi constant is related
to the gauge coupling $g$ and the $W$ mass as
\begin{equation}
\label{2:1:6:GF}
G_F=\sqrt{2} \frac{g^2}{8 M^2_W}=1.16637(1)\times 10^{-5}\,\mbox{GeV}^{-2}.
\end{equation}
   Making use of
\begin{displaymath}
\bar{q}_L\gamma^\mu {\cal W}_\mu^+ T_+ q_L
=\frac{1}{2}{\cal W}_\mu^+[V_{ud}\bar{u}\gamma^\mu(1-\gamma_5)d
+V_{us}\bar{u}\gamma^\mu(1-\gamma_5)s],
\end{displaymath}
   we see that inserting Eq.\ (\ref{2:1:6:rlw}) into Eq.\ (\ref{2:1:6:lqcdsn})
leads to the standard charged-current weak interaction in the light-quark sector,
\begin{eqnarray*}
{\cal L}_{\rm ext}&=&-\frac{g}{2\sqrt{2}}\left\{{\cal W}^+_\mu[
V_{ud}\bar{u}\gamma^\mu(1-\gamma_5)d+V_{us}\bar{u}\gamma^\mu(1-\gamma_5)s]
+\mbox{H.c.}\right\}.
\end{eqnarray*}

   The situation is slightly different for the neutral weak interaction.
Here, the three-flavor version requires a coupling to the singlet axial-vector
current which, because of the anomaly of Eq.\ (\ref{2:1:5:dsva}), we have
dropped from our discussion.
   On the other hand, in the two-flavor version the axial-vector current part
is traceless and we have
\begin{eqnarray}
\label{2:1:6:rlz}
r_\mu&=&e \tan(\theta_W) \frac{\tau_3}{2} {\cal Z}_\mu,\nonumber\\
l_\mu&=&-\frac{g}{\cos(\theta_W)}\frac{\tau_3}{2} {\cal Z}_\mu+
e \tan(\theta_W) \frac{\tau_3}{2} {\cal Z}_\mu,
\nonumber\\
v_\mu^{(s)}&=&\frac{e\tan(\theta_W)}{2}{\cal Z}_\mu,
\end{eqnarray}
where  $\theta_W$ is the weak angle.
   With these external fields, we obtain the standard weak neutral-current
interaction
\begin{displaymath}
{\cal L}_{\rm ext}=-\frac{g}{2\cos(\theta_W)}{\cal Z}_\mu\left(
\bar{u}\gamma^\mu\left\{\left[\frac{1}{2}-\frac{4}{3}\sin^2(\theta_W)\right]{\mathbbm 1}
-\frac{1}{2}\gamma_5\right\}u+\bar{d}\gamma^\mu\left\{\left[-\frac{1}{2}
+\frac{2}{3}\sin^2(\theta_W)\right]{\mathbbm 1}
+\frac{1}{2}\gamma_5\right\}d\right),
\end{displaymath}
where we made use of $e=g\sin(\theta_W)$.

\subsubsection{PCAC in the presence of an external electromagnetic field}
\label{subsec_pcacpeef}
   Finally, the technique of coupling the QCD Lagrangian to external fields
also allows us to determine the current divergences for rigid external fields,
i.e., fields which are {\em not} simultaneously transformed.
   For the sake of simplicity we restrict ourselves to the two-flavor sector.
   (The generalization to the three-flavor case is straightforward.)

   Consider a {\em global} chiral transformation only and assume that
the external fields are {\em not} simultaneously transformed.
   In this case the divergences of the currents
read \cite{Fuchs:2003vw}
\begin{eqnarray}
\label{2:1:7:divv}
\partial_\mu V^\mu_i&=&i\bar{q}\gamma^\mu[\frac{\tau_i}{2},v_\mu]q
+i\bar{q}\gamma^\mu\gamma_5[\frac{\tau_i}{2},a_\mu]q
-i\bar{q}[\frac{\tau_i}{2},s]q-\bar{q}\gamma_5[\frac{\tau_i}{2},p]q,\\
\label{2:1:7:diva}
\partial_\mu A^\mu_i&=&i\bar{q}\gamma^\mu\gamma_5[\frac{\tau_i}{2},v_\mu]q
+i\bar{q}\gamma^\mu[\frac{\tau_i}{2},a_\mu]q
+i\bar{q}\gamma_5\{\frac{\tau_i}{2},s\}q
+\bar{q}\{\frac{\tau_i}{2},p\}q.
\end{eqnarray}

   As an example, let us consider the QCD Lagrangian for a finite light quark
mass $\hat{m}=m_u=m_d$ in combination with a coupling to an external
electromagnetic field ${\cal A}_\mu$ [see Eq.\ (\ref{2:1:6:rlasu2}),
$a_\mu=0=p$].
   The expressions for the divergence of the vector and
axial-vector currents, respectively, are given by \cite{Fuchs:2003vw}
\begin{eqnarray}
\label{2:1:7:divvsc}
\partial_\mu V^\mu_i&=&-\epsilon_{3ij}e{\cal A}_\mu \bar{q}\gamma^\mu
\frac{\tau_j}{2}q=-\epsilon_{3ij}e{\cal A}_\mu V^\mu_j,\\
\label{2:1:7:divasc}
\partial_\mu A^\mu_i
&=&-e {\cal A}_\mu \epsilon_{3ij} \bar{q}\gamma^\mu
\gamma_5 \frac{\tau_j}{2} q+2\hat m i\bar{q}\gamma_5 \frac{\tau_i}{2}q
=-e {\cal A}_\mu \epsilon_{3ij} A^\mu_j+\hat m P_i,
\end{eqnarray}
with the isovector pseudoscalar density
$P_i=i\bar{q}\gamma_5 \tau_i q$.
   In fact, Eq.\ (\ref{2:1:7:divasc}) is incomplete, because the third
component of the axial-vector current, $A^\mu_3$, has an anomaly
which is related to the decay $\pi^0\to\gamma\gamma$.
   The full equation reads
\begin{equation}
\label{2:1:7:divascfull}
\partial_\mu A^\mu_i
=\hat m P_i-e {\cal A}_\mu \epsilon_{3ij} A^\mu_j
+\delta_{i3}
\frac{e^2}{32\pi^2}\epsilon_{\mu\nu\rho\sigma}{\cal F}^{\mu\nu}
{\cal F}^{\rho\sigma}, \quad \epsilon_{0123}=1,
\end{equation}
where ${\cal F}_{\mu\nu}=\partial_\mu{\cal A}_\nu-\partial_\nu{\cal A}_\mu$ is
the electromagnetic field strength tensor.

   We emphasize the formal similarity of Eq.\ (\ref{2:1:7:divasc}) to the
(pre-QCD) PCAC (Partially Conserved Axial-Vector Current) relation
obtained by Adler \cite{Adler:1965:5:5} through the inclusion of the
electromagnetic interactions with minimal electromagnetic coupling.
  Since in QCD the quarks are taken as truly elementary, their interaction
with an (external) electromagnetic field is of such a minimal type.
   In Adler's version, the right-hand side of Eq.\ (\ref{2:1:7:divascfull})
contains a renormalized field operator creating and destroying pions instead of
$\hat m P_i$.
   From a modern point of view, the combination $\hat m P_i/(M_\pi^2 F_\pi)$
serves as an interpolating pion field.
    Furthermore, the anomaly term is not yet present in Ref.\
\cite{Adler:1965:5:5}.

\subsection{Spontaneous symmetry breaking and Goldstone theorem}
\label{sec_ssbqcd}
\subsubsection{Linear sigma model}
   Spontaneous symmetry breaking occurs if the ground state has a lower symmetry
than the Hamiltonian.
   For example, in the linear sigma model \cite{Schwinger:1957em}, \cite{Gell-Mann:1960np},
the Lagrangian is constructed in terms of the O(4) multiplet $(\sigma,\pi_1,\pi_2,\pi_3)$,
\begin{equation}
\label{2:2:1:lsm}
{\cal L}=\frac{1}{2}\left(\partial_\mu\sigma\partial^\mu\sigma+
\partial_\mu\vec\pi\cdot\partial^\mu\vec\pi\right)
-\frac{m^2}{2}(\sigma^2+\vec\pi^2)-\frac{\lambda}{4}(\sigma^2+\vec\pi^2)^2,
\end{equation}
where $\lambda>0$.
   Under parity we assume $\sigma(t,\vec x)\mapsto \sigma(t,-\vec x)$ and
$\pi_i(t,\vec x)\mapsto -\pi_i(t,-\vec x)$.
   The Lagrangian is invariant under the infinitesimal transformations
\begin{displaymath}
\left(\begin{array}{c} \sigma\\ \pi_1 \\ \pi_2\\ \pi_3\end{array}\right)
\mapsto
\left(\begin{array}{c} \sigma'\\ \pi'_1\\ \pi'_2\\ \pi'_3\end{array}\right)
=\left({\mathbbm 1}-i\sum_{a=1}^6\epsilon_a T_a\right)
\left(\begin{array}{c} \sigma\\ \pi_1 \\ \pi_2\\ \pi_3\end{array}\right),
\end{displaymath}
where the six
$4\times 4$ matrices are given by
\begin{eqnarray*}
&&T_1=\left(\begin{array}{rrrr}0&0&0&0\\0&0&0&0\\0&0&0&-i\\0&0&i&0
\end{array}\right),\quad\,
T_2=\left(\begin{array}{rrrr}0&0&0&0\\0&0&0&i\\0&0&0&0\\0&-i&0&0
\end{array}\right),\quad\,
T_3=\left(\begin{array}{rrrr}0&0&0&0\\0&0&-i&0\\0&i&0&0\\0&0&0&0
\end{array}\right),\\
&&T_4=\left(\begin{array}{rrrr}0&-i&0&0\\i&0&0&0\\0&0&0&0\\0&0&0&0
\end{array}\right),\quad\,
T_5=\left(\begin{array}{rrrr}0&0&-i&0\\0&0&0&0\\i&0&0&0\\0&0&0&0
\end{array}\right),\quad\,
T_6=\left(\begin{array}{rrrr}0&0&0&-i\\0&0&0&0\\0&0&0&0\\i&0&0&0
\end{array}\right).
\end{eqnarray*}
   The linear combinations $R_i=(T_i+T_{i+3})/2$ and $L_i=(T_i-T_{i+3})/2$, $i=1,2,3$,
satisfy the commutation relations corresponding to an SU(2)$\times$SU(2) Lie group.
The  multiplet $(\sigma,\pi_1,\pi_2,\pi_3)$ transforms according to the $(\frac{1}{2},\frac{1}{2})$
representation.
   By choosing $m^2<0$, the symmetry is realized in
the Nambu-Goldstone mode \cite{Nambu:xd:1:1}, \cite{Goldstone:eq:1:1}.
   Let us assume that the ground state is characterized by the vacuum expectation values
\begin{equation}
\label{2:2:1:groundstate}
\langle \sigma\rangle = v\equiv -\sqrt{-\frac{m^2}{\lambda}},\quad \langle\pi_i\rangle=0.
\end{equation}
   Introducing $\sigma=v+\sigma'$, the Lagrangian reads
\begin{equation}
\label{2:2:1:lsmp}
{\cal L}=\frac{1}{2}\left(\partial_\mu\sigma'\partial^\mu\sigma'+
\partial_\mu\vec\pi\cdot\partial^\mu\vec\pi\right)
+\frac{1}{2}(-2m^2)\sigma'^2
+\lambda v\sigma' (\pi_1^2+\pi_2^2+\pi_3^2+\sigma'^2)
+\frac{\lambda}{4}(\pi_1^2+\pi_2^2+\pi_3^2+\sigma'^2)^2-\frac{\lambda}{4}v^4.
\end{equation}
   The ground-state configuration is no longer invariant under the full group
O(4).
   While the generators $T_1$, $T_2$, and $T_3$ annihilate the ground state of
Eq.~(\ref{2:2:1:groundstate}), the generators $T_4$, $T_5$, and $T_6$ do not.
   The model-independent feature of the above example is given by the
fact that for each of the three generators $T_4$, $T_5$, and $T_6$
which do not annihilate the ground state one obtains a {\em massless} Goldstone
boson.
   This is why Eq.~(\ref{2:2:1:lsmp}) contains no mass terms for the pions.
   In fact, the number of Goldstone bosons is determined by the structure of the
symmetry groups \cite{Goldstone:es:1:1}.
   Let $G$ denote the symmetry group of the Lagrangian with $n_G$ generators
and the subgroup $H$ the symmetry group of the ground state with $n_H$ generators.
   For each generator which does not annihilate the vacuum one obtains
a massless Goldstone boson, i.e., the total number of
Goldstone bosons equals $n_G-n_H$.

   The Lagrangians used in {\em motivating} the phenomenon
of a spontaneous symmetry breakdown are typically constructed in such a
fashion that the degeneracy of the ground states is built into
the potential at the classical level (the prototype being the ``Mexican hat''
potential).
   As in the above case, it is then argued that an
{\em elementary} Hermitian field of a multiplet transforming non-trivially
under the symmetry group $G$ acquires a vacuum expectation
value signaling a spontaneous symmetry breakdown.
   However, there also exist theories such as QCD where one cannot infer
from inspection of the Lagrangian whether the theory exhibits spontaneous
symmetry breaking.
   Rather, the criterion for spontaneous symmetry breaking is a non-vanishing
vacuum expectation value of some Hermitian operator, not an elementary field,
which emerges through the dynamics of the underlying theory.
   In particular, we will see that the quantities developing a vacuum
expectation value may also be local Hermitian operators composed of more
fundamental degrees of freedom of the theory.

   While the model of Eq.~(\ref{2:2:1:lsm}) is constructed to illustrate the concept
of a spontaneous symmetry breaking, it is not fully understood
theoretically why QCD should exhibit this phenomenon.
   We will first motivate why experimental input, the hadron spectrum
of the ``real'' world, indicates that spontaneous symmetry breaking
happens in QCD.
   Secondly, we will show that a non-vanishing singlet scalar quark
condensate is a sufficient condition for a spontaneous symmetry breaking in
QCD.

\subsubsection{The hadron spectrum}
\label{subsec_hs}
   We saw in Section \ref{subsubsec_agsl} that the QCD Lagrangian possesses
an $\mbox{SU(3)}_L\times\mbox{SU(3)}_R\times \mbox{U(1)}_V$
symmetry in the chiral limit in which the light quark masses
vanish.
   From symmetry considerations involving the Hamiltonian $H^0_{\rm QCD}$
only, one would naively expect that hadrons organize themselves into
approximately degenerate multiplets fitting the dimensionalities
of irreducible representations of the group
$\mbox{SU(3)}_L\times\mbox{SU(3)}_R\times\mbox{U(1)}_V$.
   The $\mbox{U(1)}_V$ symmetry results in
baryon number conservation and leads to a classification of
hadrons into mesons ($B=0$) and baryons ($B=1$).
   The linear combinations $Q_{aV}=Q_{aR}+Q_{aL}$ and $Q_{aA}=Q_{aR}-Q_{aL}$
of the left- and right-handed charge operators commute with
$H^0_{\rm QCD}$, have opposite parity, and thus for states of positive
parity one would expect the existence of degenerate states of negative
parity (parity doubling) which can be seen as follows.

   Let $|\alpha,+\rangle$ denote an eigenstate of $H^0_{\rm QCD}$ and parity
with eigenvalues $E_\alpha$ and $+1$, respectively,
\begin{eqnarray*}
H^0_{\rm QCD}|\alpha,+\rangle&=&E_\alpha|\alpha,+\rangle,\\
P|\alpha,+\rangle&=&|\alpha,+\rangle,
\end{eqnarray*}
such as, e.g., a member of the ground-state baryon octet (in the chiral limit).
   Defining $|\phi_{a\alpha}\rangle= Q_{aA}|\alpha,+\rangle$, because of
$[H^0_{\rm QCD},Q_{aA}]=0$, we have
\begin{eqnarray*}
H^0_{\rm QCD}|\phi_{a\alpha}\rangle
&=&H^0_{\rm QCD} Q_{aA}|\alpha,+\rangle
= Q_{aA} H^0_{\rm QCD}|\alpha,+\rangle
= E_\alpha Q_{aA}|\alpha,+\rangle
= E_\alpha |\phi_{a\alpha}\rangle,\\
P|\phi_{a\alpha}\rangle&=& PQ_{aA} P^{-1} P|\alpha,+\rangle=-Q_{aA}(+|\alpha,+\rangle)
=-|\phi_{a\alpha}\rangle.
\end{eqnarray*}
   The state $|\phi_{a\alpha}\rangle$ can be expanded in terms of the members of a
multiplet with negative parity,
\begin{displaymath}
|\phi_{a\alpha}\rangle=Q_{aA}|\alpha,+\rangle=|\beta,-\rangle\langle\beta,-|
Q_{aA}|\alpha,+\rangle={t_a}_{\beta\alpha}|\beta,-\rangle.
\end{displaymath}
   However, the low-energy spectrum of baryons does not contain a degenerate
baryon octet of negative parity.
   Naturally the question arises whether the above chain of arguments is
incomplete.
   Indeed, we have tacitly assumed that the ground state of QCD is annihilated
by the generators $Q_{aA}$.
   Let $b^\dagger_{\alpha+}$ denote an operator creating quanta with the quantum numbers
of the state $|\alpha,+\rangle$.
   Similarly, let $b^\dagger_{\alpha-}$ create degenerate quanta of opposite
parity.
   Expanding
$$
[Q_{aA},b^\dagger_{\alpha+}]= b^\dagger_{\beta-}{t_a}_{\beta\alpha},
$$
the usual chain of arguments then works as
\begin{eqnarray}
\label{2:2:2:pardoub}
Q_{aA}|\alpha,+\rangle=Q_{aA} b^\dagger_{\alpha+}|0\rangle
=\Big([Q_{aA},b^\dagger_{\alpha+}]+b_{\alpha+}^\dagger
\underbrace{Q_{aA}}_{
\mbox{$\hookrightarrow 0$}}\Big)|0\rangle
= {t_a}_{\beta\alpha} b_{\beta-}^\dagger |0\rangle.
\end{eqnarray}
   However, if the ground state is {\em not} annihilated by $Q_{aA}$, the
reasoning of Eq.\ (\ref{2:2:2:pardoub}) does no longer apply.
   In that case the ground state is not invariant under the full
symmetry group of the Lagrangian resulting in a spontaneous
symmetry breaking.
   In other words, the non-existence of degenerate multiplets
of opposite parity points to the fact that $\mbox{SU(3)}_V$ instead
of $\mbox{SU(3)}_L\times\mbox{SU(3)}_R$ is approximately realized
as a symmetry of the hadrons.
   Furthermore, the octet of the pseudoscalar mesons is special in
the sense that the masses of its members are small in comparison with
the corresponding $1^-$ vector mesons.
   They are the candidates for the Goldstone bosons of a spontaneous
symmetry breaking.

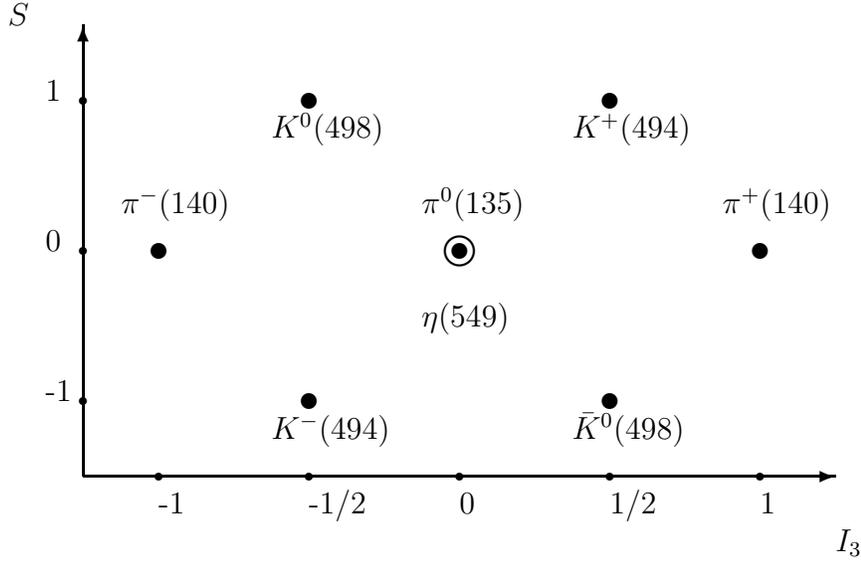
\begin{figure}[t]
\vspace{2em} \unitlength1cm
\begin{center}
\begin{picture}(10,6)
\thicklines \put(0,0){\vector(1,0){10}} \put(10,-1){$I_3$}
\put(0,0){\vector(0,1){6}} \put(-1,6){$S$} \put(3,1){\circle*{0.2}}
\put(2.5,0.5){$K^-(494)$} \put(7,1){\circle*{0.2}}
\put(6.5,0.5){$\bar{K}^0(498)$} \put(1,3){\circle*{0.2}}
\put(0.5,3.5){$\pi^-(140)$} \put(5,3){\circle*{0.2}}
\put(5,3){\circle{0.4}} \put(4.5,3.5){$\pi^0(135)$}
\put(4.5,2){$\eta(549)$} \put(9,3){\circle*{0.2}}
\put(8.5,3.5){$\pi^+(140)$} \put(3,5){\circle*{0.2}}
\put(2.5,4.5){$K^0(498)$} \put(7,5){\circle*{0.2}}
\put(6.5,4.5){$K^+(494)$} \put(0,1){\circle*{0.1}} \put(-0.5,1){-1}
\put(0,3){\circle*{0.1}} \put(-0.5,3){0} \put(0,5){\circle*{0.1}}
\put(-0.5,5){1} \put(1,0){\circle*{0.1}} \put(1,-0.5){-1}
\put(3,0){\circle*{0.1}} \put(3,-0.5){-1/2} \put(5,0){\circle*{0.1}}
\put(5,-0.5){0} \put(7,0){\circle*{0.1}} \put(7,-0.5){1/2}
\put(9,0){\circle*{0.1}} \put(9,-0.5){1}
\end{picture}
\vspace{2em}
\end{center}
\caption{\label{2:2:2:fig:meson_octet} Pseudoscalar meson octet in an
$(I_3,S)$ diagram. Baryon number $B=0$. Masses in MeV.}
\end{figure}

\begin{figure}[t]
\vspace{2em} \unitlength1cm
\begin{center}
\begin{picture}(10,6)
\thicklines \put(0,0){\vector(1,0){10}} \put(10,-1){$I_3$}
\put(0,0){\vector(0,1){6}} \put(-1,6){$S$} \put(3,1){\circle*{0.2}}
\put(2.5,0.5){$n(940)$} \put(7,1){\circle*{0.2}}
\put(6.5,0.5){$p(938)$} \put(1,3){\circle*{0.2}}
\put(0.5,3.5){$\Sigma^-(1197)$} \put(5,3){\circle*{0.2}}
\put(5,3){\circle{0.4}} \put(4.5,3.5){$\Sigma^0(1193)$}
\put(4.5,2){$\Lambda(1116)$} \put(9,3){\circle*{0.2}}
\put(8.5,3.5){$\Sigma^+(1189)$} \put(3,5){\circle*{0.2}}
\put(2.5,4.5){$\Xi^-(1321)$} \put(7,5){\circle*{0.2}}
\put(6.5,4.5){$\Xi^0(1315)$} \put(0,1){\circle*{0.1}}
\put(-0.5,1){0} \put(0,3){\circle*{0.1}} \put(-0.5,3){-1}
\put(0,5){\circle*{0.1}} \put(-0.5,5){-2} \put(1,0){\circle*{0.1}}
\put(1,-0.5){-1} \put(3,0){\circle*{0.1}} \put(3,-0.5){-1/2}
\put(5,0){\circle*{0.1}} \put(5,-0.5){0} \put(7,0){\circle*{0.1}}
\put(7,-0.5){1/2} \put(9,0){\circle*{0.1}} \put(9,-0.5){1}
\end{picture}
\vspace{2em}
\end{center}
\caption{\label{2:2:2:fig:baryon_octet} Baryon octet ($J=\frac{1}{2}$)
in an $(I_3,S)$ diagram. Masses in MeV. Baryon number $B=1$.}
\end{figure}
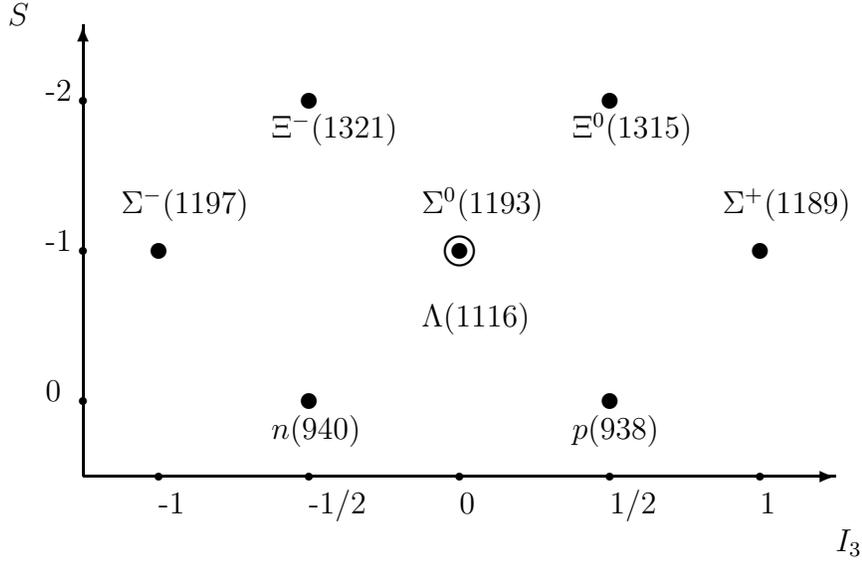

   According to the Coleman theorem \cite{Coleman:1966:1:1},  the symmetry of the
ground state determines the symmetry of the spectrum, i.e.
\begin{equation}
Q_{aV}|0\rangle =Q_V|0\rangle =0
\end{equation}
implies $\mbox{SU(3)}_V$ multiplets which can be classified according
to their baryon number.
   In the reverse conclusion, the symmetry of the ground state can be
inferred from the symmetry of
the spectrum.
   Figures \ref{2:2:2:fig:meson_octet} and
\ref{2:2:2:fig:baryon_octet} show the octets of the lowest-lying
pseudoscalar-meson states and the lowest-lying baryon states of
spin-parity $\frac{1}{2}^+$, respectively.

   The axial charges satisfy the commutation relations
\begin{eqnarray}
\label{2:2:2:crqaa}
[Q_{aA},Q_{bA}]&=&if_{abc}Q_{cV},\\
\label{2:2:2:crqva}
{[Q_{aV},Q_{bA}]}&=&if_{abc}Q_{cA}.
\end{eqnarray}
   Since the parity doubling is not observed for the low-lying states,
one assumes that the $Q_{aA}$ do {\em not} annihilate the ground
state,
\begin{equation}
\label{2:2:2:qav}
Q_{aA}|0\rangle\neq 0,
\end{equation}
i.e., the ground state of QCD is not invariant under ``axial'' transformations.
   In the present case, $G=\mbox{SU(3)}_L\times\mbox{SU(3)}_R$ with $n_G=16$
and $H=\mbox{SU(3)}_V$ with $n_H=8$ and we expect eight
Goldstone bosons.
   According to the Goldstone theorem \cite{Goldstone:eq:1:1}, \cite{Goldstone:es:1:1},
to each axial generator $Q_{aA}$, which does not annihilate the ground state,
corresponds a massless Goldstone boson field $\phi_a(x)$ with spin 0,
whose symmetry properties are tightly connected to the generator
in question.
   The Goldstone bosons have the same transformation behavior under
parity,
\begin{equation}
\label{2:2:2:parityphi}
\phi_a(t,\vec{x})\stackrel{P}{\mapsto}-\phi_a(t,-\vec{x}),
\end{equation}
i.e., they are pseudoscalars, and transform under the subgroup
$H=\mbox{SU(3)}_V$, which leaves the vacuum invariant, as
an octet [see Eq.\ (\ref{2:2:2:crqva})]:
 \begin{equation}
\label{2:2:2:transformationphiqv}
[Q_{aV},\phi_b(x)]=if_{abc}\phi_c(x).
\end{equation}

\subsubsection{The scalar singlet quark condensate}
\label{subsec_sqc}
   In the following, we will show that a non-vanishing scalar singlet quark condensate
in the chiral limit is a sufficient (but not a necessary) condition for a
spontaneous symmetry breaking in QCD. In this section all physical
quantities such as the ground state, the quark operators etc.\ are considered
in the chiral limit.

   Let us first recall the definition of the nine scalar and pseudoscalar
quark densities:
\begin{eqnarray}
\label{2:2:3:sqd}
S_a(y)&=&\bar{q}(y)\lambda_a q(y),
\quad a=0,\cdots,8,\\
\label{4:1:psqd} P_a(y)&=&i\bar{q}(y)\gamma_5\lambda_a  q(y),
\quad a=0,\cdots,8,
\end{eqnarray}
where $\lambda_0=\sqrt{2/3}\,\mathbbm 1$.
   We need the equal-time commutation relation of two quark operators of the form
$A_i(x)=q^\dagger(x)\hat{A}_i q(x)$,
where $\hat{A}_i$ symbolically denotes Dirac and flavor matrices and
a summation over color indices is implied:
\begin{equation}
\label{2:2:3:comrel}
[A_1(t,\vec{x}),A_2(t,\vec{y})]=\delta^3(\vec{x}-\vec{y})
q^\dagger(x)[\hat{A}_1,\hat{A}_2]q(x).
\end{equation}
   With the definition
\begin{displaymath}
Q_{aV}(t)=\int \mbox{d}^3 x q^\dagger(t,\vec{x}) \frac{\lambda_a}{2} q(t,\vec{x}),
\end{displaymath}
   and using
\begin{displaymath}
[\frac{\lambda_a}{2},\gamma_0\lambda_0]=0,\quad
{[}\frac{\lambda_a}{2},\gamma_0\lambda_b]=
\gamma_0 i f_{abc} \lambda_c,
\end{displaymath}
we see, after integration of Eq.\ (\ref{2:2:3:comrel})
over $\vec{x}$, that the scalar quark densities of
Eq.\ (\ref{2:2:3:sqd}) transform under
$\mbox{SU(3)}_V$ as a singlet and as an octet, respectively,
\begin{eqnarray}
\label{2:2:3:sitr}
[Q_{aV}(t),S_0(y)]&=&0,\quad a=1,\cdots,8,\\
\label{2:2:3:octr}
{[Q_{aV}(t),S_b(y)]}&=&i\sum_{c=1}^8f_{abc}S_c(y),
\quad a,b=1,\cdots,8,
\end{eqnarray}
with analogous results for the pseudoscalar quark densities.
   Using the relation
\begin{equation}
\sum_{a,b=1}^8 f_{abc}f_{abd}=3\delta_{cd}
\end{equation}
for the structure constants of SU(3), we re-express the octet components of
the scalar quark densities as
\begin{equation}
\label{2:2:3:soktett}
S_a(y)=-\frac{i}{3}\sum_{b,c=1}^8f_{abc}[Q_{bV}(t),S_c(y)].
\end{equation}

   In the chiral limit the ground state is necessarily invariant under
$\mbox{SU(3)}_V$ \cite{Vafa:1983tf}, i.e., $Q_{aV}|0\rangle=0$, and we
obtain from Eq.\ (\ref{2:2:3:soktett})
\begin{equation}
\label{2:2:3:saun}
\langle 0|S_a(y)|0\rangle
=\langle 0|S_a(0)|0\rangle
\equiv\langle S_a\rangle =0,\quad a=1,\cdots,8,
\end{equation}
where we made use of translational invariance of the ground state.
   As an intermediate result we see that the octet components of the scalar quark condensate
{\em must} vanish in the chiral limit.
   From Eq.\ (\ref{2:2:3:saun}), we obtain for $a=3$
$$\langle\bar{u}u\rangle-\langle\bar{d}d\rangle=0,$$
i.e.\ $\langle\bar{u}u\rangle=\langle\bar{d}d\rangle$
and for $a=8$
$$\langle\bar{u}u\rangle+\langle\bar{d}d\rangle
-2\langle\bar{s}s\rangle=0,
$$
i.e.\ $\langle\bar{u}u\rangle=\langle\bar{d}d\rangle=
\langle\bar{s}s\rangle$.

   Because of Eq.\ (\ref{2:2:3:sitr}) a similar argument cannot be used
for the singlet condensate, and if we assume a non-vanishing
singlet scalar quark condensate in the chiral limit, we find using
$\langle\bar{u}u\rangle=\langle\bar{d}d\rangle=
\langle\bar{s}s\rangle$:
\begin{equation}
\label{2:2:3:cqc}
0\neq \langle \bar{q}q\rangle
=\langle\bar{u}u+\bar{d}d+\bar{s}s\rangle
=3\langle\bar{u}u\rangle =
3\langle\bar{d}d\rangle
=3\langle \bar{s}s\rangle.
\end{equation}
   Finally, we make use of (no summation implied!)
$$ (i)^2 [\gamma_5 \frac{\lambda_a}{2},\gamma_0\gamma_5\lambda_a]
=\lambda^2_a\gamma_0$$ in combination with
\begin{displaymath}
\lambda_1^2=\lambda_2^2=\lambda_3^2=
\left(
\begin{array}{rrr}
1&0&0\\
0&1&0\\
0&0&0
\end{array}
\right),\,\,
\lambda_4^2=\lambda_5^2=
\left(
\begin{array}{rrr}
1&0&0\\
0&0&0\\
0&0&1
\end{array}
\right),\,\,
\lambda_6^2=\lambda_7^2=
\left(
\begin{array}{rrr}
0&0&0\\
0&1&0\\
0&0&1
\end{array}
\right),\,\,
\lambda_8^2=
\frac{1}{3}
\left(
\begin{array}{rrr}
1&0&0\\
0&1&0\\
0&0&4
\end{array}
\right)
\end{displaymath}
to obtain
\begin{equation}
\label{2:2:3:crqapsqd}
i[Q_{aA}(t), P_a(y)]
=
\left \{\begin{array}{cl}
\bar{u}u+\bar{d}d, & a=1,2,3\\
\bar{u}u+\bar{s}s, & a=4,5\\
\bar{d}d+\bar{s}s, & a=6,7\\
\frac{1}{3}(\bar{u}u+\bar{d}d+4\bar{s}s), & a=8
\end{array}
\right.
\end{equation}
where we have suppressed the $y$ dependence on the right-hand side.
   We evaluate Eq.\ (\ref{2:2:3:crqapsqd}) for a ground state which is
invariant under $\mbox{SU(3)}_V$, assuming a non-vanishing singlet scalar
quark condensate,
\begin{equation}
\label{2:2:3:crqc}
\langle 0|i[Q_{aA}(t),P_a(y)]|0\rangle
=\frac{2}{3}\langle\bar{q}q\rangle,\quad a=1,\cdots,8,
\end{equation}
   where, because of translational invariance, the right-hand side
is independent of $y$.
   Inserting a complete set of states into the commutator of
Eq.~(\ref{2:2:3:crqc}) yields that both the pseudoscalar density $P_a(y)$
as well as the axial charge operators $Q_{aA}$ must have a
non-vanishing matrix element between the vacuum and massless
one-particle states $|\phi_b\rangle$.
   In particular, because of Lorentz covariance, the
matrix element of the axial-vector current operator between the
vacuum and these massless states,
appropriately normalized,
 can be written as
\begin{equation}
\label{2:2:3:acc}
\langle 0|A_a^\mu(0)|\phi_b(p)\rangle=ip^\mu F_0 \delta_{ab},
\end{equation}
where $F_0\approx 93$ MeV denotes the ``decay'' constant of
the Goldstone bosons in the chiral limit.
   From Eq.\ (\ref{2:2:3:acc}) we see that a non-zero value of $F_0$ is a necessary
and sufficient criterion for spontaneous chiral symmetry breaking.
   On the other hand, because of Eq.\ (\ref{2:2:3:crqc})
a non-vanishing scalar quark condensate $\langle \bar{q}
q\rangle$ is a sufficient (but not a necessary) condition for
a spontaneous symmetry breakdown in QCD.

\section{Mesonic chiral perturbation theory}
\label{section_mchpt}
  Our goal is the construction of the most general theory describing the
dynamics of the Goldstone bosons associated with the spontaneous
symmetry breakdown in QCD.
   In the chiral limit, we want the effective Lagrangian to be invariant under
 $G=\mbox{SU(3)}_L\times\mbox{SU(3)}_R\times\mbox{U(1)}_V$.
   It should contain exactly eight pseudoscalar degrees of freedom transforming
as an octet under the subgroup $H=\mbox{SU(3)}_V$.
   Moreover, taking account of spontaneous symmetry breaking, the ground
state should only be invariant under $\mbox{SU(3)}_V\times\mbox{U(1)}_V$.
\subsection{Transformation properties of the Goldstone bosons}
\label{sec_tpgb}
   The purpose of this section is to discuss the transformation properties
of the field variables describing the Goldstone bosons
\cite{Weinberg:de:2:2}, \cite{Coleman:sm:2:2}, \cite{Callan:sn:2:2}.
   We will need the concept of a {\em nonlinear realization} of a group in
addition to a {\em representation} of a group which one usually encounters in
Physics.
   We will first discuss a few general group-theoretical properties before
specializing to QCD.

\subsubsection{General considerations}
\label{subsec_gc}
   Let us consider a physical system described by a Lagrangian which
is invariant under a compact Lie group $G$.
   We assume the ground state of the system to be invariant under
only a subgroup $H$ of $G$, giving rise to $n=n_G-n_H$ Goldstone bosons.
   Each of these Goldstone bosons will be described by an independent
field $\phi_i$ which is a smooth real function on Minkowski space
$M^4$.
   These fields are collected in an $n$-component vector $\Phi=(\phi_1,\cdots,\phi_n)$, defining the
vector space $M_1$.
   Our aim is to find a mapping $\varphi$ which uniquely associates with each
pair $(g,\Phi)\in G\times M_1$ an element $\varphi(g,\Phi)\in M_1$ with
the following properties:
\begin{eqnarray}
\label{4:2:condmap1}
&&\varphi(e,\Phi)=\Phi\,\,\forall\,\,\Phi\in M_1,\, e\,\,
\mbox{identity of}\,\, G,\\
\label{4:2:condmap2}
&&\varphi(g_1,\varphi(g_2,\Phi))=\varphi(g_1 g_2,\Phi)\,\,\forall\,\,
g_1,g_2\in G,\,\forall\,\Phi\in M_1.
\end{eqnarray}
   Such a mapping defines an {\em operation} of the group $G$ on $M_1$.
   The construction proceeds as follows \cite{Leutwyler:1991mz:2:2}.
   Let $\Phi=0$ denote the ``origin'' of $M_1$  which,
in a theory containing Goldstone bosons only, loosely speaking corresponds
to the ground state configuration.
   Since the ground state is supposed to be invariant under the subgroup
$H$ we require the mapping $\varphi$ to be such that all elements
$h\in H$ map the origin onto itself.
   In this context the subgroup $H$ is also known as the little group of
$\Phi=0$.

   We will establish a connection
between the Goldstone boson fields and the set of all left cosets
$\{gH|g\in G\}$ which is also referred to as the quotient $G/H$.
   For a subgroup $H$ of $G$ the set $gH=\{gh|h\in H\}$ defines the left
coset of $g$ (with an analogous definition for the right coset) which is one
element of $G/H$.
   For our purposes we need the property
that cosets either completely overlap or are
completely disjoint,
i.e, the quotient is a
set whose elements themselves are sets of group elements, and these sets
are completely disjoint.
   Under all elements of a given coset $gH$ the origin is mapped onto the
same vector in ${\mathbbm R}^n$:
$$\varphi(gh,0)=\varphi(g,\varphi(h,0))
=\varphi(g,0)\,\,\forall\,\, g\in G\,
\mbox{and}\, h\in H.$$
   Secondly, the mapping is injective with respect to the elements
of $G/H$.
Consider two elements $g$ and $g'$ of $G$ where $g'\not\in g H$.
   Let us assume $\varphi(g,0)=\varphi(g',0)$:
$$0=\varphi(e,0)=\varphi(g^{-1}g,0)
=\varphi(g^{-1},\varphi(g,0))
=\varphi(g^{-1},\varphi(g',0))=\varphi(g^{-1}g',0).$$
   However, this implies $g^{-1}g'\in H$ or $g'\in gH $
in contradiction to the assumption $g'\not\in g H$
and therefore $\varphi(g,0)=\varphi(g',0)$ cannot be true.
   In other words, the mapping can be inverted on the image of
$\varphi(g,0)$.
   The conclusion is that there exists an {\em isomorphic mapping}
between the quotient $G/H$ and and the Goldstone boson
fields.
   Of course, the Goldstone boson fields are not
constant vectors in ${\mathbbm R}^n$ but functions on Minkowski space.
   This is accomplished by allowing the cosets $gH$ to also
depend on $x$.

   Now let us discuss the transformation behavior of the Goldstone boson
fields under an arbitrary $g\in G$ in terms of the
isomorphism established above.
   To each $\Phi$ corresponds a  coset $\tilde{g}H$ with appropriate
$\tilde{g}$.
   Let $f=\tilde{g}h\in \tilde{g}H$ denote a representative of this
coset such that
$$\Phi=\varphi(f,0)=\varphi(\tilde{g}h,0).$$
   Now apply the mapping $\varphi(g)$ to $\Phi$:
$$\varphi(g,\Phi)=\varphi(g,\varphi(\tilde{g}h,0))
=\varphi(g\tilde{g}h,0)=\varphi(f',0)=\Phi',\quad
f'\in g(\tilde{g}H).
$$
   In order to obtain the transformed $\Phi'$ from a given
$\Phi$ we simply need to multiply the left coset $\tilde{g}H$ representing
$\Phi$ by $g$  in order to obtain the new left coset representing $\Phi'$:
\begin{eqnarray*}
\Phi &\stackrel{g}{\to} &\Phi'\\
\downarrow&&\uparrow\\
\tilde g H&\stackrel{g}{\to}&g\tilde g H
\end{eqnarray*}
   This procedure uniquely determines the transformation behavior of
the Goldstone bosons up to an appropriate choice of variables
parameterizing the elements of the quotient $G/H$.

\subsubsection{Application to QCD}
\label{subsec_aqcd}

   The symmetry groups relevant to the application in QCD are
\begin{displaymath}
G=\mbox{SU($N$)}\times\mbox{SU($N$)}=\{(L,R)|
L\in \mbox{SU($N$)}, R\in
\mbox{SU($N$)}\}\,\,\mbox{and}\,\, H=\{(V,V)|V\in \mbox{SU($N$)}\}
\cong\mbox{SU($N$)}.
\end{displaymath}
Let $\tilde{g}=(\tilde{L},\tilde{R})\in G$.
   We characterize the left coset $\tilde{g}H$
through the SU($N$) matrix $U=\tilde{R}\tilde{L}^\dagger$
\cite{Balachandran:zj:2:2} such that $\tilde{g}H=({\mathbbm 1},\tilde{R}\tilde{L}^\dagger)H$.
   This corresponds to the convention of choosing as the representative of the
coset the element which has the unit matrix in its first argument.
   The transformation behavior of $U$ under $g=(L,R)\in G$ is obtained by
multiplication in the left coset:
$$g \tilde{g}H=(L, R\tilde{R}\tilde{L}^\dagger)H=
({\mathbbm 1},R\tilde{R}\tilde{L}^\dagger L^\dagger)(L,L)H=
({\mathbbm 1},R(\tilde{R}
\tilde{L}^\dagger)L^\dagger)H,$$
i.e.\
\begin{equation}
\label{3:1:2:utrafo}
U=\tilde{R}\tilde{L}^\dagger \mapsto U'=R(\tilde{R}\tilde{L}^\dagger)L^\dagger
=RUL^\dagger.
\end{equation}
   As mentioned above, we need to introduce an $x$ dependence to account for
   the fact that we are dealing with fields:
\begin{equation}
\label{3:1:2:utrfafo}
U(x)\mapsto R U(x) L^\dagger.
\end{equation}
   For the physically relevant cases
the corresponding unitary matrices may be parameterized as
\begin{equation}
\label{3:1:2:upar}
U(x)=\exp\left(i\frac{\phi(x)}{F_0}\right),
\end{equation}
where, for $N=2$,
\begin{equation}
\label{3:1:2:phisu2}
\phi=\sum_{i=1}^3\phi_i\tau_i
\equiv
\left(\begin{array}{cc}
\pi^0&\sqrt{2}\pi^+\\
\sqrt{2}\pi^-&-\pi^0
\end{array}\right)
\end{equation}
and, for $N=3$,
\begin{equation}
\label{3:1:2:phisu3}
\phi=\sum_{a=1}^8  \phi_a\lambda_a \equiv
\left(\begin{array}{ccc}
\pi^0+\frac{1}{\sqrt{3}}\eta &\sqrt{2}\pi^+&\sqrt{2}K^+\\
\sqrt{2}\pi^-&-\pi^0+\frac{1}{\sqrt{3}}\eta&\sqrt{2}K^0\\
\sqrt{2}K^- &\sqrt{2}\bar{K}^0&-\frac{2}{\sqrt{3}}\eta
\end{array}\right).
\end{equation}

   The origin $\phi(x)=0$, i.e.\ $U_0=\mathbbm 1$, denotes the ground state of
the system.
   Under transformations of the subgroup $H=\{(V,V)|V\in \mbox{SU($N$)}\}$
corresponding to rotating both left- and right-handed quark fields in
QCD by the same $V$, the ground state remains invariant,
$$U_0\mapsto VU_0 V^\dagger=V V^\dagger={\mathbbm 1}=U_0.$$
   On the other hand, under ``axial transformations,'' i.e.\ rotating
the left-handed quarks by $A$ and the right-handed quarks by $A^\dagger$,
the ground state does {\em not} remain invariant,
$$U_0\mapsto A^\dagger U_0 A^\dagger=A^\dagger A^\dagger
\neq U_0,$$
   which is consistent with the assumed spontaneous symmetry
breakdown.

   The traceless and Hermitian matrices of Eqs.\ (\ref{3:1:2:phisu2}) and
(\ref{3:1:2:phisu3}) contain the Goldstone boson fields.
   We want to discuss their transformation behavior under
the subgroup $H=\{(V,V)|V\in \mbox{SU($N$)}\}$.
   Expanding
$$U={\mathbbm 1}+i\frac{\phi}{F_0}-\frac{\phi^2}{2F_0^2}+\cdots,$$
we immediately see that the transformation behavior of Eq.\ (\ref{3:1:2:utrfafo})
restricted to the subgroup $H$,
\begin{displaymath}
{\mathbbm 1}+i\frac{\phi}{F_0}-\frac{\phi^2}{2F_0^2}+\cdots\mapsto
V({\mathbbm 1}+i\frac{\phi}{F_0}-\frac{\phi^2}{2F_0^2}+\cdots)V^\dagger
={\mathbbm 1}+i\frac{V \phi V^\dagger}{F_0}-\frac{V\phi V^\dagger V\phi V^\dagger}{2F_0^2}
+\cdots,
\end{displaymath}
implies
\begin{equation}
\label{3:1:2:phihtrafo}
\phi\mapsto V\phi V^\dagger.
\end{equation}
   However, this corresponds exactly to the fact that the Goldstone
   bosons transform according to the adjoint representation under SU(3)$_V$,
   i.e.~they transform as an octet.

   For group elements of $G$ of the form $(A,A^\dagger)$ one may
proceed in a completely analogous fashion.
   However, one finds that the fields $\phi_a$ do {\em not} have
a simple transformation behavior under these group elements.

\subsection{Effective Lagrangian and power-counting scheme}
   The application of effective field theory (EFT) to strong interaction
processes has become one of the most important theoretical tools
in the low-energy regime.
   The basic idea consists of writing down the most general possible
Lagrangian, including {\em all} terms consistent with assumed symmetry
principles, and then calculating matrix elements with this Lagrangian
within some perturbative scheme \cite{Weinberg:1978kz}.
   A successful application of this program thus requires two main ingredients:
\begin{enumerate}
\item[(1)]
A knowledge of the most general effective Lagrangian;
\item[(2)]
an expansion scheme for observables in terms of a consistent power-counting
method.
\end{enumerate}
\subsubsection{The lowest-order effective Lagrangian}
\label{sec_loel}
   In terms of the SU(3) matrix $U(x)$ of Eqs.~(\ref{3:1:2:upar}) and
(\ref{3:1:2:phisu3}) the most general, chirally invariant, effective Lagrangian
with the minimal number of derivatives reads
\begin{equation}
\label{3:2:1:l2}
{\cal L}_{\rm eff}
=\frac{F^2_0}{4}\mbox{Tr}\left(\partial_\mu U \partial^\mu U^\dagger
\right),
\end{equation}
   where $F_0\approx 93$ MeV is a free parameter which later on will be
related to the pion decay $\pi^+\to\mu^+\nu_\mu$.
   Because of the trace property $\mbox{Tr}(AB)=\mbox{Tr}(BA)$,
the Lagrangian is invariant under the {\em global}\,
$\mbox{SU(3)}_L\times\mbox{SU(3)}_R$ transformation $U \mapsto R U L^\dagger$
of Eq.~(\ref{3:1:2:utrfafo}).
   The global $\mbox{U(1)}_V$ invariance is trivially satisfied, because
the Goldstone bosons have baryon number zero, thus transforming
as $\phi\mapsto\phi$ under $\mbox{U(1)}_V$ which also implies $U \mapsto U$.

   The substitution $\phi_a(t,\vec{x})\mapsto -\phi_a(t,\vec{x})$ or,
equivalently, $U(t,\vec{x})\mapsto U^\dagger(t,\vec{x})$ provides a simple
method of testing, whether an expression is of so-called even or odd
{\em intrinsic} parity,
i.e., even or odd in the number of Goldstone boson fields.
   For example, the Lagrangian of Eq.\ (\ref{3:2:1:l2}) is even.
   Since the Goldstone bosons of QCD are pseudoscalars, the true parity transformation
is given by $\phi_a(t,\vec{x})\mapsto -\phi_a(t,-\vec{x})$ or,
equivalently, $U(t,\vec{x})\mapsto U^\dagger(t,-\vec{x})$.

   The purpose of the multiplicative constant $F^2_0/4$ in Eq.\ (\ref{3:2:1:l2})
is to generate the standard form of the kinetic term
$\frac{1}{2}\partial_\mu \phi_a\partial^\mu \phi_a$, which can be seen by
expanding the exponential
$U={\mathbbm 1}+i\phi/F_0+\cdots$, $\partial_\mu U=i\partial_\mu\phi/F_0
+\cdots$, resulting in
\begin{displaymath}
{\cal L}_{\rm eff}=
\frac{F^2_0}{4}\mbox{Tr}\left[\frac{i\partial_\mu\phi}{F_0}
\left(-\frac{i\partial^\mu\phi}{F_0}\right)\right]+\cdots
=\frac{1}{4}\partial_\mu\phi_a\partial^\mu
\phi_b\mbox{Tr}(\lambda_a\lambda_b)+\cdots
=\frac{1}{2}\partial_\mu\phi_a\partial^\mu\phi_a +{\cal L}_{\rm int},
\end{displaymath}
where we made use of $\mbox{Tr}(\lambda_a \lambda_b)=2\delta_{ab}$.
   Since there are no other terms containing only two fields,
the eight fields $\phi_a$ indeed describe eight independent {\em massless}
particles.

   There are no other independent, chirally invariant terms containing only
two derivatives.
   A term of the type $\mbox{Tr}[(\partial_\mu\partial^\mu U) U^\dagger]$ may
be re-expressed as
$$\mbox{Tr}[(\partial_\mu\partial^\mu U) U^\dagger]
=\partial_\mu[\mbox{Tr}(\partial^\mu U U^\dagger)]
-\mbox{Tr}(\partial^\mu U \partial_\mu U^\dagger),$$
i.e., up to a total derivative it is proportional to the Lagrangian
of Eq.\ (\ref{3:2:1:l2}).
   However, in the present context, total derivatives do not have a dynamical
significance, i.e.\ they leave the equations of motion unchanged and can thus
be dropped.
   The product of two invariant traces is excluded at lowest order,
because $\mbox{Tr}(\partial_\mu U U^\dagger)=0$.

   Let us turn to the vector and axial-vector currents associated with
the global $\mbox{SU(3)}_L\times\mbox{SU(3)}_R$ symmetry of the effective
Lagrangian of Eq.\ (\ref{3:2:1:l2}).
   To that end, we parameterize
\begin{eqnarray}
\label{3:2:1:L}
L&=&{\mathbbm 1}-i\epsilon^L_a\frac{\lambda_a}{2},\\
\label{3:2:1:R}
R&=&{\mathbbm 1}-i\epsilon^R_a\frac{\lambda_a}{2}.
\end{eqnarray}
   In order to construct $J^{\mu}_{aL}$, set $\epsilon^R_a=0$ and choose
$\epsilon^L_a=\epsilon^L_a(x)$.
   Using
\begin{equation}
\label{3:2:1:utrick}
U^\dagger U={\mathbbm 1}\quad\Rightarrow\quad\partial^\mu(U^\dagger U)=0\quad
\Rightarrow\quad \partial^\mu U^\dagger U=-U^\dagger \partial^\mu U,
\end{equation}
the variation of the Lagrangian can be brought into the form
\begin{displaymath}
\label{4:3:dll}
\delta{\cal L}_{\rm eff}=\frac{F^2_0}{4}i\partial_\mu \epsilon^L_a\mbox{Tr}\left(\lambda_a
\partial^\mu U^\dagger U\right).
\end{displaymath}
   Applying the method of Ref.\ \cite{Gell-Mann:1960np},
we obtain for the left currents
\begin{equation}
\label{3:2:1:jl}
J^\mu_{aL}=\frac{\partial \delta {\cal L}_{\rm eff}}{
\partial \partial_\mu \epsilon_a^L}
=i\frac{F^2_0}{4}\mbox{Tr}\left(\lambda_a \partial^\mu U^\dagger U\right),
\end{equation}
   and, completely analogously, choosing $\epsilon_a^L=0$ and
$\epsilon_a^R=\epsilon_a^R(x)$, for the right currents
\begin{equation}
\label{3:2:1:jr}
J^\mu_{aR}=\frac{\partial \delta {\cal L}_{\rm eff}
}{\partial \partial_\mu \epsilon^R_a}
=-i\frac{F^2_0}{4}\mbox{Tr}\left(\lambda_a U \partial^\mu U^\dagger\right).
\end{equation}
   Combining Eqs.\ (\ref{3:2:1:jl}) and (\ref{3:2:1:jr}) the vector and
axial-vector currents read
\begin{eqnarray}
\label{3:2:1:jv}
J^\mu_{aV}&=&J^\mu_{aR}+J^\mu_{aL}=-i\frac{F^2_0}{4}
\mbox{Tr}\left(\lambda_a[U,\partial^\mu U^\dagger]\right),\\
\label{3:2:1:ja}
J^\mu_{aA}&=&J^\mu_{aR}-J^\mu_{aL}=-i\frac{F^2_0}{4}
\mbox{Tr}\left(\lambda_a\{U,\partial^\mu U^\dagger\}\right).
\end{eqnarray}
   Furthermore, because of the symmetry of ${\cal L}_{\rm eff}$ under
$\mbox{SU(3)}_L\times\mbox{SU(3)}_R$, both vector and axial-vector
currents are conserved.
   Using the substitution $U\leftrightarrow U^\dagger$ and Eq.~(\ref{3:2:1:utrick}),
the vector current densities $J^\mu_{aV}$ of Eq.\ (\ref{3:2:1:jv})
contain only terms with an even number of Goldstone bosons and the axial-vector
current densities $J^\mu_{aA}$ of Eq.\ (\ref{3:2:1:ja}) only terms with an odd
number of Goldstone bosons.
   To find the leading term let us expand Eq.\ (\ref{3:2:1:ja}) in the fields,
\begin{displaymath}
J^\mu_{aA}=-i\frac{F^2_0}{4}\mbox{Tr}\left(\lambda_a\left\{{\mathbbm 1}+\cdots,
-i\frac{\lambda_b\partial^\mu \phi_b}{F_0}+\cdots\right\}\right)=
-F_0\partial^\mu\phi_a+\cdots.
\end{displaymath}
   We conclude that the axial-vector current has a non-vanishing
matrix element when evaluated between the vacuum and a one-Goldstone boson
state:
\begin{equation}
\label{3:2:1:jamatrixelemtn}
\langle 0|J^\mu_{aA}(x)|\phi_b(p)\rangle
=\langle 0|-F_0\partial^\mu\phi_a(x)|\phi_b(p)\rangle
=-F_0\partial^\mu \exp(-ip\cdot x)\delta_{ab}
=ip^\mu F_0\exp(-ip\cdot x)\delta_{ab}.
\end{equation}
   Equation (\ref{3:2:1:jamatrixelemtn}) is the manifestation of
Eq.~(\ref{2:2:3:acc}) at lowest order in the effective field theory.

\subsubsection{Symmetry breaking through the quark masses}
    Up to now, we have assumed a perfect
$\mbox{SU(3)}_L\times\mbox{SU(3)}_R$ symmetry.
   As has been discussed in Section \ref{subsec_csbdqm},
the quark-mass term of QCD results
in an explicit symmetry breaking,
\begin{equation}
\label{3:2:2:qmt}
{\cal L}_{\cal M}=-\bar{q}_R{\cal M} q_L-\bar{q}_L {\cal M}^\dagger q_R,\quad
{\cal M}=\left(\begin{array}{ccc}m_u&0&0\\0&m_d&0\\0&0&m_s\end{array}\right).
\end{equation}
   In order to incorporate the consequences of Eq.\ (\ref{3:2:2:qmt})
into the effective-Lagrangian framework, one makes use of the
following argument \cite{Georgi}:
   Although ${\cal M}$ is in reality just a constant matrix and does not
transform along with the quark fields, ${\cal L}_{\cal M}$ of Eq.\ (\ref{3:2:2:qmt})
{\em would be} invariant {\em if} ${\cal M}$ transformed as
\begin{equation}
\label{3:2:2:mgtrafo}
{\cal M}\mapsto R {\cal M} L^\dagger.
\end{equation}
   One then constructs the most general Lagrangian ${\cal L}(U,{\cal M})$ which is
invariant under Eqs.\ (\ref{3:1:2:utrfafo}) and
(\ref{3:2:2:mgtrafo}) and expands this function in powers of ${\cal M}$.
   At lowest order in ${\cal M}$ one obtains
\begin{equation}
\label{3:2:2:lqm}
{\cal L}_{\rm s.b.}=\frac{F^2_0 B_0}{2}\mbox{Tr}({\cal M}U^\dagger+U{\cal M}^\dagger),
\end{equation}
   where the subscript s.b.\ refers to symmetry breaking.
   In order to interpret the new parameter $B_0$, let us consider the Hamiltonian
density corresponding to the sum of the Lagrangians of Eq.~(\ref{3:2:1:l2}) and (\ref{3:2:2:lqm}):
\begin{eqnarray*}
{\cal H}_{\rm eff}
&=&\frac{F_0^2}{4}\mbox{Tr}(\dot U \dot U^\dagger)
+\frac{F_0^2}{4}\mbox{Tr}(\vec{\nabla}U\cdot\vec{\nabla}U^\dagger)
-{\cal L}_{\rm s.b.}.
\end{eqnarray*}
   Since the first two terms are always larger or equal to zero,
${\cal H}_{\rm eff}$ is minimized by constant and uniform fields.
   Using the ansatz
\begin{displaymath}
\phi=\phi_0+\frac{1}{F_0^2}\phi_2+\frac{1}{F_0^4}\phi_4+\cdots
\end{displaymath}
for the minimizing field values and organizing the individual terms in
powers of $1/F_0^2$, one finds $\phi=0$ as the classical solution even
in the presence of quark-mass terms.
   Now consider the energy density of the ground state ($U_{\rm min}=U_0={\mathbbm 1}$),
\begin{equation}
\label{2:2:3:Heff}
\langle{\cal H}_{\rm eff}\rangle_{\rm min}=-F_0^2 B_0(m_u+m_d+m_s),
\end{equation}
   and compare its derivative with respect to (any of) the light quark masses
$m_q$ with the corresponding quantity in QCD,
\begin{displaymath}
\left.\frac{\partial \langle 0|{\cal H}_{\rm QCD}|0\rangle}{\partial m_q}
\right|_{m_u=m_d=m_s=0}=\frac{1}{3}\langle 0|\bar{q}{q}|0\rangle_0
=\frac{1}{3}\langle \bar{q}q\rangle_0,
\end{displaymath}
   where $\langle\bar{q}{q}\rangle_0$ is the chiral quark condensate of
Eq.\ (\ref{2:2:3:cqc}).
   Within the framework of the lowest-order effective
Lagrangian, the constant $B_0$ is thus related to the chiral quark condensate
as
\begin{equation}
\label{3:2:2:b0}
3 F^2_0B_0=-\langle\bar{q}q\rangle_0.
\end{equation}

   Let us add a few remarks.
\begin{enumerate}
\item A term $\mbox{Tr}({\cal M})$ by itself is not invariant.
\item The combination $\mbox{Tr}({\cal M}U^\dagger-U {\cal M}^\dagger)$ has the wrong
behavior under parity $\phi(t,\vec{x})\to-\phi(t,-\vec{x})$.
\item Because ${\cal M}={\cal M}^\dagger$, ${\cal L}_{\rm s.b.}$ contains only terms even
in $\phi$.
\end{enumerate}
   In order to determine the masses of the Goldstone bosons, we identify the
terms of second order in the fields in ${\cal L}_{\rm s.b.}$,
\begin{equation}
\label{3:2:2:lmzo}
{\cal L}_{\rm s.b}=-\frac{B_0}{2}\mbox{Tr}(\phi^2{\cal M}) +\cdots.
\end{equation}
   For the sake of simplicity we consider the isospin-symmetric limit
$m_u=m_d=\hat m$ so that the $\pi^0\eta$ term vanishes and there is no
$\pi^0$-$\eta$ mixing.
   The masses of the Goldstone bosons, to lowest order in
the quark masses, are then given by
\begin{eqnarray}
\label{3:2:2:mpi2}
M^2_\pi&=&2 B_0 \hat m,\\
\label{3:2:2:mk2}
M^2_K&=&B_0(\hat m+m_s),\\
\label{3:2:2:meta2}
M^2_\eta&=&\frac{2}{3} B_0\left(\hat m+2m_s\right).
\end{eqnarray}
   These results, in combination with Eq.\ (\ref{3:2:2:b0}),
$B_0=-\langle\bar{q}q\rangle_0/(3 F_0^2)$, correspond to relations
obtained in Ref.\ \cite{Gell-Mann:rz:3} and are referred to as the
Gell-Mann, Oakes, and Renner relations.
   Without additional input regarding the numerical value of $B_0$,
Eqs.\ (\ref{3:2:2:mpi2}) - (\ref{3:2:2:meta2}) do not allow for an extraction of
the  absolute values of the quark masses $\hat m$ and $m_s$, because re-scaling
$B_0\to \lambda B_0$ in combination with $m_q\to m_q/\lambda$ leaves the
relations invariant.
   For the ratio of the quark masses one obtains, using the empirical values
$M_\pi=135$ MeV, $M_K=496$ MeV,
and $M_\eta=547$ MeV,
\begin{eqnarray}
\frac{M^2_K}{M^2_\pi}=\frac{\hat m+m_s}{2\hat m}&\Rightarrow&\frac{m_s}{\hat m}=25.9,
\nonumber\\
\frac{M^2_\eta}{M^2_\pi}=\frac{2m_s+\hat m}{3\hat m}&\Rightarrow&\frac{m_s}{\hat m}=24.3.
\end{eqnarray}

   Let us conclude this section with a remark on $\langle\bar{q}q\rangle_0$.
   A non-vanishing quark condensate in the chiral
limit is a sufficient but not a necessary condition for a spontaneous
chiral symmetry breaking.
   The effective Lagrangian term of Eq.\ (\ref{3:2:2:lqm}) not only results in
a shift of the vacuum energy but also in finite Goldstone boson
masses and both effects are proportional to the parameter
$B_0$.
   We recall that it was
a symmetry argument which excluded a term $\mbox{Tr}({\cal M})$ which,
at leading order in ${\cal M}$, would decouple the vacuum energy shift
from the Goldstone boson masses.
   The scenario underlying ${\cal L}_{\rm s.b.}$ of Eq.\ (\ref{3:2:2:lqm})
is similar to that of a Heisenberg ferromagnet which exhibits a
spontaneous magnetization $\langle \vec{M}\rangle$, breaking the
O(3) symmetry of the Heisenberg Hamiltonian down to O(2).
   In the present case the analogue of the order parameter
$\langle \vec{M}\rangle$ is the quark condensate $\langle \bar{q} q\rangle_0$.
   In the case of the ferromagnet, the interaction with an external magnetic
field $\vec{H}$ is given by $-\langle \vec{M}\rangle\cdot \vec{H}$, which corresponds
to Eq.\ (\ref{2:2:3:Heff}), with the quark masses playing the role of the
external field $\vec{H}$ (see Table \ref{3:2:2:table_analogy}).
\begin{table}[t]
\begin{center}
{\renewcommand{\baselinestretch}{1.5}\small\normalsize
\begin{tabular}{c|c|c}
& Heisenberg ferromagnet & QCD\\
\hline
Symmetry of Hamiltonian & O(3) & $\mbox{SU(3)}_L\times\mbox{SU(3)}_R$\\
Symmetry of $|0\rangle$ & O(2) & $\mbox{SU(3)}_V$\\
Vacuum expectation value & $\langle \vec{M}\rangle$ & $\langle
\bar{q} q\rangle_0$\\
Explicit symmetry breaking & External magnetic field & Quark
masses\\
Interaction &$-\langle \vec{M}\rangle\cdot \vec{H}$&$\langle{\cal
H}_{\rm eff}\rangle$ of Eq.~(\ref{2:2:3:Heff})
\end{tabular}
} \caption{\label{3:2:2:table_analogy} Comparison between the
symmetry-breaking patterns of a Heisenberg ferromagnet and QCD.}
\end{center}
\end{table}
   However, in principle, it is also possible that $B_0$ vanishes or is
rather small.
   In such a case  the quadratic masses of the Goldstone bosons might
be dominated by terms which are nonlinear in the quark masses, i.e., by
higher-order terms in the expansion of ${\cal L}(U,{\cal M})$.
   Such a scenario is the origin of the so-called generalized chiral
perturbation theory \cite{Knecht:1995tr:3}.
   The analogue would be an antiferromagnet which shows a spontaneous
symmetry breaking but with $\langle \vec{M}\rangle=0$.
    The analysis of the $s$-wave $\pi\pi$-scattering lengths
\cite{Colangelo:2000jc}, \cite{Colangelo:2001sp}
supports the conjecture that the quark condensate is indeed the leading order
parameter of the spontaneously broken chiral symmetry
(see also Sec. \ref{subsubsection_pion_decay_pipi_scattering}).

\subsubsection{Construction of invariants}
\label{subsubsection_ci}
   So far, we have discussed the lowest-order effective Lagrangian for a {\em global}
$\mbox{SU(3)}_L\times\mbox{SU(3)}_R$ symmetry.
   In Sec.\ \ref{subsubsection_gfcwigf} we stated that the Ward identities of QCD
are obtained from a {\em locally} invariant generating functional involving a coupling to
external fields.
   Therefore, following Refs.\ \cite{Gasser:1983yg}, \cite{Gasser:1984gg}, we
will promote the global symmetry of the effective Lagrangian to a local one,
\begin{displaymath}
L\rightarrow V_L(x),\, R\rightarrow V_R(x),
\end{displaymath}
and introduce a coupling to the {\em same} external fields $v$, $a$, $s$, and $p$ as in
QCD [see Eq.\ (\ref{2:1:6:lqcdsn})].
   The transformation behavior of the special unitary matrix $U$ of Eqs.~(\ref{3:1:2:upar}) and
(\ref{3:1:2:phisu3}) under
$G=\mbox{SU(3)}_L\times\mbox{SU(3)}_R$, parity $P$, and charge
conjugation $C$ is given by
\begin{displaymath}
U\stackrel{G}{\mapsto}V_R U V_L^{\dagger},\quad
U(\vec{x},t)\stackrel{P}{\mapsto}U^\dagger(-\vec{x},t),\quad
U\stackrel{C}{\mapsto}U^T.
\end{displaymath}
  Given an object $A$ transforming as $V_R A V_L^\dagger$, such as $U$ or $\chi$,
the covariant derivative of $A$, $D_\mu A$, is defined as
\begin{equation}
\label{3:2:3:kaa}
D_\mu A\equiv\partial_\mu A -i r_\mu A+iA l_\mu
\mapsto D'_\mu A'
= V_R(D_\mu A) V_L^\dagger.
\end{equation}
The defining property is that the covariant derivative should transform as the object
it acts on. In particular, the covariant derivative of $U$ is given by
\begin{equation}
\label{3:2:3:kdU}
 D_{\mu} U =\partial_{\mu}U-ir_{\mu}U+iU l_{\mu}.
\end{equation}
   For the external fields we introduce corresponding field strength tensors in matrix form
as
\begin{eqnarray}
\label{3:2:3:fr}
f_{\mu\nu}^R&\equiv&\partial_\mu r_\nu-\partial_\nu r_\mu-i{[r_\mu,r_\nu]}
\stackrel{G}{\mapsto}V_R f_{\mu\nu}^R V_R^\dagger,\\
\label{3:2:3:fl} f_{\mu\nu}^L&\equiv&\partial_\mu l_\nu-\partial_\nu
l_\mu-i{[l_\mu,l_\nu]}\stackrel{G}{\mapsto}V_L f_{\mu\nu}^L
V_L^\dagger.
\end{eqnarray}
   They are traceless, because $\mbox{Tr}(l_\mu)=\mbox{Tr}(r_\mu)=0$ and the trace of any
commutator vanishes.
   Finally, we introduce the linear combination $\chi=2B_0(s+ip)$, where,
e.g., pure QCD is given by $\chi=2B_0\mbox{diag}(m_u,m_d,m_s)$.

   The effective Lagrangian is constructed in terms of $U$,
$U^\dagger$, $\chi$, $\chi^\dagger$, $f^{R}_{\mu\nu}$,
$f^{L}_{\mu\nu}$ and covariant derivatives of these
objects.
   Suppose we have matrices $A,B,C,\cdots$, all of which transform as
\begin{displaymath}
A\stackrel{G}{\mapsto} V_R A V_L^\dagger, \quad
B\stackrel{G}{\mapsto} V_R B V_L^\dagger, \quad
\cdots.
\end{displaymath}
   Invariants may be formed by ``multiplying'' in the following way:
\begin{eqnarray*}
\mbox{Tr}(A B^\dagger)& \stackrel{G}{\mapsto} & \mbox{Tr}(V_R A
\underbrace{V_L^\dagger\, V_L}_{\mbox{$\mathbbm 1$}}  B^\dagger V_R^{\dagger})
=\mbox{Tr}( V_R^{\dagger}\, V_R A B^\dagger)=\mbox{Tr}(A B^\dagger),
\end{eqnarray*}
where the generalization to a longer string of terms is obvious
and the product of invariant traces is also invariant:
\begin{equation}
\label{3:2:3:chains} \mbox{Tr}(AB^\dagger C D^\dagger),\quad
\mbox{Tr}(A B^\dagger)\mbox{Tr}(C D^\dagger),\quad \cdots.
\end{equation}
   In the chiral counting scheme the elements count as
\begin{displaymath}
U={\cal O}(q^0),\quad
D_{\mu} U={\cal O}(q),\quad
r_{\mu},l_{\mu}={\cal O}(q),\quad
f^{L/R}_{\mu\nu}={\cal O}(q^2), \quad
\chi = {\cal O}(q^2).
\end{displaymath}
   Any additional covariant derivative counts as ${\cal O}(q)$.
   The list of objects $A$ up to and including order $q^2$ which
transform as $A'=V_R A V_L^{\dagger}$ reads
\begin{displaymath}
U,\, D_{\mu} U,\, D_{\mu} D_{\nu}U,\, \chi,\, U f^L_{\mu\nu},\,
f^R_{\mu\nu} U.
\end{displaymath}
   The construction of chirally invariant expressions up to and including
order $q^2$ proceeds as follows.
   At ${\cal O}(q^0)$ the only invariant term is a constant,
$\mbox{Tr}\left ( U U^{\dagger} \right ) =
\mbox{Tr} ( {\mathbbm 1} ) = const$.
   Because of $\mbox{Tr}\left( D_{\mu} U U^{\dagger} \right) = 0$,
terms of the type $\mbox{Tr}[{\cal O}(q)] \times \mbox{Tr}(\cdots)$ are excluded.
   At ${\cal O}(q^2)$ we have
\begin{eqnarray*}
&&\mbox{Tr} \left (D_{\mu} D_{\nu} U U^{\dagger}
\right)
=-\mbox{Tr}\left[D_{\nu} U (D_{\mu} U)^{\dagger}\right],\quad
\mbox{Tr} \left [ D_{\mu} U (D_{\nu} U)^{\dagger} \right ],\quad
\mbox{Tr} \left [ U (D_{\mu} D_{\nu} U)^{\dagger} \right ]
=-\mbox{Tr}\left[D_{\mu} U (D_{\nu} U)^{\dagger}\right], \\
& & \mbox{Tr} \left ( \chi U^{\dagger} \right ),\quad
\mbox{Tr} \left ( U \chi^{\dagger} \right ),\quad
\mbox{Tr} \left [ (U f^L_{\mu\nu}) U^{\dagger} \right ]
= \mbox{Tr} \left ( f^L_{\mu\nu} \right )=0,\quad
\mbox{Tr} \left ( f^R_{\mu\nu} \right )=0.
\end{eqnarray*}
   Because of Lorentz invariance, indices have to be contracted and the remaining three
candidates are
\begin{displaymath}
\mbox{Tr}\left[D_{\mu} U (D^{\mu} U)^{\dagger} \right], \quad
\mbox{Tr}\left ( \chi U^{\dagger} \pm U \chi^{\dagger} \right ).
\end{displaymath}
   Finally, due to parity conservation,
\begin{displaymath}
{\cal L}(\vec{x},t)  \stackrel{P}{\mapsto}  {\cal L}(-\vec{x},t).
\end{displaymath}
$\mbox{Tr}(\chi U^\dagger-U\chi^\dagger)$ has to be excluded because of the wrong parity.
   At ${\cal O}(q^2)$, charge conjugation does not generate any additional constraint.

   The locally invariant lowest-order Lagrangian ${\cal L}_2$ is given by
\begin{equation}
\label{3:2:3:lol} {\cal L}_2 = \frac{F_0^2}{4} \mbox{Tr} \left[D_{\mu}
U (D^{\mu}U)^{\dagger} \right] +\frac{F_0^2}{4} \mbox{Tr} \left (
\chi U^{\dagger}+ U \chi^{\dagger} \right ).
\end{equation}
   At ${\cal O}(q^2)$ it contains two parameters: the SU(3) chiral limit of the Goldstone
boson decay constant $F_0\approx 93\, \mbox{MeV},$ and, hidden in the definition of
$\chi$, $ B_0=-\langle 0|\bar{q}{q}|0\rangle_0/(3 F^2_0)$.

   The lowest-order equation of motion corresponding to Eq.\ (\ref{3:2:3:lol})
is obtained by considering small variations of the SU(3) matrix,
\begin{equation}
\label{3:2:3:deltau} U'(x)=U(x)+\delta U(x)=\left({\mathbbm 1}+i\sum_{a=1}^8
\Delta_a(x)\lambda_a\right) U(x),
\end{equation}
where the $\Delta_a(x)$ are real functions.
  The matrix $U'$ satisfies both conditions
$U' U'^\dagger= {\mathbbm 1}$ and $\mbox{det}(U')=1$
up to and including terms linear in $\Delta_a$.
   Applying the principle of stationary action, the variation of the action reads
\begin{displaymath}
\delta S=i\frac{F^2_0}{4}\int_{t_1}^{t_2}\mbox{d}t
\int \mbox{d}^3 x \sum_{a=1}^8
\Delta_a(x) \mbox{Tr}\left\{\lambda_a[
D_\mu D^\mu U U^\dagger- U(D_\mu D^\mu U)^\dagger -\chi U^\dagger
+ U\chi^\dagger]\right\},
\end{displaymath}
where we made use of partial integration, the standard boundary conditions
$\Delta_a(t_1,\vec{x})=\Delta_a(t_2,\vec{x})=0$, the divergence theorem,
and the definition of the covariant derivative of Eq.\ (\ref{3:2:3:kaa}).
   Since the test functions $\Delta_a(x)$ may be chosen arbitrarily, we obtain
eight Euler-Lagrange equations
\begin{equation}
\label{3:2:3:eom8} \mbox{Tr}\left\{\lambda_a[D^2U U^\dagger - U (D^2
U)^\dagger -\chi U^\dagger + U\chi^\dagger]\right\}=0, \quad
a=1,\cdots, 8,
\end{equation}
which may be combined into a compact matrix form
\begin{equation}
\label{3:2:3:eom} {\cal O}_{\rm EOM}^{(2)}(U)\equiv D^2 U U^\dagger -
U (D^2 U)^\dagger -\chi U^\dagger + U \chi^\dagger
+\frac{1}{3}\mbox{Tr}(\chi U^\dagger- U \chi^\dagger)=0.
\end{equation}
   The trace term in Eq.~(\ref{3:2:3:eom}) appears, because Eq.~(\ref{3:2:3:eom8})
contains eight and not nine independent equations.

\subsubsection{Two simple applications: Pion decay and $\pi\pi$ scattering}
\label{subsubsection_pion_decay_pipi_scattering}
   The Lagrangian ${\cal L}_2$ of Eq.\ (\ref{3:2:3:lol}) has
predictive power, once the low-energy coupling constant $F_0$ is identified.
   This LEC may be obtained from the weak decay of the pion,
$\pi^+\to\mu^+\nu_\mu$.
   For that purpose we insert the corresponding external fields
of Eq.~(\ref{2:1:6:rlw}), describing the interaction of quarks with
the massive charged weak bosons, into ${\cal L}_2$.
   The coupling of a single $W$ boson to a single Goldstone boson
originates from the covariant derivatives in ${\cal L}_2$,
\begin{displaymath}
\frac{F^2_0}{4}\mbox{Tr}[D_\mu U (D^\mu U)^\dagger]
=i\frac{F^2_0}{2}\mbox{Tr}(l_\mu \partial^\mu U^\dagger U)+\cdots
=\frac{F_0}{2}\mbox{Tr}(l_\mu \partial^\mu \phi)+\cdots,
\end{displaymath}
and is given by
\begin{equation}
\label{3:2:4:lwphi} {\cal L}_{W\phi}
= -\frac{g}{\sqrt{2}}\frac{F_0}{2}\mbox{Tr}
[({\cal W}_\mu^+T_+ + {\cal W}^-_\mu T_-)\partial^\mu\phi]
= -g \frac{F_0}{2} [{\cal
W}_\mu^+(V_{ud}\partial^\mu\pi^-+V_{us}\partial^\mu K^-) +{\cal
W}_\mu^-(V_{ud}\partial^\mu \pi^++V_{us}\partial^\mu K^+)].
\end{equation}
   The invariant amplitude of the weak pion decay is of the structure
``leptonic vertex $\times$ $W$ propagator $\times$ hadronic vertex,''
\begin{equation}
\label{3:2:4:minv}
{\cal M}=i\left[-\frac{g}{2\sqrt{2}}\bar{u}_{\nu_\mu} \gamma^\rho
(1-\gamma_5)v_{\mu^+}\right]\,\frac{ig_{\rho\sigma}}{M^2_W}
\,i\left[-g\frac{F_0}{2}V_{ud}
(-ip^\sigma)\right]
=-G_F V_{ud} F_0 \bar{u}_{\nu_\mu} p\hspace{-.4em}/(1-\gamma_5)
v_{\mu^+},
\end{equation}
where $G_F$ is the Fermi constant of Eq.\ (\ref{2:1:6:GF}) and
$p$ denotes the four-momentum of the pion.
   In the gauge-boson propagator, momenta $p$ have been neglected in
comparison to the gauge-boson mass $M_W$.
   The corresponding decay rate is
\begin{displaymath}
\frac{1}{\tau}=\frac{G^2_F V_{ud}^2}{4\pi} F^2_0 M_\pi m_\mu^2
\left(1-\frac{m_\mu^2}{M_\pi^2}\right)^2.
\end{displaymath}
   The constant $F_0$ is referred to as the pion-decay constant in the
chiral limit.
   It measures the strength of the matrix element of the axial-vector
current operator between a one-Goldstone-boson state and the vacuum
[see Eq.\ (\ref{2:2:3:acc})].
   Since the interaction of the $W$ boson with the quarks is of the
$V-A$ type and the vector current operator does not contribute to the
matrix element between a single pion and the vacuum, pion decay
is completely determined by the axial-vector current.
   The degeneracy of a single coupling constant $F_0$ is removed at
next-to-leading order, ${\cal O}(q^4)$ \cite{Gasser:1984gg},
once SU(3) symmetry breaking is taken into account.
   The empirical numbers for $F_\pi$ and $F_K$ are
$92.4\,\mbox{MeV}$ and $113\,\mbox{MeV}$, respectively
\cite{PDG_2008}.

\begin{figure}[t]
\begin{center}
\epsfig{file=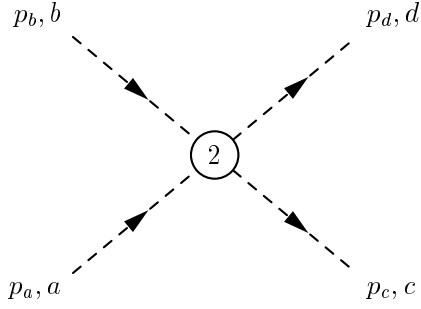,width=0.3\textwidth}
\caption{\label{3:2:4:pipi:fig} Lowest-order Feynman diagram for
$\pi\pi$ scattering. The vertex is derived from ${\cal L}_2$, denoted by
2 in the interaction blob.}
\end{center}
\end{figure}

   Now that the LEC $F_0$ has been identified, we will show
how the lowest-order Lagrangian predicts the prototype of a
Goldstone-boson reaction, namely, $\pi\pi$ scattering.
   We consider ${\cal L}_2$ in the $\mbox{SU}(2)_L\times\mbox{SU}(2)_R$
sector with $r_\mu=l_\mu=0$,
\begin{displaymath}
{\cal L}_2=\frac{F^2}{4}\mbox{Tr} \left(\partial_\mu U
\partial^\mu U^\dagger\right) +\frac{F^2}{4}\mbox{Tr}\left(\chi
U^\dagger+ U \chi^\dagger\right),
\end{displaymath}
where
\begin{displaymath}
\chi=2B\left(\!\!\begin{array}{cc} \hat m &0\\0&\hat
m\end{array} \!\!\right),\quad
U=\exp\left(i\frac{\phi}{F}\right),\quad \phi=\sum_{i=1}^3  \phi_i \tau_i
\equiv\left(\begin{array}{cc}
\pi^0 &\sqrt{2}\pi^+\\
\sqrt{2}\pi^-&-\pi^0
\end{array}\right).
\end{displaymath}
   In the $\mbox{SU}(2)_L\times\mbox{SU}(2)_R$ sector it is common to express quantities in
the chiral limit without index 0, e.g., $F$ and $B$. By this one
means the $\mbox{SU}(2)_L\times\mbox{SU}(2)_R$ chiral limit, i.e., $m_u=m_d=0$ but $m_s$ at its
physical value.
   In the $\mbox{SU}(3)_L\times\mbox{SU}(3)_R$  sector the quantities $F_0$ and $B_0$
denote the chiral limit for all three quarks: $m_u=m_d=m_s=0$.
   Using the substitution $U\leftrightarrow U^\dagger$, we see that
${\cal L}_2$ contains even powers of $\phi$ only:
\begin{displaymath}
{\cal L}_2= {\cal L}_2^{2\phi} +{\cal L}_2^{4\phi} +\cdots.
\end{displaymath}
   Since ${\cal L}_2$ does not produce a vertex with three Goldstone
bosons, there are no $s$-, $u$-, and $t$-channel pole diagrams, i.e.,
   at $D=2$, $\pi\pi$ scattering is entirely generated by a four-Goldstone-boson-interaction
term.
   Expanding
$$
U={\mathbbm 1}+i\frac{\phi}{F}-\frac{1}{2}\frac{\phi^2}{F^2}-\frac{i}{6}
\frac{\phi^3}{F^3}+\frac{1}{24}\frac{\phi^4}{F^4}+\cdots,
$$
the interaction term ${\cal L}_2^{4\phi}$ is identified as
$$
{\cal L}_2^{4\phi}=\frac{1}{48 F^2}\left[\mbox{Tr}(
[\phi,\partial_\mu \phi][\phi, \partial^\mu \phi])
+2B\mbox{Tr}({\cal M}\phi^4)\right].
$$
   We note that substituting $F\to F_0$, $B\to B_0$ and the
relevant expressions for $\phi$ and the quark-mass matrix $\cal M$
the corresponding formula for $\mbox{SU}(3)_L\times\mbox{SU}(3)_R$ looks identical.
   Inserting  $\phi=\phi_i\tau_i$ and working out the traces yields
\begin{eqnarray*}
{\cal L}^{4\phi}_2
&=&\frac{1}{6F^2}(\phi_i\partial^\mu\phi_i\partial_\mu\phi_j\phi_j
-\phi_i\phi_i\partial_\mu\phi_j\partial^\mu\phi_j) +\frac{M^2}{24
F^2}\phi_i\phi_i\phi_j\phi_j,
\end{eqnarray*}
where $M^2=2 B \hat m$.
   The Feynman rule derived from ${\cal L}^{4\phi}_2$ for Cartesian isospin
indices $a$, $b$, $c$, and $d$ reads (see Fig.\ \ref{3:2:4:pipi:fig})
\begin{eqnarray}
\label{3:2:4:mpipi} {\cal M}
&=&i\left[\delta_{ab}\delta_{cd}\frac{s-M^2}{F^2}
+\delta_{ac}\delta_{bd}\frac{t-M^2}{F^2}
+\delta_{ad}\delta_{bc}\frac{u-M^2}{F^2}\right]\nonumber\\
&&-\frac{i}{3F^2}
\left(\delta_{ab}\delta_{cd}+\delta_{ac}\delta_{bd}+\delta_{ad}\delta_{bc}
\right) \left(\Lambda_a+\Lambda_b+\Lambda_c+\Lambda_d\right),
\end{eqnarray}
where $\Lambda_k=p_k^2-M^2$ and $s$, $t$, and $u$ are the usual
Mandelstam variables,
\begin{displaymath}
s=(p_a+p_b)^2,\quad t=(p_a-p_c)^2,\quad u=(p_a-p_d)^2.
\end{displaymath}
    In general, the $T$-matrix element for the scattering
process $\pi_a(p_a)+\pi_b(p_b)\to\pi_c(p_c)+\pi_d(p_d)$
can be parameterized as
\begin{equation}
\label{3:2:4:tpipi}
T^{ab;cd}(p_a,p_b;p_c,p_d)=\delta^{ab}\delta^{cd}A(s,t,u)
                  +\delta^{ac}\delta^{bd}A(t,s,u)
                  +\delta^{ad}\delta^{bc}A(u,t,s),
\end{equation}
where the function $A$ satisfies
$A(s,t,u)=A(s,u,t)$ \cite{Weinberg:1966kf}.
   Since the last line of the Feynman rule of Eq.~(\ref{3:2:4:mpipi} ) disappears, if the external lines
satisfy on-mass-shell conditions, at ${\cal O}(q^2)$ the prediction
for the function $A$ is given by
\begin{equation}
\label{3:2:4:Ato2}
A(s,t,u)=\frac{s-M_\pi^2}{F_\pi^2}.
\end{equation}
   In Eq.~(\ref{3:2:4:Ato2}) we substituted $F_\pi$ for $F$ and $M_\pi$ for $M$,
because the difference is of ${\cal O}(q^4)$ in $T$.
   Equation (\ref{3:2:4:Ato2}) illustrates an important general property
of Goldstone-boson interactions.
   If we consider the (theoretical) limit $M_\pi^2,s,t,u\to 0$, the $T$ matrix
vanishes, $T\to 0$.
   In other words, the strength of Goldstone-boson interactions vanishes
in the zero-energy and mass limit.

   Usually, $\pi\pi$ scattering is discussed in terms of its isospin decomposition.
   Since the pions form an isospin triplet, the two isovectors of both
the initial and final states may be coupled to $I=0,1,2$.
   For $m_u=m_d=\hat m$ the strong interactions are invariant under isospin
transformations, implying that scattering matrix elements can be decomposed
as
\begin{equation}
\label{3:2:4:tdec}
\langle I',I_3'|T|I,I_3\rangle=T^I \delta_{II'}\delta_{I_3 I_3'}.
\end{equation}
    For the case of $\pi\pi$ scattering the three isospin amplitudes
are given in terms of the invariant amplitude $A$ of Eq.\ (\ref{3:2:4:tpipi})
by \cite{Gasser:1983yg}
\begin{eqnarray}
\label{3:2:4:isk}
T^{I=0}&=&3A(s,t,u)+A(t,u,s)+A(u,s,t),\nonumber\\
T^{I=1}&=&A(t,u,s)-A(u,s,t),\nonumber\\
T^{I=2}&=&A(t,u,s)+A(u,s,t).
\end{eqnarray}
   For example, the physical $\pi^+\pi^+$ scattering process is described by
$T^{I=2}$.
   Other physical processes are obtained using the appropriate Clebsch-Gordan
coefficients.

   Evaluating the $T$ matrices at threshold, one obtains the $s$-wave
$\pi\pi$-scattering lengths
\begin{equation}
\label{$:10:swsl}
T^{I=0}|_{\rm thr}=32\pi a^0_0,\quad T^{I=2}|_{\rm thr}=32\pi a^2_0,
\end{equation}
   where the subscript $0$ refers to $s$ wave and the superscript to
the isospin.
   ($T^{I=1}|_{\rm thr}$ vanishes because of Bose symmetry).
    The convention in
ChPT differs by a factor $(-M_\pi)$ from the usual definition of a
scattering length in the effective range expansion.
   The current-algebra prediction of Ref.\ \cite{Weinberg:1966kf} is identical
with the lowest-order result obtained from Eq.\ (\ref{3:2:4:Ato2}),
\begin{equation}
\label{4:10:a00a02lo}
a_0^0=\frac{7 M_\pi^2}{32 \pi F_\pi^2}=0.159,\quad
a_0^2=-\frac{M_\pi^2}{16 \pi F_\pi^2}=-0.0454,
\end{equation}
   where we made use of the numerical values
$F_\pi=92.4$ MeV and $M_\pi=M_{\pi^+}=139.57$ MeV.
   In order to obtain the results of Eq.\ (\ref{4:10:a00a02lo}), use has
been made of $s_{\rm thr}=4 M_\pi^2$ and $t_{\rm thr}=u_{\rm thr}=0$.
   Equations (\ref{4:10:a00a02lo}) represent an absolute prediction of chiral symmetry.
   Once $F_\pi$ is known (from pion decay), the scattering lengths are predicted.
   The $s$-wave $\pi\pi$-scattering lengths have been calculated at next-to-leading (NL) order
\cite{Gasser:1983yg} and at next-to-next-to-leading order \cite{Bijnens:1995yn},
\cite{Bijnens:1997vq}.
   By matching the chiral representation of the scattering amplitude with a dispersive
representation \cite{Roy:1971tc}, \cite{Ananthanarayan:2000ht},
the predictions for the $s$-wave $\pi\pi$-scattering lengths are
\cite{Colangelo:2000jc}, \cite{Colangelo:2001df}
\begin{equation}
\label{4:2:4:a00a02disp}
a_0^0=0.220\pm 0.005,\quad
a_0^2=-0.0444\pm0.0010.
\end{equation}
   The empirical results for the $s$-wave $\pi\pi$-scattering lengths
have been obtained from various sources.
   In the $K_{e4}$ decay $K^+\to\pi^+\pi^-e^+\nu_e$, the connection with
low-energy $\pi\pi$ scattering stems from a partial-wave analysis of the form factors
relevant for the $K_{e4}$ decay in terms of $\pi\pi$ angular momentum eigenstates.
   In the low-energy regime the phases of these form factors are related by
(a generalization of) Watson's theorem \cite{Watson:1954uc} to the
corresponding phases of $I=0$ $s$-wave and $I=1$ $p$-wave elastic
scattering \cite{Colangelo:2001sp}.
   Using effective field theory techniques, isospin-breaking effects generated
by real and virtual photons, and by the mass difference of the up and down quarks
were discussed in Ref.~\cite{Colangelo:2008sm}.
   Performing a combined analysis of the Geneva-Saclay data \cite{Rosselet:1976pu},
the BNL-E865 data \cite{Pislak:2001bf:3}, \cite{Pislak:2003sv}, and the
NA48/2 data \cite{Batley:2007zz} results in
\cite{Colangelo:2008sm}
\begin{equation}
\label{4:10:swslexpold}
a_0^0=0.217\pm 0.008_{\rm exp} \pm 0.006_{\rm th}
\end{equation}
which is in excellent agreement with the prediction of Eq.~(\ref{4:2:4:a00a02disp}).
   The $\pi^\pm p\to \pi^\pm \pi^+ n$ reactions require an extrapolation to the pion
pole to extract the $\pi\pi$ amplitude
and are thus regarded to contain more model dependence,
$a_0^0=0.204\pm 0.014\, (\mbox{stat}) \pm 0.008\, (\mbox{syst})$
\cite{Kermani:1998gq}.
   The DIRAC Collaboration \cite{Adeva:2005pg} makes use of a lifetime measurement of
pionium to extract $|a_0^0-a_0^2|=0.264^{+0.033}_{-0.020}$.
   Finally, in the $K^\pm\to\pi^\pm\pi^0\pi^0$ decay, isospin-symmetry breaking
leads to a cusp structure $\sim a_0-a_2$ in the $\pi^0\pi^0$ invariant mass distribution
near $s_{\pi^0\pi^0}\approx 4M_{\pi^+}^2$ \cite{Cabibbo:2004gq},
\cite{Cabibbo:2005ez}.
   Based on the model of \cite{Cabibbo:2005ez}, the
NA48/2 Collaboration extract $a_0^0-a_0^2=0.268\pm 0.010\,(\mbox{stat})\pm 0.004\,(\mbox{syst})
\pm 0.013\,(\mbox{ext})$.
   A more sophisticated analysis of the cusps in $K\to 3\pi$ within an effective
   field theory framework can be found in Refs.~\cite{Colangelo:2006va}, \cite{Bissegger:2007yq},
   and \cite{Bissegger:2008ff}.

   In particular, when analyzing the data of Ref.\ \cite{Pislak:2001bf:3}
in combination with the Roy equations, an upper limit $|\bar{l}_3|\leq 16$
was obtained in Ref.\ \cite{Colangelo:2001sp} for the
scale-independent low-energy coupling constant which is related to
$l_3$ of the $\mbox{SU}(2)_L\times\mbox{SU}(2)_R$ Lagrangian of
Gasser and Leutwyler \cite{Gasser:1983yg}.
   The great interest generated by this result is to be understood in
the context of the pion mass at ${\cal O}(q^4)$
\begin{equation}
\label{4:10:Mpi2}
M_\pi^2=M^2-\frac{\bar{l}_3}{32\pi^2 F^2} M^4+ {\cal O}(M^6),
\end{equation}
   where $M^2=2\hat m B$. Recall that the constant $B$ is related
to the scalar quark condensate in the chiral limit
and that a non-vanishing quark condensate is a sufficient criterion
for spontaneous chiral symmetry breakdown in QCD.
   If the expansion of $M_\pi^2$ in powers of the quark masses is dominated
by the linear term in Eq.\ (\ref{4:10:Mpi2}), the result is often referred
to as the Gell-Mann-Oakes-Renner relation \cite{Gell-Mann:rz:3}.
   If the terms of order $\hat m^2$ were comparable or even larger than the linear
terms, a different power counting or bookkeeping
in ChPT would be required \cite{Knecht:1995tr:3}.
   The estimate $|\bar{l}_3|\leq 16$ implies that the
Gell-Mann-Oakes-Renner relation is indeed a decent starting
point, because the contribution of the second term of Eq.\ (\ref{4:10:Mpi2})
to the pion mass is approximately given by
\begin{displaymath}
-\frac{\bar{l}_3 M_\pi^2}{64\pi^2 F_\pi^2} M_\pi
=-0.054 M_\pi\,\,\mbox{for}\,\, \bar{l}_3=16,
\end{displaymath}
i.e., more than 94 \% of the pion mass must stem from the quark
condensate \cite{Colangelo:2001sp}.

\subsubsection{Primer to dimensional regularization}
\begin{figure}[t]
\begin{center}
\resizebox{0.1\textwidth}{!}{%
\includegraphics{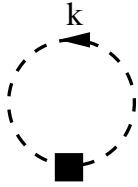}
}
\end{center}
\caption{Generic one-loop diagram. The black box denotes some unspecified
vertex structure which is irrelevant for the discussion.}
\label{3:2:5:fig:dimregexample}
\end{figure}

\begin{figure}[t]
\begin{center}
\epsfig{file=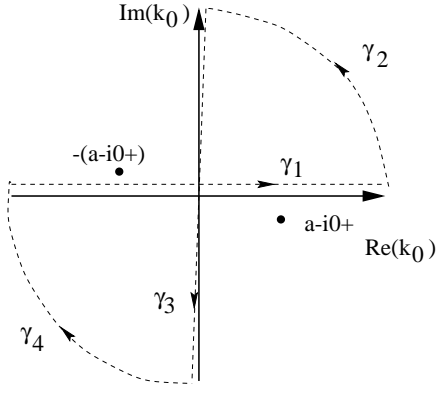,width=0.3\textwidth}
\end{center}
\caption{\label{3:2:4:wickrotation_fig}
Path of integration in the complex $k_0$ plane.}
\end{figure}

    If we want to use the Lagrangian of Eq.~(\ref{3:2:3:lol}) beyond the tree level,
we will encounter ultraviolet divergences from loop integrals.
   For the regularization of the loop diagrams we will make use of dimensional
regularization \cite{'tHooft:fi:10}, \cite{Leibbrandt:1975dj:10}, \cite{'tHooft:1978xw:10},
because it preserves algebraic relations between the Green
functions (Ward identities).
   We will illustrate the method by considering the following simple example,
\begin{equation}
\label{3:2:5:drb:int}
I=\int\frac{\mbox{d}^4k}{(2\pi)^4}\frac{i}{k^2-M^2+i0^+},
\end{equation}
which shows up in the generic diagram of Fig.\ \ref{3:2:5:fig:dimregexample}.
   Introducing $a\equiv\sqrt{\vec{k}\,^2+M^2}>0$, we
define $f(k_0)=\{[k_0+(a-i0^+)][k_0-(a-i0^+)]\}^{-1}$.
   In order to determine $\int_{-\infty}^{\infty} \mbox{d}k_0 f(k_0)$ as part of
the calculation of $I$, we consider
$f$ in the complex $k_0$ plane and make use of Cauchy's theorem
$\oint_C \mbox{d}z f(z)=0$
for functions which are differentiable in every point inside the closed
contour $C$.
   Choosing the path as shown in Fig.~\ref{3:2:4:wickrotation_fig} and taking
account of the fact that the quarter circles at infinity do not contribute, we obtain
the so-called Wick rotation
\begin{equation}
\label{3:2:4:wickrotation}
\int_{-\infty}^\infty \mbox{d}t\,f(t)=-i\int_{\infty}^{-\infty}\mbox{d}t\, f(it)
=i\int_{-\infty}^\infty \mbox{d}t\, f(it).
\end{equation}
   As an intermediate result, the integral of
Eq.\ (\ref{3:2:5:drb:int}) reads
$$I=\int \frac{\mbox{d}^4 l}{(2\pi)^4} \frac{1}{l^2+M^2},
$$
   where $l^2=l_1^2+l_2^2+l_3^2+l_4^2$ denotes a Euclidian scalar
product in four dimensions.
   Performing the angular integration in four dimensions and introducing a cutoff
$\Lambda$ for the radial integration, the integral $I$ diverges quadratically for
large values of $l$ (ultraviolet divergence):
\begin{eqnarray}
\label{2:2:4:Icutoff}
I(\Lambda)&=&\frac{1}{8\pi^2}\int_0^\Lambda\mbox{d}l\frac{l^3}{l^2+M^2}
=\frac{\Lambda^2}{(4\pi)^2}+\frac{M^2}{(4\pi)^2}\ln\left(\frac{M^2}{\Lambda^2+M^2}\right)\nonumber\\
&=&\frac{M^2}{(4\pi)^2}\left[\frac{1}{x}+\ln(x)-\ln(1+x)\right],
\end{eqnarray}
where $x=M^2/\Lambda^2\to 0$ as $\Lambda\to\infty$.
     In dimensional regularization, we generalize the integral from 4 to $n$
dimensions and introduce polar coordinates
\begin{eqnarray*}
l_1&=& l\cos(\theta_1),\quad l_2= l\sin(\theta_1)\cos(\theta_2),\quad
l_3=l\sin(\theta_1)\sin(\theta_2)\cos(\theta_3),\nonumber\\
&\vdots&\nonumber\\
l_{n-1}&=&l\sin(\theta_1)\sin(\theta_2)\cdots\cos(\theta_{n-1}),\quad
l_{n}=l\sin(\theta_1)\sin(\theta_2)\cdots\sin(\theta_{n-1}),
\end{eqnarray*}
where $0\leq l$, $\theta_i\in[0,\pi]$ ($i=1,\cdots,n-2$), and $\theta_{n-1}\in [0,2\pi]$.
    A general integral is then symbolically of the form
\begin{eqnarray*}
\int {\rm d}^n l\cdots = \int_0^\infty {\rm d}l\,l^{n-1}
\int_0^{2\pi}{\rm d}\theta_{n-1} \int_0^\pi
{\rm d}\theta_{n-2}\sin(\theta_{n-2})\cdots\int_0^\pi {\rm d}\theta_1
\sin^{n-2}(\theta_1)\cdots.
\end{eqnarray*}
   If the integrand does not depend on the angles, the angular integration
can explicitly be carried out:
\begin{displaymath}
\int {\rm d}\Omega_n=
2\frac{\pi^{\frac{n}{2}}}{\Gamma\left(\frac{n}{2}\right)}.
\end{displaymath}
   We define the integral for $n$ dimensions ($n$ integer) as
\begin{displaymath}
I_n(M^2,\mu^2)=\mu^{4-n}\int\frac{{\rm d}^nk}{(2\pi)^n}\frac{i}{k^2-M^2+i0^+},
\end{displaymath}
where the scale $\mu$ ('t Hooft parameter, renormalization scale) has been introduced
so that the integral has the same dimension for arbitrary $n$.
   The integral formally reads
\begin{eqnarray}
\label{2:2:4:iint}
I_n(M^2,\mu^2)&=& \mu^{4-n}2\frac{\pi^{\frac{n}{2}}}{\Gamma\left(
\frac{n}{2}\right)} \frac{1}{(2\pi)^n}
\int_0^\infty {\rm d}l
\frac{l^{n-1}}{l^2+M^2}\nonumber\\[0.5em]
&=&\mu^{4-n}\,\,2\frac{\pi^{\frac{n}{2}}}{\Gamma\left( \frac{n}{2}\right)}\,\,
\frac{1}{(2\pi)^n}\,\, \frac{1}{2} (M^2)^{\frac{n}{2}-1}\frac{\Gamma
\left(\frac{n}{2}\right)\Gamma\left(1-\frac{n}{2}\right)}{
\Gamma(1)}\nonumber\\
&=&\frac{\mu^{4-n}}{(4\pi)^{\frac{n}{2}}} (M^2)^{\frac{n}{2}-1}
\Gamma\left(1-\frac{n}{2}\right).
\end{eqnarray}
      Since $\Gamma(z)$ is an analytic function in the complex plane except
for poles of first order in $0,-1,-2,\cdots$, and
$a^z=\exp[\ln(a)z]$, $a\in {\mathbbm R}^+$ is an analytic function in $\mathbbm C$,
the right-hand side of Eq.\ (\ref{2:2:4:iint}) can be thought of
as a function of a {\em complex} variable $n$ which is
analytic in $\mathbbm C$ except for poles of first order
for $n=2,4,6,\cdots$.
   The analytic continuation for complex $n$ reads
\begin{eqnarray}
\label{3:2:4:IMmun}
I(M^2,\mu^2,n)&=&
\frac{M^2}{(4\pi)^2}\left(\frac{4\pi\mu^2}{M^2}\right)^{2-\frac{n}{2}}
\Gamma\left(1-\frac{n}{2}\right)
=\frac{M^2}{16\pi^2}\left[
R+\ln\left(\frac{M^2}{\mu^2}\right)\right]+O(n-4),
\end{eqnarray}
where
\begin{equation}
\label{3:2:5:R}
R=\frac{2}{n-4}
-[\mbox{ln}(4\pi)+\Gamma'(1)+1].
\end{equation}
   The comparison between Eqs.~(\ref{3:2:4:IMmun}) and (\ref{2:2:4:Icutoff})
illustrates the following general observations:
   in dimensional regularization power-law divergences are
analytically continued to zero and
   logarithmic ultraviolet divergences of one-loop integrals show up as
single poles in $\epsilon=4-n$.

\subsubsection{Power-counting scheme}
   The Lagrangian ${\cal L}_{\rm eff}$ of mesonic chiral perturbation theory is
organized as a string of terms with an increasing number of derivatives and
quark-mass terms,
\begin{equation}
\label{3:2:3:ll2l4}
{\cal L}_{\rm eff}={\cal L}_2 + {\cal L}_4 + {\cal L}_6 +\cdots,
\end{equation}
where the subscripts refer to the order in the momentum and quark-mass expansion.
   The index 2, for example, denotes either two derivatives or one quark-mass term.
   In terms of Feynman rules, derivatives generate four-momenta.
   A quark-mass term counts as two derivatives because of Eqs.\
(\ref{3:2:2:mpi2}) - (\ref{3:2:2:meta2}) ($M^2\sim m_q$) in combination with the
on-shell condition $p^2=M^2$.
   We will generically count a small four-momentum---or the corresponding
derivative---and a Goldstone-boson mass as of ${\cal O}(q)$.
   The chiral orders in Eq.\ (\ref{3:2:3:ll2l4}) are all even [${\cal O}(q^{2k})$,
$k\geq 1$], because Lorentz indices of derivatives always have to be
contracted and quark-mass terms count as ${\cal O}(q^2)$.

\begin{figure}[t]
\begin{minipage}[t]{0.25\textwidth}
\begin{center}
\epsfig{file=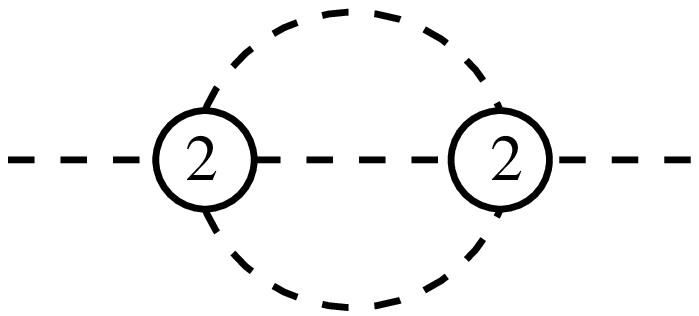,width=0.8\textwidth}
\end{center}
\begin{displaymath}
D=4\cdot 2-2\cdot 3+2\cdot 2=6.
\end{displaymath}
\end{minipage}
\hspace{0.1\textwidth}
\begin{minipage}[t]{0.25\textwidth}
\begin{center}
\epsfig{file=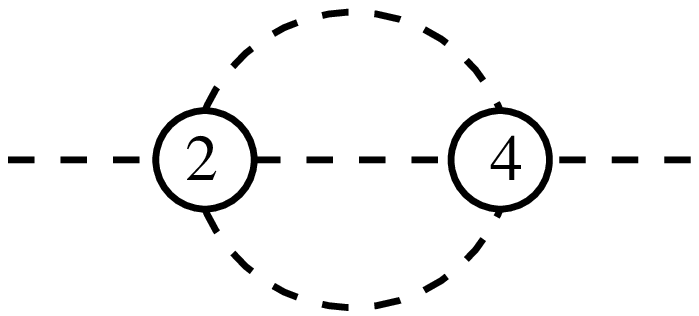,width=0.8\textwidth}
\end{center}
\begin{displaymath}
D=4\cdot 2-2\cdot 3+2\cdot 1 +4 \cdot 1=8.
\end{displaymath}
\end{minipage}
\hspace{0.1\textwidth}
\begin{minipage}[t]{0.25\textwidth}
\begin{center}
\epsfig{file=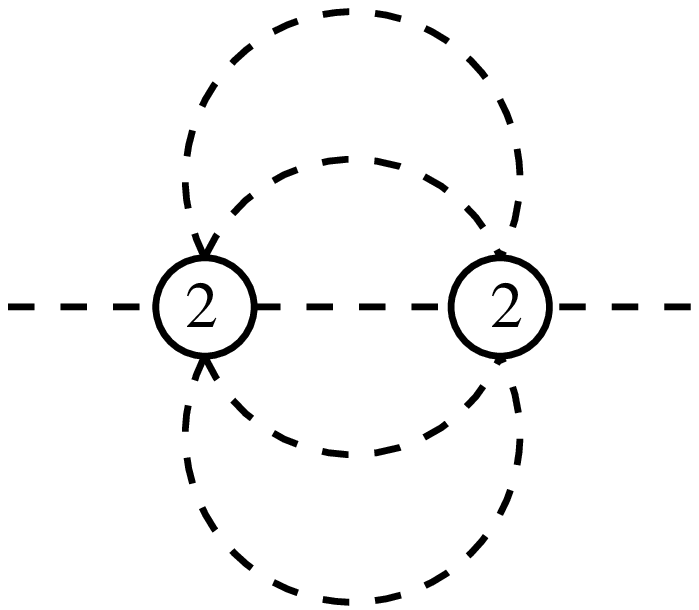,width=0.8\textwidth}
\end{center}
\begin{displaymath}
D=4\cdot 4 -2\cdot 5+2\cdot 2 = 10.
\end{displaymath}
\end{minipage}
\caption{\label{3:2:3:figure:power_counting} Application of the
power-counting formula of Eq.~(\ref{3:2:3:mr2a}) in
$n=4$ dimensions.}
\end{figure}

   Besides the knowledge of the most general Lagrangian, we need a method which
allows one to assess the importance of different renormalized diagrams contributing
to a given process.
   For that purpose we analyze a given diagram under a simultaneous re-scaling of
all {\em external} momenta, $p_i\mapsto t p_i$, and
the light-quark masses, $m_q\mapsto t^2 m_q$ (corresponds to $M^2\mapsto t^2 M^2$).
   The chiral dimension $D$ of a given diagram is defined as
\begin{equation}
\label{3:2:3:mr1} {\cal M}(tp_i, t^2 m_q,t\mu)=t^D {\cal M}(p_i,m_q,\mu)={\cal
O}(q^D).
\end{equation}
  For small enough momenta (and masses) contributions
with increasing $D$ become less important.
   The chiral dimension is given by
\begin{eqnarray}
\label{3:2:3:mr2a} D&=&n N_L-2N_I+\sum_{k=1}^\infty 2k N_{2k}\\&=&
2+(n-2)N_L+\sum_{k=1}^\infty 2(k-1)N_{2k} \label{3:2:3:mr2b}
\\
&\geq&2\,\,\,{\rm in\,\, 4\,\, dimensions,}\nonumber
\end{eqnarray}
where $n$ is the number of space-time dimensions,
$N_L$ the number of independent loops,
$N_I$ the number of internal Goldstone boson lines, and
$N_{2k}$ the number of vertices from ${\cal L}_{2k}$.
   Equation (\ref{3:2:3:mr2b}) establishes a relation
between the momentum and loop expansion, because at each chiral
order, the maximum number of loops is bounded from above.
   In other words, we have a perturbative scheme in terms of external momenta
and masses which are small compared to some scale
[here: $\Lambda_\chi=4\pi F_0={\cal O}$ (1 GeV)].
   Examples of the application of the power-counting formula are shown in
Fig.\ \ref{3:2:3:figure:power_counting}.

   In order to prove the power-counting formula we start from the
Feynman rules for evaluating the S-matrix element and investigate the
behavior of the individual building blocks.
   Internal lines are described by a propagator in $n$ dimensions
which under re-scaling behaves as
\begin{displaymath}
\int \frac{\mbox{d}^n k}{(2\pi)^n} \frac{i}{k^2-M^2+i0^+}
\to \int \frac{\mbox{d}^n k}{(2\pi)^n}
\frac{i}{t^2(k^2/t^2-M^2+i0^+)}
\stackrel{\mbox{$k=tk'$}}{=}t^{n-2}
\int \frac{\mbox{d}^n k'}{(2\pi)^n} \frac{i}{k'^2-M^2+i0^+}.
\end{displaymath}
   Vertices with $2k$ derivatives or $k$ quark-mass terms re-scale as
\begin{displaymath}
\delta^n(q) q^{2k}\to t^{2k-n}\delta^n(q) q^{2k},
\end{displaymath}
since $p\to tp$ if $q$ is an external momentum,  and
$k=tk'$ if $q$ is an internal momentum (see above).
   These are the rules to calculate $S \sim \delta^n(P) \cal M$.
   We need to add $n$ to compensate for the overall momentum-conserving
delta function.
   Applying these rules, the scaling behavior of the contribution to $\cal M$ of
a given diagram reads
\begin{displaymath}
D=n+(n-2)N_I+\sum_{k=1}^\infty N_{2k} (2k-n).
\end{displaymath}
   The relation between the number of independent loops, the number of internal
lines, and the total number of vertices $N_V=\sum_{k=1}^\infty N_{2k}$ is given
by $N_L=N_I-(N_V-1)$.
   The product of $N_V$ momentum-conserving $\delta$ functions
contains overall momentum conservation.
   Therefore, one has $N_V-1$ rather
than $N_V$ restrictions on the internal momenta.
    Applying
\begin{displaymath}
-n \sum_{k=1}^\infty N_{2k}=-n N_V=n(N_L-N_I-1)
\end{displaymath}
results in Eq.~(\ref{3:2:3:mr2a}):
\begin{displaymath}
D=n N_L-2N_I+\sum_{k=1}^\infty 2k N_{2k}.
\end{displaymath}
On the other hand, applying
\begin{displaymath}
-n \sum_{k=1}^\infty N_{2k}= -2 \sum_{k=1}^\infty N_{2k}
+(n-2)(N_L-N_I-1),
\end{displaymath}
results in Eq.~(\ref{3:2:3:mr2b}):
\begin{displaymath}
D=2+\sum_{k=1}^\infty2(k-1)N_{2k}+(n-2)N_L.
\end{displaymath}
In particular, diagrams containing loops are suppressed due to the
term $2 N_L$ in four dimensions.

\begin{figure}[t]
\begin{minipage}[t]{0.45\textwidth}
\begin{center}
\epsfig{file=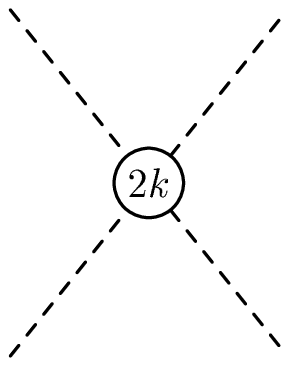,width=0.3\textwidth}
$$D=0-0+2k=2k=\left\{
\begin{array}{l}
0\,\,\mbox{for}\,\,k=0,\\
2\,\,\mbox{for}\,\,k=1.
\end{array}\right.
$$
\end{center}
\end{minipage}
\hspace{0.1\textwidth}
\begin{minipage}[t]{0.3\textwidth}
\begin{center}
\epsfig{file=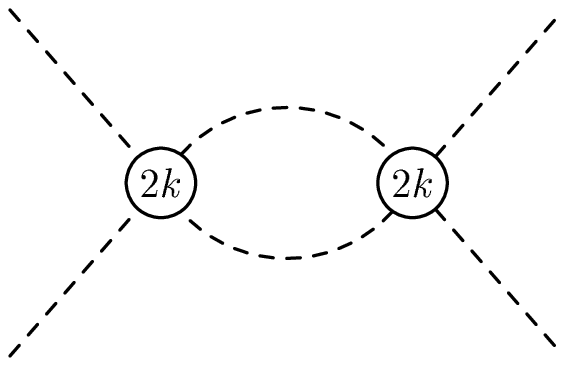,width=\textwidth}
$$D=4-2\cdot 2+2\cdot (2k)=4k=\left\{
\begin{array}{l}
0\,\,\mbox{for}\,\,k=0,\\
4\,\,\mbox{for}\,\,k=1.
\end{array}\right.
$$
\end{center}
\end{minipage}
\caption{\label{3:2:5:minimalk:fig}
The loop diagram is only suppressed if $k_{\rm min} > 0$.}
\end{figure}

   Note that a minimal $k>0$ is important.
   Otherwise, an infinite number of diagrams containing vertices from
${\cal L}_0$ would have to be summed (see Fig.\ \ref{3:2:5:minimalk:fig}).
   This is for example the case when dealing with the nucleon-nucleon
interaction.

\subsection{Next-to-leading order}
   Already in 1967 it was shown by Weinberg \cite{Weinberg:1966fm} that an
effective Lagrangian is a convenient tool for reproducing
the results of current algebra in terms of tree-level calculations.
   In the purely mesonic sector, ${\cal L}_2$ of Eq.\ (\ref{3:2:3:lol})
represents the corresponding Lagrangian.
   It was noted by Li and Pagels \cite{Li:1971vr} that a perturbation theory around
a symmetry which is realized in the Nambu-Goldstone mode, in general, leads
to observables which are non-analytic functions of the symmetry-breaking
parameters, here the quark masses.
   In 1979 Weinberg initiated the application of an effective-field-theory
program beyond the tree level allowing for a systematic
calculation of corrections to the chiral limit \cite{Weinberg:1978kz}.
   When calculating one-loop graphs, using vertices from ${\cal L}_2$, one generates ultraviolet
divergences which in the framework of dimensional regularization appear
as poles at space-time dimension $n=4$.
   The loop diagrams are renormalized by absorbing the infinite parts
into the redefinition of the fields and the parameters of the most
general Lagrangian.
   Since ${\cal L}_2$ is not renormalizable in
the traditional sense, the infinities cannot be absorbed by a renormalization
of the coefficients $F_0$ and $B_0$.
   However, to quote from Ref.\ \cite{Weinberg:1995mt}:
   ``... the cancelation of ultraviolet divergences does
not really depend on renormalizability; as long as we include every one of the
infinite number of interactions allowed by symmetries, the so-called
non-renormalizable theories are actually just as renormalizable as renormalizable
theories.''
   According to Weinberg's power counting of Eq.\ (\ref{3:2:3:mr2b}),
one-loop graphs with vertices from ${\cal L}_2$ are of ${\cal O}(q^4)$.
   The conclusion is that one needs to construct the most general Lagrangian
${\cal L}_4$ and to adjust (renormalize) its parameters to cancel one-loop infinities.

   Beyond the quantum corrections to processes already described by ${\cal L}_2$,
at next-to-leading order we will encounter another important feature,
namely, the effective Wess-Zumino-Witten (WZW) action.
   The WZW action provides an effective description of the
constraints due to the anomalous Ward identities.
   In general, anomalies arise if the symmetries of the Lagrangian at the classical
level are not supported by the quantized theory after renormalization.

\subsubsection{The ${\cal O}(q^4)$ Lagrangian of Gasser and Leutwyler}
\label{subsubsec_clop4}
   The most general $\mbox{SU(3)}_L\times\mbox{SU(3)}_R$-invariant Lagrangian at ${\cal O}(q^4)$
is given by \cite{Gasser:1984gg}
\begin{eqnarray}
\label{3:3:1:l4gl}
\lefteqn{{\cal L}_4=
L_1 \left\{\mbox{Tr}[D_{\mu}U (D^{\mu}U)^{\dagger}] \right\}^2
+ L_2 \mbox{Tr} \left [D_{\mu}U (D_{\nu}U)^{\dagger}\right]
\mbox{Tr} \left [D^{\mu}U (D^{\nu}U)^{\dagger}\right]}\nonumber\\
& & + L_3 \mbox{Tr}\left[
D_{\mu}U (D^{\mu}U)^{\dagger}D_{\nu}U (D^{\nu}U)^{\dagger}
\right ]
+ L_4 \mbox{Tr} \left [ D_{\mu}U (D^{\mu}U)^{\dagger} \right ]
\mbox{Tr} \left( \chi U^{\dagger}+ U \chi^{\dagger} \right )
\nonumber \\
& & +L_5 \mbox{Tr} \left[ D_{\mu}U (D^{\mu}U)^{\dagger}
(\chi U^{\dagger}+ U \chi^{\dagger})\right]
+ L_6 \left[ \mbox{Tr} \left ( \chi U^{\dagger}+ U \chi^{\dagger} \right )
\right]^2
\nonumber \\
& & + L_7 \left[ \mbox{Tr} \left ( \chi U^{\dagger} - U \chi^{\dagger} \right )
\right]^2
+ L_8 \mbox{Tr} \left ( U \chi^{\dagger} U \chi^{\dagger}
+ \chi U^{\dagger} \chi U^{\dagger} \right )
\nonumber \\
& & -i L_9 \mbox{Tr} \left [ f^R_{\mu\nu} D^{\mu} U (D^{\nu} U)^{\dagger}
+ f^L_{\mu\nu} (D^{\mu} U)^{\dagger} D^{\nu} U \right ]
+ L_{10} \mbox{Tr} \left ( U f^L_{\mu\nu} U^{\dagger} f_R^{\mu\nu} \right )
\nonumber \\
& & + H_1 \mbox{Tr} \left ( f^R_{\mu\nu} f^{\mu\nu}_R +
f^L_{\mu\nu} f^{\mu\nu}_L \right )
+ H_2 \mbox{Tr} \left ( \chi \chi^{\dagger} \right ).
\end{eqnarray}
   The numerical values of the low-energy coupling constants $L_i$ are not
determined by chiral symmetry.
   In analogy to $F_0$ and $B_0$ of ${\cal L}_2$ they are parameters
containing information on the underlying dynamics
and should,
in principle, be calculable in terms of the (remaining) parameters of
QCD, namely, the heavy-quark masses and the QCD scale $\Lambda_{\rm QCD}$.
   In practice, they parameterize our inability to solve the
dynamics of QCD in the non-perturbative regime.
   So far they have either been fixed using empirical input
or theoretically using QCD-inspired models, meson-resonance
saturation \cite{Ecker:1988te} \cite{Pich:2008xj}, and lattice QCD (see Ref.\ \cite{Necco:2009cq} for
a recent overview).

   By construction Eq.\ (\ref{3:3:1:l4gl}) represents the most general
Lagrangian at ${\cal O}(q^4)$, and it is thus possible to absorb the one-loop
divergences by an appropriate renormalization of the coefficients
$L_i$ and $H_i$:
\begin{equation}
\label{3:3:1:lihi}
L_i=L_i^r+\frac{\Gamma_i}{32\pi^2}R \quad (i=1,\cdots,10),\quad
H_i=H^r_i+\frac{\Delta_i}{32\pi^2}R\quad (i=1,2),
\end{equation}
   where $R$ has already been defined in Eq.~(\ref{3:2:5:R}):
\begin{displaymath}
R=\frac{2}{n-4}
-[\mbox{ln}(4\pi)+\Gamma'(1)+1],
\end{displaymath}
   with $n$ denoting the number of space-time dimensions and
$\gamma_E=-\Gamma'(1)$ being Euler's constant.
   The constants $\Gamma_i$ and $\Delta_i$ are given in Table
\ref{4:8:tableli}.
   Except for $L_3$ and $L_7$, the low-energy coupling constants $L_i$ and the
``contact terms''---i.e., pure external field terms---$H_1$ and $H_2$
are required in the renormalization of the one-loop graphs.
   Since $H_1$ and $H_2$ contain only external fields, they are of no
physical relevance.
\begin{table}[t]
\begin{center}
\begin{tabular}{|c|r|r|}
\hline
Coefficient &Empirical Value &$\Gamma_i$\\
\hline
$L_1^r$ &    $ 0.4\pm 0.3$  &$\frac{3}{32}$\\
$L_2^r$ &    $ 1.35\pm 0.3$ &$\frac{3}{16}$\\
$L_3^r$ &    $-3.5\pm 1.1$  &$0$\\
$L_4^r$ &    $-0.3\pm 0.5$  &$\frac{1}{8}$\\
$L_5^r$ &    $ 1.4\pm 0.5$  &$\frac{3}{8}$\\
$L_6^r$ &    $-0.2\pm 0.3$  &$\frac{11}{144}$\\
$L_7^r$ &    $-0.4\pm 0.2$  &$0$\\
$L_8^r$ &    $ 0.9\pm 0.3$  &$\frac{5}{48}$\\
$L_9^r$ &    $ 6.9\pm 0.7$  &$\frac{1}{4}$\\
$L_{10}^r$ & $-5.5\pm 0.7$  &$-\frac{1}{4}$\\
\hline
\end{tabular}
\end{center}
\caption[test]{\label{4:8:tableli} Renormalized low-energy
coupling constants $L_i^r$ in units of $10^{-3}$ at the scale
$\mu=M_\rho$, see \cite{Bijnens:1994qh:9}.
$\Delta_1=-1/8$, $\Delta_2=5/24$. }
\end{table}
   The idea of renormalization consists of adjusting the parameters of the
counter terms of the most general effective Lagrangian
so that they cancel the divergences of (multi-) loop diagrams.
   In doing so, one still has the freedom of choosing a suitable
renormalization condition.
   For example, in the minimal subtraction scheme (MS) one would fix the
parameters of the counter term Lagrangian such that they would precisely absorb
the contributions proportional to $2/(n-4)$ in $R$,
while the modified minimal subtraction scheme of ChPT ($\widetilde{\rm MS}$) would,
in addition, cancel the term in the square brackets.

   The renormalized coefficients $L_i^r$ depend on the scale $\mu$
introduced by dimensional regularization [see Eq.\ (\ref{3:2:4:IMmun})] and
their values at two different scales $\mu_1$ and $\mu_2$
are related by
\begin{equation}
\label{3:3:1:limu1mu2}
L^r_i(\mu_2)=L^r_i(\mu_1)
+\frac{\Gamma_i}{16\pi^2}\ln\left(\frac{\mu_1}{\mu_2}\right).
\end{equation}
   We will see that the scale dependence of the coefficients and
the finite part of the loop-diagrams compensate each other in
such a way that physical observables are scale independent.

   A discussion of the two-flavor Lagrangian at ${\cal O}(q^4)$
\cite{Gasser:1983yg} can be found in Appendix D of Ref.\ \cite{Scherer:2002tk}.
   For the construction of the ${\cal O}(q^6)$ Lagrangian of even intrinsic
parity, see Refs.\ \cite{Scherer:1994wi}, \cite{Fearing:1994ga} and \cite{Bijnens:1999sh}.
   For a status report on mesonic chiral perturbation theory beyond the one-loop level, we
refer the reader to Ref.\ \cite{Bijnens:2006zp}.

\subsubsection{The effective Wess-Zumino-Witten action}
   The Lagrangians  discussed so far have a larger symmetry than QCD \cite{Witten:1983tw}.
   For example, if we consider the case of ``pure'' QCD, i.e., no external
fields except for the quark-mass term $\chi=2B_0 {\cal M}$, ${\cal L}_2$ and ${\cal L}_4$
contain interaction terms with an even number of Goldstone bosons
only (even intrinsic parity).
   In other words, they cannot describe, e.g, $K^+K^-\to\pi^+\pi^-\pi^0$.
   Analogously, ${\cal L}_2$ and ${\cal L}_4$ including a coupling to electromagnetic
fields cannot describe $\pi^0\to\gamma\gamma$.

   In order to overcome this shortcoming, Witten suggested to add the simplest term
possible which breaks the symmetry of having only an even number of Goldstone
bosons at the Lagrangian level.
   For the case of massless Goldstone bosons without any
external fields the modified equation of motion reads
\begin{equation}
\label{3:3:2:modeom}
\partial_\mu\left(\frac{F_0^2}{2}U\partial^\mu U^\dagger\right)
+\lambda \epsilon^{\mu\nu\rho\sigma} U\partial_\mu U^\dagger
U\partial_\nu U^\dagger U\partial_\rho U^\dagger U\partial_\sigma
U^\dagger=0,
\end{equation}
where $\lambda$ is a (purely imaginary) constant.
   For the purpose of writing down an action corresponding to Eq.~(\ref{3:3:2:modeom}),
we extend the range of definition of the fields to a
hypothetical fifth dimension,
\begin{equation}
\label{3:2:ualpha}
U(y)=\exp\left(i\alpha\frac{\phi(x)}{F_0}\right),
\quad y^i=(x^\mu,\alpha),\,\,i=0,\cdots, 4,\,\, 0\leq\alpha\leq 1,
\end{equation}
   where Minkowski space is defined as the surface of the five-dimensional
space for $\alpha =1$.
   The action in in the absence of external fields (denoted by a superscript 0)
is given by \cite{Witten:1983tw}
\begin{equation}
\label{3:2:2:sano}
S_{\rm ano}^0=n S_{\rm WZW}^0,\quad
S_{\rm WZW}^0=-\frac{i}{240\pi^2}\int_0^1 \mbox{d} \alpha \int \mbox{d}^4 x
\epsilon^{ijklm} \mbox{Tr}\left( {\cal U}^L_i\cdots {\cal U}^L_m
\right),
\end{equation}
where
\begin{displaymath}
\epsilon_{01234}=-\epsilon^{01234}=1,\quad {\cal U}^L_i=U^\dagger
\frac{\partial U}{\partial y^i},\quad \lambda=\frac{in}{48\pi^2}.
\end{displaymath}
   A rather unusual and surprising feature of Eq.~(\ref{3:2:2:sano})
is that the action functional corresponding to the new term cannot be
written as the four-dimensional integral of a Lagrangian expressed
in terms of $U$ and its derivatives.
   Wess and Zumino derived consistency or integrability
relations which are satisfied by the anomalous Ward identities
and then explicitly constructed a functional involving the pseudoscalar octet
which satisfies the anomalous Ward identities \cite{Wess:1971yu}.
   In particular, Wess and Zumino emphasized that their interaction
Lagrangians cannot be obtained as part of a chiral invariant Lagrangian.
   Using topological arguments Witten showed that the constant $n$
appearing in Eq.\ (\ref{3:2:2:sano}) must be an integer.
   However, it was pointed out in Ref.~\cite{Bar:2001qk} that the traditional
argument relating $n$ with the number of colors $N_c$ is
incomplete. Before discussing this argument, let us investigate the consequences
of $S^0_{\rm WZW}$.

   Expanding the SU(3) matrix $U(y)$ in terms of the Goldstone
boson fields, $U(y)={\mathbbm 1}+i\alpha \phi(x)/F_0+O(\phi^2)$, one obtains
an infinite series of terms, each involving an odd number of
Goldstone bosons, i.e., the WZW action $S_{\rm WZW}^0$
is of odd intrinsic parity.
   For each individual term the $\alpha$ integration can be performed
explicitly resulting in an ordinary action in terms of a four-dimensional
integral of a local Lagrangian.
   For example, the term with the smallest number of Goldstone bosons
reads
\begin{equation}
\label{3:2:2:swzw5phi}
S_{\rm WZW}^{5\phi}
=\frac{1}{240\pi^2 F^5_0}\int \mbox{d}^4 x
\epsilon^{\mu\nu\rho\sigma}\mbox{Tr}(\phi\partial_\mu\phi\partial_\nu\phi
\partial_\rho\phi\partial_\sigma\phi),
\end{equation}
which describes, e.g., $K^+K^-\to\pi^+\pi^-\pi^0$.
   In particular, the WZW action without external fields involves at
least five Goldstone bosons \cite{Wess:1971yu}.

   The connection to the number of colors $N_c$ is established by
introducing a coupling to electromagnetism \cite{Wess:1971yu}, \cite{Witten:1983tw}.
   In the presence of external fields there will be an additional
term in the anomalous action,
\begin{equation}
\label{4:8:sanofull}
S_{\rm ano}=
S_{\rm ano}^0+
S_{\rm ano}^{\rm ext}=n(S_{\rm WZW}^0+S_{\rm WZW}^{\rm ext}),
\end{equation}
given by (see, e.g., Ref.\ \cite{Bijnens:1993xi})
\begin{equation}
\label{3:3:2:deltaswzw}
S_{\rm WZW}^{\rm ext}= -\frac{i}{48\pi^2}\int \mbox{d}^4 x\,
\epsilon^{\mu\nu\rho\sigma}\mbox{Tr}(Z_{\mu\nu\rho\sigma})
\end{equation}
with
\begin{eqnarray}
\label{3:3:2:zmunialphabeta}
Z_{\mu\nu\rho\sigma}&=&
\frac{1}{2}\ U l_\mu U^\dagger r_\nu U l_\rho U^\dagger r_\sigma
+U l_\mu l_\nu l_\rho U^\dagger r_\sigma
 - U^\dagger r_\mu r_\nu r_\rho U l_\sigma\nonumber\\
&&+i  U\partial_\mu l_\nu l_\rho U^\dagger r_\sigma
 - i U^\dagger\partial_\mu r_\nu r_\rho U l_\sigma
 +i \partial_\mu r_\nu U l_\rho U^\dagger r_\sigma
 - i \partial_\mu l_\nu U^\dagger r_\rho U l_\sigma\nonumber\\
&&-i {\cal U}_\mu^L l_\nu U^\dagger r_\rho U l_\sigma
 + i {\cal U}_\mu^R r_\nu U l_\rho U^\dagger r_\sigma
 -i {\cal U}_\mu^L l_\nu l_\rho l_\sigma
 + i {\cal U}_\mu^R r_\nu r_\rho r_\sigma\nonumber\\
&&+\frac{1}{2}{\cal U}_\mu^L U^\dagger\partial_\nu r_\rho U l_\sigma
 -\frac{1}{2}{\cal U}_\mu^R U\partial_\nu l_\rho U^\dagger r_\sigma
 +\frac{1}{2}{\cal U}_\mu^L U^\dagger r_\nu U\partial_\rho l_\sigma
 -\frac{1}{2}{\cal U}_\mu^R U l_\nu U^\dagger\partial_\rho r_\sigma\nonumber\\
&&-{\cal U}_\mu^L{\cal U}_\nu^L U^\dagger r_\rho U l_\sigma
 + {\cal U}_\mu^R{\cal U}_\nu^R U l_\rho U^\dagger r_\sigma
+\frac{1}{2}\ {\cal U}_\mu^L l_\nu {\cal U}_\rho^L l_\sigma
 - \frac{1}{2}\ {\cal U}_\mu^R r_\nu {\cal U}_\rho^R r_\sigma\nonumber\\
&&+{\cal U}_\mu^L l_\nu\partial_\rho l_\sigma
 - {\cal U}_\mu^R r_\nu\partial_\rho r_\sigma+{\cal U}_\mu^L\partial_\nu  l_\rho l_\sigma
 - {\cal U}_\mu^R\partial_\nu  r_\rho r_\sigma\nonumber\\
&&-i {\cal U}_\mu^L{\cal U}_\nu^L{\cal U}_\rho^L l_\sigma
 + i {\cal U}_\mu^R{\cal U}_\nu^R{\cal U}_\rho^R r_\sigma,
\end{eqnarray}
where we defined the abbreviations ${\cal U}_\mu^L=U^\dagger\partial_\mu U$
and ${\cal U}_\mu^R=U\partial_\mu U^\dagger$.

   As a special case, let us consider a coupling to external electromagnetic
fields by inserting
\begin{displaymath}
r_\mu=l_\mu=-e Q {\cal A}_\mu,
\end{displaymath}
where $Q$ is the quark-charge matrix.
   The terms involving three and four electromagnetic four-potentials
vanish upon contraction with the totally antisymmetric tensor
$\epsilon^{\mu\nu\rho\sigma}$, because their contributions to
$Z_{\mu\nu\rho\sigma}$ are symmetric in at least two indices, and
we obtain
\begin{equation}
\label{3:3:2:lanoelm}
n{\cal L}^{\rm ext}_{\rm WZW}=-e n {\cal A}_\mu J^\mu
+i \frac{n e^2 }{48\pi^2}\epsilon^{\mu\nu\rho\sigma}
\partial_\nu {\cal A}_\rho{\cal A}_\sigma
\mbox{Tr}[2Q^2(U\partial_\mu U^\dagger - U^\dagger \partial_\mu U)
- Q U^\dagger Q \partial_\mu U
+Q U Q \partial_\mu U^\dagger].
\end{equation}
   We note that the current
\begin{equation}
\label{3:3:2:jmu}
J^\mu=\frac{\epsilon^{\mu\nu\rho\sigma}}{48\pi^2}
\mbox{Tr}(Q\partial_\nu U U^\dagger \partial_\rho U U^\dagger
\partial_\sigma U U^\dagger
+Q U^\dagger \partial_\nu U U^\dagger \partial_\rho U U^\dagger
\partial_\sigma U),\quad \epsilon_{0123}=1,
\end{equation}
by itself is not gauge invariant and the additional terms of
Eq.\ (\ref{3:3:2:lanoelm}) are required to obtain a gauge-invariant action.
   The standard procedure of determining $n$ is to investigate
the interaction Lagrangian which is relevant to the decay
$\pi^0\to\gamma\gamma$ by expanding
$
U={\mathbbm 1}+i\mbox{diag}(\pi^0,-\pi^0,0)/F_0+\cdots.
$
   However, as pointed out by B\"ar and Wiese, when considering the electromagnetic
interaction for an arbitrary number of colors one should replace the ordinary
quark charge matrix by
\begin{displaymath}
Q=\left(
\begin{array}{rrr}
\frac{2}{3}&0&0\\
0&-\frac{1}{3}&0\\
0&0&-\frac{1}{3}
\end{array}
\right)
\to
\left(
\begin{array}{ccc}
\frac{1}{2N_c}+\frac{1}{2}&0&0\\
0&\frac{1}{2N_c}-\frac{1}{2}&0\\
0&0&\frac{1}{2N_c}-\frac{1}{2}
\end{array}
\right).
\end{displaymath}
   The corresponding effective Lagrangian for $\pi^0\to\gamma\gamma$ decay,
\begin{displaymath}
{\cal L}_{\pi^0\gamma\gamma}=-\frac{n}{N_c}
\frac{e^2}{32\pi^2}\epsilon^{\mu\nu \rho\sigma}{\cal
F}_{\mu\nu}{\cal F}_{\rho\sigma}\frac{\pi^0}{F_0},
\end{displaymath}
results in the decay rate
\begin{displaymath}
\Gamma_{\pi^0\to\gamma\gamma}= \frac{\alpha^2 M^3_{\pi^0}}{64 \pi^3
F^2_0}\,\frac{n^2}{N_c^2}
=7.6\,\mbox{eV}\times\left(\frac{n}{N_c}\right)^2
\end{displaymath}
   in good agreement with the experimental value
$(7.7\pm 0.6)$ eV for $n=N_c$.
   However, the result is no indication for $N_c=3$ \cite{Bar:2001qk}.
   The conclusion from their analysis is that one should rather consider
three-flavor processes such as $\eta\to\pi^+\pi^-\gamma$ or
$K\gamma\to K\pi$ to test the expected $N_c$ dependence in a low-energy reaction.
   For example, the Lagrangian relevant to the decay
$\eta\to\pi^+\pi^-\gamma$ is given by
\begin{displaymath}
{\cal L}_{\eta\pi^+\pi^-\gamma}=\frac{ien}{12\sqrt{3}\pi^2 F_0^3}(Q_u-Q_d)
\epsilon^{\mu\nu\rho\sigma}{\cal A}_\mu\partial_\nu\eta\partial_\rho\pi^+
\partial_\sigma\pi^-,
\end{displaymath}
where the quark-charge difference $Q_u-Q_d=1$ is independent of $N_c$.
   However, by investigating the corresponding $\eta$ and $\eta'$
decays up to next-to-leading order in the framework of the combined $1/N_c$ and
chiral expansions, Borasoy and Lipartia have concluded that the number of
colors cannot be determined from these decays due to the importance of
sub-leading terms which are needed to account for the experimental decay
widths and photon spectra \cite{Borasoy:2004mf}.

   For a discussion of the ${\cal O}(q^6)$ Lagrangian of odd intrinsic parity
see Refs.\ \cite{Ebertshauser:2001nj} and \cite{Bijnens:2001bb}.

\subsubsection{Masses of the Goldstone bosons}
\label{sec_aop4}
   A discussion of the masses at ${\cal O}(q^4)$ is one of the simplest applications
of chiral perturbation theory beyond the tree level.
   For that purpose let us consider ${\cal L}_2 + {\cal L}_4$ for QCD with finite quark
masses but in the absence of external fields.
   We restrict ourselves to the limit of isospin symmetry, i.e.,
$m_u=m_d=\hat m$.
   In order to determine the masses we calculate the self energies
$\Sigma(p^2)$ of the Goldstone bosons.

\begin{figure}[t]
\begin{center}
\epsfig{file=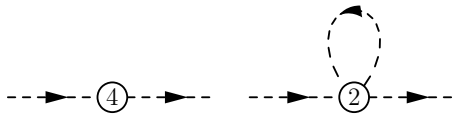,width=6cm}
\caption{\label{3:3:3:selfenergy}
Self-energy diagrams at ${\cal O}(q^4)$.
   Vertices derived from ${\cal L}_{2n}$ are denoted by $2n$ in the
interaction blobs.}
\end{center}
\end{figure}

   Let
\begin{equation}
\label{3:3:3:propF}
\Delta_{\phi F}(p)=\frac{1}{p^2-M^2_{\phi,2}+i0^+},\quad \phi=\pi,K,\eta,
\end{equation}
denote the Feynman propagator containing the lowest-order masses,
\begin{displaymath}
M^2_{\pi,2}=2 B_0 \hat m,\quad
M^2_{K,2}=B_0(\hat m+m_s),\quad
M^2_{\eta,2}=\frac{2}{3} B_0\left(\hat m+2m_s\right).
\end{displaymath}
   (The subscript 2 refers to chiral order 2.)
   The proper self-energy insertions,
$-i\Sigma_\phi(p^2)$, consist of one-particle-irreducible
diagrams only, i.e., diagrams which do not fall apart into
two separate pieces when cutting an arbitrary internal line.
   At chiral order $D=4$, the contributions to
$-i\Sigma_{\phi,4}(p^2)$ are those shown in Fig.\ \ref{3:3:3:selfenergy}.
   In general, the full (unrenormalized) propagator may be summed using a geometric series (see
Fig.\ \ref{3:3:3:fullprop}):
\begin{figure}[t]
\begin{center}
\epsfig{file=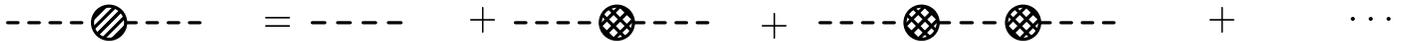,width=\textwidth}
\caption{\label{3:3:3:fullprop}
Unrenormalized propagator as the sum of irreducible self-energy diagrams.
Hatched and cross-hatched ``vertices'' denote one-particle-reducible
and one-particle-irreducible contributions, respectively.}
\end{center}
\end{figure}
\begin{eqnarray}
\label{3:3:3:prop1}
i\Delta_\phi(p)&=&\frac{i}{p^2-M^2_{\phi,2}+i0^+}+\frac{i}{p^2-M^2_{\phi,2}+i0^+}
[-i\Sigma_\phi(p^2)]\frac{i}{p^2-M^2_{\phi,2}+i0^+}+\cdots\nonumber\\
&=&
\frac{i}{p^2-M^2_{\phi,2}-\Sigma_\phi(p^2)+i0^+}.
\end{eqnarray}
   The physical mass, including the interaction, is defined as the pole of Eq.\ (\ref{3:3:3:prop1}),
\begin{equation}
\label{3:3:3:mdef}
M^2_\phi-M^2_{\phi,2}-\Sigma_\phi(M^2_\phi)\stackrel{!}{=}0,
\end{equation}
where the precision of the determination of $M^2_\phi$ depends on the precision of
the calculation of $\Sigma_\phi$.

   For our particular application with exactly two external meson lines,
the relevant interaction Lagrangians can be written as
\begin{equation}
{\cal L}_{\rm int}={\cal L}_2^{4\phi}+{\cal L}_4^{2\phi},
\end{equation}
where
\begin{eqnarray}
\label{3:3:3:l24phi} {\cal L}^{4\phi}_2&=&\frac{1}{24
F^2_0}\left\{\mbox{Tr}( [\phi,\partial_\mu \phi]\phi \partial^\mu
\phi) +B_0\mbox{Tr}({\cal M}\phi^4)\right\},\\
\label{3:3:3:l42phi}
{\cal L}_4^{2\phi}&=&
-\frac{1}{2}\left(a_\pi\pi^0\pi^0+b_\pi \partial_\mu\pi^0\partial^\mu\pi^0\right)
-a_\pi\pi^+\pi^--b_\pi\partial_\mu\pi^+\partial^\mu\pi^-\nonumber\\
&&-a_KK^+K^- - b_K\partial_\mu K^+\partial^\mu K^-
-a_KK^0\bar{K}^0 -b_K\partial_\mu K^0\partial^\mu\bar{K}^0\nonumber\\
&&-\frac{1}{2}\left(a_\eta\eta^2+b_\eta\partial_\mu\eta\partial^\mu\eta\right).
\end{eqnarray}
The constants $a_\phi$ and $ b_\phi$ are given by
\begin{eqnarray}
\label{3:3:3:ab}
a_\pi&=&\frac{64 B^2_0}{F^2_0}\left[(2\hat m+m_s)\hat mL_6+\hat m^2L_8\right],\nonumber\\
b_\pi&=&-\frac{16 B_0}{F^2_0}\left[(2\hat m+m_s)L_4+\hat mL_5\right],\nonumber\\
a_K&=&\frac{32B^2_0}{F^2_0}\left[(2\hat m+m_s)(\hat m+m_s)L_6+\frac{1}{2}(\hat m+m_s)^2L_8
\right],
\nonumber\\
b_K&=&-\frac{16 B_0}{F^2_0}\left[(2\hat m+m_s)L_4+\frac{1}{2}(\hat m+m_s)L_5\right]\nonumber\\
a_\eta&=&\frac{64 B^2_0}{3F^2_0}\left[(2\hat m+m_s)(\hat m+2m_s)L_6+2(\hat m-m_s)^2L_7
+(\hat m^2+2m_s^2)L_8\right],\nonumber\\
b_\eta&=&-\frac{16B_0}{F^2_0}\left[(2\hat m+m_s)L_4+\frac{1}{3}(\hat m+2m_s)L_5\right].
\end{eqnarray}
   At ${\cal O}(q^4)$ the self energies are of the form
\begin{equation}
\label{3:3:3:sigmaphi}
\Sigma_{\phi,4}(p^2)=A_\phi+B_\phi p^2,
\end{equation}
   where the constants $A_\phi$ and $B_\phi$ receive a tree-level
contribution from ${\cal L}_4$ and a one-loop contribution with a
vertex from ${\cal L}_2$ (see Fig.\ \ref{3:3:3:selfenergy}).
   For the tree-level contribution of ${\cal L}_4$ this is easily seen, because
the Lagrangians of Eq.\ (\ref{3:3:3:l42phi}) contain either exactly
two derivatives of the fields or no derivatives at all.
   For example, the contact contribution for the $\eta$ reads
$$-i\Sigma_{\eta,4}^{\rm tree}(p^2)=-i(a_\eta + b_\eta p^2).$$

   For the one-loop contribution the argument is as follows.
   The Lagrangian ${\cal L}_2^{4\phi}$ contains either two derivatives
or no derivatives at all which, symbolically, can be written as
$\phi\phi\partial\phi\partial\phi$ and $\phi^4$, respectively.
   The first term results in $M^2$ or $p^2$, depending on whether the
$\phi$ or the $\partial \phi$ are contracted with the external fields.
   The ``mixed'' situation vanishes upon integration.
   The second term, $\phi^4$, does not generate a momentum dependence.

   As a specific example, we evaluate the pion-loop contribution to the
$\pi^0$ self energy (see Fig.\ \ref{4:9:pi0seloop}) by applying
the Feynman rule of Eq.~(\ref{3:2:4:mpipi})
for $a=c=3$, $p_a=p_c=p$, $b=d=j$, and $p_b=p_d=k$:
\begin{figure}
\begin{center}
\epsfig{file=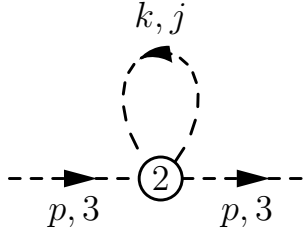,width=4cm}
\caption{\label{4:9:pi0seloop} Contribution of the pion loops to
the $\pi^0$ self energy.}
\end{center}
\end{figure}
\begin{displaymath}
\label{3:3:3:diag}
\frac{1}{2}\int \frac{\mbox{d}^4k}{(2\pi)^4}\frac{i}{3F_0^2}
[-4p^2-4k^2+5 M^2_{\pi,2}]
\frac{i}{k^2-M_{\pi,2}^2+i0^+},
\end{displaymath}
   where $1/2$ is a symmetry factor.
   Since the integral diverges, we consider its extension to $n$ dimensions.
   In addition to the loop-integral of Eq.\ (\ref{3:2:4:IMmun}),
we need
\begin{displaymath}
\mu^{4-n}i\int \frac{\mbox{d}^n k}{(2\pi)^n}\frac{k^2}{k^2-M^2+i0^+}=
\mu^{4-n}i\int \frac{\mbox{d}^n k}{(2\pi)^n}\frac{k^2-M^2+M^2}{k^2-M^2+i0^+},
\end{displaymath}
   where we have added $0=-M^2+M^2$ in the numerator.
We make use of
$$\mu^{4-n}i\int \frac{\mbox{d}^n k}{(2\pi)^n}=0$$
in dimensional regularization which is ``shown'' as follows.
   Consider the (more general) integral
\begin{equation}
\label{3:3:3:dnkk2p}
\int \mbox{d}^n k (k^2)^p,
\end{equation}
substitute $k=\lambda k'\ (\lambda > 0)$, and relabel $k'=k$
\begin{displaymath}
=\lambda^{n+2p}\int \mbox{d}^n k (k^2)^p.
\end{displaymath}
   Since $\lambda>0$ is arbitrary and, for fixed $p$, the result
is to hold for arbitrary $n$, Eq.\ (\ref{3:3:3:dnkk2p})
is set to zero in dimensional regularization.
   We emphasize that the vanishing of Eq.\ (\ref{3:3:3:dnkk2p}) has
the character of a prescription.
   The integral does not depend on any scale and its analytic continuation
is ill defined in the sense that there is no dimension $n$ where it is
meaningful.
  It is ultraviolet divergent for $n+2p\geq 0$ and infrared divergent
for $n+2p\leq 0$.

   We then obtain
$$
\mu^{4-n}i\int \frac{\mbox{d}^n k}{(2\pi)^n}\frac{k^2}{k^2-M^2+i0^+}=
M^2 I(M^2,\mu^2,n),
$$
with $I(M^2,\mu^2,n)$ of Eq.\ (\ref{3:2:4:IMmun}).
   The pion-loop contribution to the $\pi^0$ self energy is thus
$$\frac{i}{6 F^2_0}(-4p^2+M^2_{\pi,2})I(M^2_{\pi,2},\mu^2,n),$$
which is indeed of the type discussed in Eq.\ (\ref{3:3:3:sigmaphi})
and diverges as $n\to 4$.

   After analyzing all loop contributions and combining them with the tree-level
contributions of Eqs.\ (\ref{3:3:3:ab}), the constants
$A_\phi$ and $B_\phi$ of Eq.\ (\ref{3:3:3:sigmaphi}) are given by
\begin{eqnarray}
\label{3:3:3:AB}
A_\pi&=&\frac{M^2_{\pi,2}}{F^2_0}\Bigg\{
\underbrace{-\frac{1}{6}I(M^2_{\pi,2})
-\frac{1}{6}I(M^2_{\eta,2})-\frac{1}{3}I(M^2_{K,2})}_{\mbox{loop contribution}}
\underbrace{+32[(2\hat m+m_s)B_0L_6+\hat mB_0L_8]}_{\mbox{tree-level contribution}}
\Bigg\},\nonumber\\
B_\pi&=&\frac{2}{3}\frac{I(M^2_{\pi,2})}{F^2_0}+\frac{1}{3}
\frac{I(M^2_{K,2})}{F^2_0}-\frac{16B_0}{F^2_0}\left[
(2\hat m+m_s)L_4+\hat m L_5\right],\nonumber\\
A_K&=&\frac{M^2_{K,2}}{F^2_0}\Bigg\{\frac{1}{12}I(M^2_{\eta,2})
-\frac{1}{4}I(M^2_{\pi,2})-\frac{1}{2}I(M^2_{K,2})+32\left[(2\hat m+m_s)B_0L_6+\frac{1}{2}(\hat m+m_s)
B_0L_8\right]\Bigg\},\nonumber\\
B_K&=&\frac{1}{4}\frac{I(M^2_{\eta,2})}{F^2_0}
+\frac{1}{4}\frac{I(M^2_{\pi,2})}{F^2_0}
+\frac{1}{2}\frac{I(M^2_{K,2})}{F^2_0}-16 \frac{B_0}{F^2_0}\left[(2\hat m+m_s)L_4+\frac{1}{2}(\hat m
+m_s)L_5\right],
\nonumber\\
A_\eta&=&\frac{M^2_{\eta,2}}{F^2_0}\left[-\frac{2}{3}I(M^2_{\eta,2})\right]
+\frac{M^2_{\pi,2}}{F^2_0}\left[\frac{1}{6}I(M^2_{\eta,2})
-\frac{1}{2}I(M^2_{\pi,2})
+\frac{1}{3}I(M^2_{K,2})\right]\nonumber\\
&&+\frac{M^2_{\eta,2}}{F^2_0}[16M^2_{\eta,2} L_8+32(2\hat m+m_s)
B_0L_6]+\frac{128}{9}\frac{B^2_0(\hat m-m_s)^2}{F^2_0}(3L_7+L_8),\nonumber\\
B_\eta&=&\frac{I(M^2_{K,2})}{F^2_0}-\frac{16}{F^2_0}(2\hat m+m_s)B_0L_4
-8\frac{M^2_{\eta,2}}{F^2_0}L_5,
\end{eqnarray}
   where, for simplicity, we have suppressed the dependence on the
scale $\mu$ and the number of dimensions $n$ in
the integrals $I(M^2,\mu^2,n)$ [see Eq.\ (\ref{3:2:4:IMmun})].
   Both the integrals $I$ and the bare coefficients $L_i$
(with the exception of $L_7$) have $1/(n-4)$ poles and finite pieces.
   In particular, the coefficients $A_\phi$ and $B_\phi$ are {\em not}
finite as $n\to 4$ showing that they do not correspond to observables.

   The masses at ${\cal O}(q^4)$ are determined by solving Eq.~(\ref{3:3:3:mdef})
with the predictions of Eq.\ (\ref{3:3:3:sigmaphi}) for the self energies,
$$
M^2_\phi=M_{\phi,2}^2+A_\phi+B_\phi M^2_\phi,
$$
from which we obtain
\begin{displaymath}
M^2_\phi=\frac{M_{\phi,2}^2+A_\phi}{1-B_\phi}=M_{\phi,2}^2(1+B_\phi)+A_\phi+
{\cal O}(q^6),
\end{displaymath}
   because $A_\phi={\cal O}(q^4)$ and $\{B_\phi, M_{\phi,2}^2\}={\cal O}(q^2)$.
   Expressing the bare coefficients $L_i$ in Eq.\ (\ref{3:3:3:AB}) in terms of
the renormalized coefficients by using Eq.\ (\ref{3:3:1:lihi}),
the results for the masses of the Goldstone bosons at ${\cal O}(q^4)$
read \cite{Gasser:1984gg}
\begin{eqnarray}
\label{3:3:3:mpi24}
M^2_{\pi,4}&=&M^2_{\pi,2}\Bigg\{1+\frac{M^2_{\pi,2}}{32\pi^2F^2_0}
\ln\left(\frac{M^2_{\pi,2}}{\mu^2}\right)-\frac{M^2_{\eta,2}}{96\pi^2F^2_0}
\ln\left(\frac{M^2_{\eta,2}}{\mu^2}\right)\nonumber\\
&&+\frac{16}{F^2_0}\left[(2\hat m+m_s)B_0(2L^r_6-L^r_4)
+\hat mB_0(2L^r_8-L^r_5)\right]\Bigg\},\\
\label{3:3:3:mk24}
M^2_{K,4}&=&M^2_{K,2}\Bigg\{1+\frac{M^2_{\eta,2}}{48\pi^2F^2_0}
\ln\left(\frac{M^2_{\eta,2}}{\mu^2}\right)\nonumber\\
&&+\frac{16}{F^2_0}\left[(2\hat m+m_s)B_0(2L^r_6-L^r_4)
+\frac{1}{2}(\hat m+m_s)B_0(2L^r_8-L^r_5)\right]\Bigg\},\nonumber\\
&&\\
\label{3:3:3:meta24}
M^2_{\eta,4}&=&M^2_{\eta,2}\left[1+\frac{M^2_{K,2}}{16\pi^2F^2_0}
\ln\left(\frac{M^2_{K,2}}{\mu^2}\right)
-\frac{M^2_{\eta,2}}{24\pi^2F^2_0}\ln\left(\frac{M^2_{\eta,2}}{\mu^2}\right)
\right.\nonumber\\
&&\left.+\frac{16}{F^2_0}(2\hat m+m_s)B_0(2L^r_6-L^r_4)
+8\frac{M^2_{\eta,2}}{F^2_0}(2L^r_8-L^r_5)\right]\nonumber\\
&&+M^2_{\pi,2}\left[\frac{M^2_{\eta,2}}{96\pi^2F^2_0}
\ln\left(\frac{M^2_{\eta,2}}{\mu^2}\right)
-\frac{M^2_{\pi,2}}{32\pi^2F^2_0}
\ln\left(\frac{M^2_{\pi,2}}{\mu^2}\right)
+\frac{M^2_{K,2}}{48\pi^2F^2_0}
\ln\left(\frac{M^2_{K,2}}{\mu^2}\right)\right]\nonumber\\
&&+\frac{128}{9}\frac{B^2_0(\hat m-m_s)^2}{F^2_0}
(3L^r_7+L^r_8).
\end{eqnarray}
   First of all, we note that the expressions for the masses are finite.
   The infinite parts of the coefficients $L_i$ of the Lagrangian
of Gasser and Leutwyler exactly cancel the divergent terms
resulting from the integrals.
   This is the reason why the bare coefficients $L_i$ must be infinite.
   Furthermore, at ${\cal O}(q^4)$ the masses of the Goldstone bosons
vanish, if the quark masses are sent to zero.
   This is, of course, what we had expected from QCD in the chiral limit
but it is comforting to see that the self interaction in ${\cal L}_2$
(in the absence of quark masses) does not generate Goldstone
boson masses at higher order.
   At ${\cal O}(q^4)$, the squared Goldstone boson masses contain terms
which are analytic in the quark masses, namely, of the form $m^2_q$
multiplied by the renormalized low-energy coupling constants $L_i^r$.
   However, there are also non-analytic terms  of the
type $m^2_q \ln(m_q)$---so-called  chiral logarithms---which do not involve
new parameters.
   Such a behavior is an illustration of the mechanism found by Li and
Pagels \cite{Li:1971vr}, who noticed that a perturbation theory
around a symmetry which is realized in the Nambu-Goldstone mode
results in both analytic as well as non-analytic expressions in
the perturbation.
   Finally, the scale dependence of the renormalized coefficients
$L_i^r$ of Eq.\ (\ref{3:3:1:lihi}) is by construction such that it cancels
the scale dependence of the chiral logarithms.
   Thus, physical observables do not depend on the scale $\mu$.
   It is straightforward to verify this statement by differentiating
Eqs.\ (\ref{3:3:3:mpi24}) - (\ref{3:3:3:meta24}) with respect to $\mu$
and by making use of
$$
\frac{\mbox{d} L_i^r(\mu)}{\mbox{d}\mu}=-\frac{\Gamma_i}{16\pi^2\mu},
$$
where the $\Gamma_i$ are given in Table \ref{4:8:tableli}.

\subsubsection{Electromagnetic polarizabilities of the pion}
\label{sec:2}

   Another strong constraint provided by chiral symmetry is the connection
between the electromagnetic polarizabilities of the charged pion and
the radiative pion beta decay.
   In the framework of classical electrodynamics, the electric and magnetic
polarizabilities $\alpha$ and $\beta$ describe the response of a system to a
static, uniform, external electric and magnetic field in terms of induced
electric and magnetic dipole moments.
   In principle, empirical information on the pion polarizabilities can be obtained
from the differential cross section of low-energy Compton scattering on a charged
pion,
\begin{displaymath}
\frac{\mbox{d}\sigma}{\mbox{d}\Omega_{\rm lab}}= \left(\frac{\omega'}{\omega}\right)^2
\frac{e^2}{4\pi M_\pi}\left\{\frac{e^2}{4\pi M_\pi}
\frac{1+z^2}{2}-\frac{\omega\omega'}{2} \left[(\alpha+\beta)_{\pi^+}(1+z)^2
+(\alpha-\beta)_{\pi^+}(1-z)^2\right]\right\}
+\cdots,
\end{displaymath}
where $z=\hat{q}\cdot\hat{q}\,'$ and $\omega'/\omega=[1+\omega(1-z)/M_\pi]$.
   The forward and backward differential cross sections are
sensitive to $(\alpha+\beta)_{\pi^+}$ and $(\alpha-\beta)_{\pi^+}$, respectively.

   Within the framework of the partially conserved axial-vector (PCAC)
hypothesis and current algebra the electromagnetic
polarizabilities of the charged pion are related to the
radiative charged-pion beta decay $\pi^+\to e^+\nu_e\gamma$~\cite{Terentev:1972ix}.
   The result obtained using ChPT at leading non-trivial order (${\cal O}(q^4)$)~\cite{Bijnens:1987dc}
is equivalent to the original PCAC result,
\begin{displaymath}
\alpha_{\pi^+}=-\beta_{\pi^+}=2 \frac{e^2}{4\pi} \frac{1}{(4\pi F_\pi)^2
M_\pi}\frac{\bar l_\Delta}{6},
\end{displaymath}
where $\bar l_\Delta\equiv(\bar l_6-\bar l_5)$ is a linear
combination of scale-independent parameters of the two-flavor ${\cal O}(q^4)$
Lagrangian~\cite{Gasser:1983yg}.
   At ${\cal O}(q^4)$ this difference is related to the ratio $\gamma=F_A/F_V$ of the
pion axial-vector form factor $F_A$ and the vector form factor
$F_V$ of radiative pion beta decay~\cite{Gasser:1983yg},
$\gamma={\bar{l}}_\Delta/6$.
   Once this ratio is known, chiral symmetry
makes an {\em absolute} prediction for the polarizabilities.
   This situation is similar to the $s$-wave $\pi\pi$-scattering
lengths of Eq.\ (\ref{4:10:a00a02lo}) which are predicted
once $F_\pi$ is known.
   Using the most recent determination $\gamma=0.443\pm 0.015$ by
the PIBETA Collaboration~\cite{Frlez:2003pe} (assuming
$F_V=0.0259$ obtained from the conserved vector current
hypothesis) results in the ${\cal O}(q^4)$ prediction
$\alpha_{\pi^+}=(2.64\pm 0.09)\times 10^{-4}\, \mbox{fm}^3$,
where the estimate of the error is only the one
due to the error of $\gamma$ and does not include effects from
higher orders in the quark-mass expansion.

   Corrections to the leading-order PCAC result have been
calculated at ${\cal O}(q^6)$ and turn out to be rather small~\cite{Burgi:1996qi},
\cite{Gasser:2006qa}.
   Using updated values for the LECs, the predictions of
\cite{Gasser:2006qa} are
\begin{eqnarray}
(\alpha + \beta)_{\pi^+} &=& 0.16 \times 10^{-4}\, \mbox{fm}^3,\label{eq:1.1}\\
(\alpha - \beta)_{\pi^+} &=& (5.7 \pm 1.0)\times 10^{-4}\, \mbox{fm}^3.\label{eq:1.2}
\end{eqnarray}
   The corresponding corrections to the ${\cal O}(q^4)$ result
indicate a similar rate of convergence as for the $\pi\pi$-scattering lengths
\cite{Gasser:1983yg}, \cite{Bijnens:1995yn}.
   The error for $(\alpha + \beta)_{\pi^+}$ is
of the order $0.1\times 10^{-4}\, \mbox{fm}^3$, mostly from the dependence on the scale at which the
${\cal O}(q^6)$ low-energy coupling constants are estimated by resonance saturation.

   As there is no stable pion target, empirical information about the pion
polarizabilities is not easy to obtain.
   For that purpose, one has to consider reactions which contain
the Compton scattering amplitude as a building block, such as, e.g., the
Primakoff effect in high-energy pion-nucleus bremsstrahlung, $\pi^-Z\to \pi^-Z
\gamma$, radiative pion photoproduction on the nucleon,
$\gamma p\to \gamma \pi^+n$, and pion pair
production in $e^+e^-$ scattering, $e^+e^-\to e^+e^-\pi^+\pi^-$.
   Unfortunately, at present, the experimental situation looks rather contradictory
(see Refs.~\cite{Ahrens:2004mg}, \cite{Gasser:2006qa} for recent reviews of the
data and further references to the experiments).

\begin{figure}[t]
\begin{center}
\epsfig{file=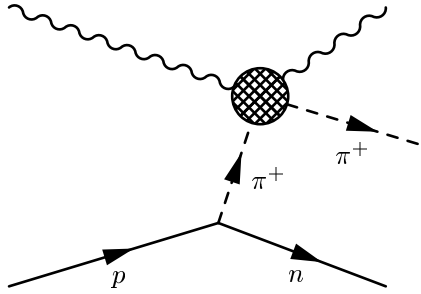,width=0.3\textwidth}
\end{center}
\caption{The reaction $\gamma p\to\gamma\pi^+n$ contains Compton scattering on
a pion as a sub diagram in the $t$ channel, where $t=(p_n-p_p)^2$.}
\label{fig:tchannel}
\end{figure}

\begin{figure}[t]
\begin{center}
\resizebox{0.4\textwidth}{!}{%
\includegraphics{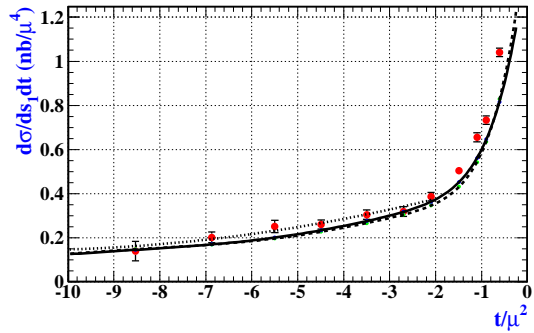}
}
\end{center}
\caption{Differential cross section averaged over 537 MeV $< E_\gamma <$ 817 MeV
and 1.5 $M_\pi^2<s_1<5 M_\pi^2$. Solid line: model 1; dashed line: model 2;
dotted line: fit to experimental data.} \label{fig:cros1c}
\end{figure}

  The potential of studying the influence of the pion polarizabilities on radiative pion
photoproduction from the proton was extensively studied in
\cite{Drechsel:1994kh}.
   In terms of Feynman diagrams, the reaction $\gamma p\to\gamma\pi^+n$ contains
real Compton scattering on a charged pion as a pion pole diagram (see
Fig.~\ref{fig:tchannel}).
   In the recent experiment on $\gamma p\to\gamma\pi^+n$
at the Mainz Microtron MAMI \cite{Ahrens:2004mg}, the cross section was obtained
in the kinematic region 537 MeV $< E_\gamma <$ 817 MeV,
$140^{\circ}\le\theta^{\rm cm}_{\gamma\gamma'}\leq 180^{\circ}$.
  Figure \ref{fig:cros1c} shows the
experimental data, averaged over the full photon beam energy interval and over
the squared pion-photon center-of-mass energy $s_1$ from 1.5 $M_\pi^2$ to 5
$M_\pi^2$ as a function of the squared pion momentum transfer $t$ in units of
$M_\pi^2$.
   For such small values of $s_1$, the differential cross section is expected to
be insensitive to the pion polarizabilities.
   Also shown are two model calculations: model 1 (solid curve) is a simple Born
approximation using the pseudoscalar pion-nucleon interaction including the
anomalous magnetic moments of the nucleon; model 2 (dashed curve) consists of
pole terms without the anomalous magnetic moments but including contributions
from the resonances $\Delta (1232)$, $P_{11}(1440)$, $D_{13}(1520)$ and
$S_{11}(1535)$.
   The dotted curve is a fit to the experimental data.

\begin{figure}[t]
\begin{center}
\resizebox{0.4\textwidth}{!}{%
\includegraphics{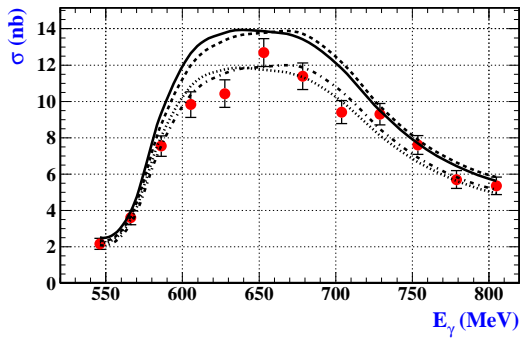}
}
\end{center}
\caption{
The cross section of the process $\gamma p\to\gamma\pi^+n$ integrated over $s_1$
and $t$ in the region where the contribution of the pion polarizability is
biggest and the difference between the predictions of the theoretical models
under consideration does not exceed 3 \%. The dashed and dashed-dotted lines are
predictions of model 1 and the solid and dotted lines of model 2 for
$(\alpha-\beta)_{\pi^+}=0$ and $(\alpha-\beta)_{\pi^+}=14\times 10^{-4}\,
\mbox{fm}^3$, respectively.} \label{fig:cros2c}
\end{figure}

   The kinematic region where the polarizability contribution
is biggest is given by $5M_\pi^2< s_1<15M_\pi^2$ and $-12M_\pi^2<t<-2M_\pi^2$.
   Figure \ref{fig:cros2c} shows the cross section as a function of the beam
energy integrated over $s_1$ and $t$ in this second region.
   The dashed and solid lines (dashed-dotted and dotted lines) refer to
models 1 and 2, respectively, each with $(\alpha-\beta)_{\pi^+}=0$
($(\alpha-\beta)_{\pi^+}=14\times 10^{-4}\, \mbox{fm}^3$).
   By comparing the experimental data of the 12 points with the predictions of
the models, the corresponding values of $(\alpha-\beta)_{\pi^+}$ for each data
point have been determined in combination with the corresponding statistical and
systematic errors.
    The result extracted from the combined analysis of the 12 data points
reads \cite{Ahrens:2004mg}
\begin{equation}
\label{alphambeta} (\alpha-\beta)_{\pi^+}=(11.6\pm 1.5_{\rm stat}\pm 3.0_{\rm
syst}\pm 0.5_{\rm mod}) \times 10^{-4}\, \mbox{fm}^3
\end{equation}
and has to be compared with the ChPT result of $(5.7 \pm 1.0)\times 10^{-4}\, \mbox{fm}^3$,
which deviates by 2 standard deviations from the experimental result.

   Clearly, the model-dependent input to the result of Eq.~(\ref{alphambeta}) deserves
further study.
   In particular, the model error was estimated by comparing the analysis with two specific
models.
   In Ref.\ \cite{Kao:2004xh} radiative pion photoproduction was studied in the framework of
heavy-baryon chiral perturbation theory at the one-loop level.
   Unfortunately, the kinematical conditions of the MAMI experiment were not explicitly
considered.
   It was argued that the extraction of pion polarizabilities is, in principle, possible
and that the main uncertainty in the extraction arises
from the effect of two structures of the ${\cal O}(q^3)$ Lagrangian.

   The Primakoff method was used at Serpukhov with the result~\cite{Antipov:1982kz}
\begin{equation}
(\alpha - \beta)_{\pi^+} = (13.6 \pm 2.8_{\rm{stat}} \pm
2.4_{\rm{syst}}) \times 10^{-4}\, \mbox{fm}^3, \label{eq:_1.4}
\end{equation}
in agreement with the value from MAMI.
   Recently, also the COMPASS Collaboration
at CERN has investigated this reaction, and the data analysis is
underway \cite{Guskov:2008zz}.
   Unfortunately, the third method based on the
reactions $e^+e^- \rightarrow \gamma \gamma \rightarrow \pi^+\pi^-$, has led to
even more contradictory results (see Ref.~\cite{Gasser:2006qa}).

   Also on the theoretical side there has been a long-standing problem.
   The application of dispersion sum rules as performed in
\cite{Fil'kov:1998np}, \cite{Fil'kov:2005ss} yields $(\alpha-\beta)_{\pi^+}=(13.0^{+2.6}_{-1.9}) \times
10^{-4}\, \mbox{fm}^3$ which provides an even more pronounced discrepancy with the predictions
of chiral perturbation theory than the MAMI result \cite{Gasser:2006qa}.
   These dispersion relations are based on specific forms for the absorptive part
of the Compton amplitudes.
   In Ref.\ \cite{Pasquini:2008ep}, the analytic properties of these forms have been examined
and the strong enhancement of intermediate-meson contributions was shown to be
connected with spurious singularities.
   It was shown that the results of dispersion theory and effective field theory are
not in conflict, once the basic requirements of dispersion relations
are taken into account.

   Clearly, the electromagnetic polarizabilities of the charged pion remain one of
the challenging topics of hadronic physics in the low-energy domain.
   Chiral symmetry provides a strong constraint in terms of radiative pion beta
decay and mesonic chiral perturbation theory makes a firm prediction beyond the
current algebra result at the two-loop level.
   Both the experimental determination as well as the theoretical extraction from
experiment require further efforts.

\section{Baryonic chiral perturbation theory}
\label{section_bchpt}
\subsection{Lagrangian}
   So far we have considered the purely mesonic sector involving the interaction of Goldstone
bosons with each other and with the external fields.
   Now we want to describe matrix elements with a single baryon in the initial and
final states.

\subsubsection{Transformation properties of the fields}
\label{sec_tpf}
   Our aim is the most general description of the interaction of baryons with Goldstone
bosons and external fields at low energies.
   For that purpose we not only need to specify the transformation behavior of the Goldstone
bosons and external fields but also of the remaining dynamical fields entering the Lagrangian.
   Our discussion follows Refs.\ \cite{Georgi}, \cite{Gasser:1987rb}.
   Consider the nucleon doublet and the octet of $\frac{1}{2}^+$ baryons (see Fig.~\ref{2:2:2:fig:baryon_octet}),
\begin{eqnarray}
\label{4:1:1:nucleondoublet}
\Psi&=&\left(\begin{array}{c}p\\n\end{array}\right),\\
\label{4:1:1:su3oktett}
B&=&\sum_{a=1}^8 \frac{B_a\lambda_a}{\sqrt{2}}=
\left(\begin{array}{ccc}
\frac{1}{\sqrt{2}}\Sigma^0+\frac{1}{\sqrt{6}}\Lambda&\Sigma^+&p\\
\Sigma^-&-\frac{1}{\sqrt{2}}\Sigma^0+\frac{1}{\sqrt{6}}\Lambda&n\\
\Xi^-&\Xi^0&-\frac{2}{\sqrt{6}}\Lambda
\end{array}\right).
\end{eqnarray}
   Each entry of $\Psi$ and $B$, respectively, is a complex, four-component Dirac field.
   In contradistinction to the case of the Goldstone-boson matrix $\phi$ of Eq.~(\ref{3:1:2:phisu3}),
we have $B\neq B^\dagger$.
   The representation of the isospin group $\mbox{SU}(2)_V$ and the flavor group $\mbox{SU}(3)_V$ on
$\{\Psi\}$ and $\{B\}$, respectively, is given by
\begin{eqnarray}
\label{4:1:1:su2hom}
\Psi&\mapsto& V \Psi,\quad V\in \mbox{SU}(2)_V,\\
\label{4:1:1:su3hom}
B&\mapsto& V B V^\dagger,\quad V\in\mbox{SU}(3)_V,
\end{eqnarray}
i.e., $\Psi$ transforms under the fundamental representation of SU(2) and
$B$ transforms under the adjoint representation of SU(3).
   Starting from Eqs.~(\ref{4:1:1:su2hom}) and (\ref{4:1:1:su3hom}) we will discuss
realizations of $\mbox{SU(2)}_L\times\mbox{SU(2)}_R$ and  $\mbox{SU(3)}_L\times\mbox{SU(3)}_R$
on $\{\Psi\}$ and $\{B\}$, respectively.

   Let us begin with $G=\mbox{SU(2)}_L\times\mbox{SU(2)}_R$.
   Recall that the transformation of Eq.~(\ref{3:1:2:utrfafo}),
\begin{displaymath}
U\mapsto RUL^\dagger,
\end{displaymath}
defines a nonlinear realization of $G$ on $\{U\}$.
   Introducing $u^2=U$, we define the
SU(2)-valued function $K(L,R,U)$ by
\begin{equation}
\label{4:1:1:kdef}
u\mapsto u'=\sqrt{RUL^\dagger}\equiv RuK^{-1}(L,R,U),\quad\mbox{i.e.,}\quad
K(L,R,U)=u'^{-1}Ru=\sqrt{RUL^\dagger}^{-1}R \sqrt{U}.
\end{equation}
   The transformation
\begin{equation}
\label{4:1:1:su2real}
\varphi(g):\left(\begin{array}{c}U\\ \Psi\end{array}\right)\mapsto
\left(\begin{array}{c}U'\\ \Psi'\end{array}\right)
=\left(\begin{array}{c}RUL^\dagger\\K(L,R,U)\Psi\end{array}\right)
\end{equation}
defines an operation of the group $G$ on the set $\{(U,\Psi)\}$.
   This is true, because (a) the identity of $G$ leaves any pair $(U,\Psi)$ invariant and
(b) the transformation satisfies the homomorphism property
\begin{eqnarray*}
\varphi(g_1)\varphi(g_2)
\left(\begin{array}{c}U\\ \Psi\end{array}\right)&=&\varphi(g_1)
\left(\begin{array}{c}R_2UL_2^\dagger\\K(L_2,R_2,U)\Psi\end{array}\right)
=\left(\begin{array}{c}R_1R_2UL_2^\dagger L_1^\dagger\\
K(L_1,R_1,R_2UL_2^\dagger)K(L_2,R_2,U)\Psi
\end{array}\right)\\
&=&\left(\begin{array}{c}R_1 R_2U(L_1L_2)^\dagger\\
K(L_1L_2,R_1R_2,U)\Psi\end{array}\right)=\varphi(g_1g_2)\left(\begin{array}{c}U\\ \Psi\end{array}\right),
\end{eqnarray*}
where we made use of
\begin{displaymath}
K(L_1, R_1, R_2 U L_2^\dagger)K (L_2,R_2,U)
=K((L_1 L_2), (R_1 R_2),U).
\end{displaymath}
   Note that for a general group element $g=(L,R)\in G$ the transformation behavior
of $\Psi$ depends on $U$.
   The exception to this rule is the case of an isospin transformation $R=L=V$,
where, because of $U'=u'^2=VuV^\dagger VuV^\dagger=Vu^2V^\dagger=VUV^\dagger$, one
has $u'=VuV^\dagger$.
   Comparing with Eq.\ (\ref{4:1:1:kdef}), we obtain $K^{-1}(V,V,U)=V^\dagger$ or
$K(V,V,U)=V$.
   This is consistent with our starting point that $\Psi$ transforms linearly as
an isospin doublet under the isospin subgroup $H=\mbox{SU(2)}_V$ of
$G=\mbox{SU(2)}_L\times\mbox{SU(2)}_R$.
   Recall that the symmetry of the vacuum determines the multiplet structure of
the spectrum \cite{Coleman:1966:1:1}.

   For $G=\mbox{SU(3)}_L\times \mbox{SU(3)}_R$ one uses
\begin{equation}
\label{4:1:1:su3real}
\varphi(g):\left(\begin{array}{c}U\\ B\end{array}\right)\mapsto
\left(\begin{array}{c}U'\\ B'\end{array}\right)
=\left(\begin{array}{c}RUL^\dagger\\K(L,R,U)B K^\dagger(L,R,U)
\end{array}\right),
\end{equation}
where $K$ is defined completely analogously to Eq.\ (\ref{4:1:1:kdef}) after
inserting the corresponding SU(3) matrices.

   The generalization to other multiplets is straightforward. One first specifies
the transformation behavior under the subgroup $H$ in terms of $V$ and $V^\dagger$.
   In order to find the transformation behavior under $G$ one simply replaces $V\to K$
and $V^\dagger\to K^\dagger$.

\subsubsection{Baryonic effective Lagrangian at lowest order}
\label{sec_loebl}

   Given the dynamical fields of Eqs.\ (\ref{4:1:1:su2real}) and
(\ref{4:1:1:su3real}) and their transformation properties, we will now discuss
the most general effective baryonic Lagrangian at lowest order.
   We will start with the effective $\pi N$ Lagrangian
${\cal L}^{(1)}_{\pi N}$ which we demand to have a {\em local}
$\mbox{SU}(2)_L\times\mbox{SU(2)}_R\times\mbox{U(1)}_V$ symmetry.
   The transformation behavior of the external fields is given in
Eq.\ (\ref{2:1:6:sg}), whereas $U$ and the nucleon doublet transform as
\begin{equation}
\label{4:1:2:psitrans}
\left(\begin{array}{c}U(x)\\ \Psi(x)\end{array}\right)\mapsto
\left(\begin{array}{c}V_R(x)U(x)V_L^\dagger(x)\\
\exp[-i\Theta(x)]K[V_L(x),V_R(x),U(x)]\Psi(x)\end{array}\right).
\end{equation}
   The local character of the transformation implies that we need to
introduce a covariant derivative $D_\mu \Psi$ with the usual property that
it transforms in the same way as $\Psi$,
\begin{equation}
\label{4:1:2:kovder}
D_\mu\Psi=(\partial_\mu+\Gamma_\mu-iv_\mu^{(s)})\Psi,
\end{equation}
where the so-called connection is given by
\begin{equation}
\label{4:1:2:gamma}
\Gamma_\mu=\frac{1}{2}\left[u^\dagger(\partial_\mu-ir_\mu)u
+u(\partial_\mu-il_\mu)u^\dagger\right].
\end{equation}
   Since $K$ not only depends on $V_L$ and $V_R$ but also on $U$,
the covariant derivative contains besides the external fields also
$u$ and $u^\dagger$ and their derivatives.
   At ${\cal O}(q)$ there exists another Hermitian building block,
the so-called vielbein,
\begin{equation}
\label{4:1:2:chvi}
u_\mu\equiv i\left[u^\dagger(\partial_\mu-i r_\mu)u-u(\partial_\mu-i
l_\mu)u^\dagger\right],
\end{equation}
which under parity transforms as an axial vector, $u_\mu\mapsto -u^\mu$,
and under $\mbox{SU(2)}_L\times\mbox{SU(2)}_R\times \mbox{U}(1)_V$,
transforms as $u_\mu \mapsto Ku_\mu K^\dagger$.

   The structure of the most general effective $\pi N$ Lagrangian describing
processes with a single nucleon in the initial and final states is of the type
$\bar{\Psi} \widehat{O} \Psi$, where $\widehat{O}$ is an operator acting
in Dirac and flavor space, transforming under
$\mbox{SU(2)}_L\times\mbox{SU(2)}_R\times \mbox{U}(1)_V$
as $K\widehat{O}K^\dagger$.
   The Lagrangian must be a Hermitian Lorentz scalar
which is even under the discrete symmetries $C$, $P$, and $T$.
   The most general such Lagrangian with the smallest number of derivatives
is given by \cite{Gasser:1987rb}
\begin{equation}
\label{4:1:2:l1pin}
{\cal L}^{(1)}_{\pi N}=\bar{\Psi}\left(iD\hspace{-.6em}/ -m
+\frac{\texttt{g}_A}{2}\gamma^\mu \gamma_5 u_\mu\right)\Psi.
\end{equation}
   It contains two parameters not determined by chiral symmetry:
the chiral limit $m$ of the nucleon mass $m_N$
and the chiral limit $\texttt{g}_A$ of the axial-vector coupling constant
$g_A$.
   The physical value of $g_A$ is determined from neutron beta decay and is
given by $g_A=1.2695\pm 0.0029$.
   The overall normalization of the Lagrangian is chosen such that in the
case of no external fields and no pion fields it reduces to that
of a free nucleon of mass $m$.

   Similarly as in the mesonic case, the Lagrangian ${\cal L}^{(1)}_{\pi N}$
has predictive power once the two parameters have been identified.
   For example, a tree-level calculation of pion-nucleon scattering produces
the famous Weinberg-Tomozawa relation for the $s$-wave $\pi N$-scattering
lengths \cite{Weinberg:1966kf}, \cite{Tomozawa:1966jm},
\begin{equation}
\label{5:3:weinbergtomozawa}
a^I=-\frac{M_\pi}{8\pi (1+\mu)F_\pi^2}\left[I(I+1)-\frac{3}{4}-2\right],
\end{equation}
where $I=1/2$ or $I=3/2$ refers to the total isospin of the $\pi N$ system.
   As in $\pi\pi$ scattering, the $s$-wave $\pi N$-scattering lengths vanish
in the chiral limit, i.e., Goldstone bosons interact ``weakly'' with other
hadrons in the zero-energy and mass limit.

   Since the nucleon mass $m_N$ does not vanish in the chiral limit, the
zeroth component $\partial^0$ of the partial derivative acting on the nucleon
field does not produce a ``small'' quantity.
   This results in new features of the chiral power counting in the
baryonic sector.
   The counting of the external fields as well as of covariant derivatives
acting on the mesonic fields remains the same as in mesonic chiral
perturbation theory.
   On the other hand, the counting of bilinears $\bar{\Psi}\Gamma\Psi$ is
probably easiest understood by investigating the matrix elements of
positive-energy plane-wave solutions to the free Dirac equation in the
Dirac representation:
\begin{equation}
\psi^{(+)}(t,\vec{x}\,)=\exp(-ip\cdot x) \sqrt{E+m_N}\left(
\begin{array}{c}
\chi\\
\frac{\vec{\sigma}\cdot\vec{p}}{E+m_N}\chi
\end{array}
\right),
\end{equation}
where $\chi$ denotes a two-component Pauli
spinor and $p^\mu=(E,\vec{p})$ with $E=\sqrt{\vec{p}\,^2+m_N^2}$.
   In the low-energy limit, i.e., for non-relativistic kinematics, the lower
(small) component is suppressed as $|\vec{p}\,|/m_N$ in comparison with the
upper (large) component.
   For the analysis of the bi-linears it is convenient
to divide the 16 Dirac matrices into even and odd ones, ${\cal E}=
\{{\mathbbm 1}, \gamma_0,\gamma_5 \gamma_i,\sigma_{ij}\}$ and ${\cal
O}=\{\gamma_5,\gamma_5 \gamma_0,\gamma_i,\sigma_{i0}\}$
\cite{Foldy:1949wa}, respectively, where odd
matrices couple large and small components but not large with
large, whereas even matrices do the opposite.
   Finally, $i\partial^\mu$ acting on the nucleon solution produces $p^\mu$
which we write symbolically as $p^\mu=(m_N,\vec{0})+(E-m_N,\vec{p}\,)$,
where we count the second term as ${\cal O}(q)$, i.e., as a small quantity.
   We are now in the position to summarize the chiral counting scheme for
the (new) elements of baryon chiral perturbation theory \cite{Krause:1990xc}:
\begin{eqnarray}
\label{4:1:2:powercounting}
&&\Psi,\bar{\Psi} =  {\cal O}(q^0),\, D_{\mu} \Psi = {\cal  O}(q^0),\,
(iD\hspace{-.6em}/ - m)\Psi={\cal O}(q),\nonumber\\
&&{\mathbbm 1},\gamma_\mu,\gamma_5\gamma_\mu,\sigma_{\mu\nu}={\cal O}(q^0),\,
\gamma_5 ={\cal O}(q),
\end{eqnarray}
   where the order given is the minimal one.
   For example, $\gamma_\mu$ has both an ${\cal O}(q^0)$ piece, $\gamma_0$,
as well as an ${\cal O}(q)$ piece, $\gamma_i$.
   Note that because of the additional spin degree of freedom the baryonic
effective Lagrangian contains both odd and even chiral orders.
   A rigorous non-relativistic reduction may be achieved in the framework
of the Foldy-Wouthuysen method \cite{Foldy:1949wa}, \cite{Fearing:1993ii} or the
heavy-baryon approach \cite{Jenkins:1990jv}, \cite{Bernard:1992qa}.

   The construction of the $\mbox{SU(3)}_L\times\mbox{SU(3)}_R$ Lagrangian
proceeds similarly except for the fact that the baryon fields are contained
in the $3\times 3$ matrix of Eq.\ (\ref{4:1:1:su3oktett})
transforming as $K B K^\dagger$.
   As in the mesonic sector, the building blocks are written as products
transforming as $K\cdots K^\dagger$ with a trace taken at the end.
   The lowest-order Lagrangian reads \cite{Georgi}, \cite{Krause:1990xc}
\begin{equation}
\label{4:1:2:l1su3}
{\cal L}^{(1)}_{MB}=\mbox{Tr}\left[\bar{B}\left(iD\hspace{-.7em}/\hspace{.2em}
-M_0\right)B\right]
-\frac{D}{2}\mbox{Tr}\left(\bar{B}\gamma^\mu\gamma_5\{u_\mu,B\}\right)
-\frac{F}{2}\mbox{Tr}\left(\bar{B}\gamma^\mu\gamma_5[u_\mu,B]\right),
\end{equation}
where $M_0$ denotes the mass of the baryon octet in the chiral limit.
  The covariant derivative of $B$ is defined as
\begin{equation}
\label{4:1:2:kovderb}
D_\mu B=\partial_\mu B +[\Gamma_\mu,B],
\end{equation}
with $\Gamma_\mu$ of Eq.\ (\ref{4:1:2:gamma}) [for $\mbox{SU(3)}_L\times\mbox{
SU(3)}_R$].
   The constants $D$ and $F$ may be determined by fitting the semi-leptonic
decays $B\to B'+e^-+\bar{\nu}_e$ at tree level \cite{Borasoy:1998pe}:
\begin{equation}
\label{5:2:df}
D=0.80,\quad
F=0.50.
\end{equation}

\subsection{Renormalization and power counting}
  In the following discussion we will restrict ourselves to the two-flavor case.
  The effective Lagrangian relevant to the one-nucleon sector
consists of the sum of the purely mesonic and $\pi N$ Lagrangians,
respectively,
\begin{equation}
\label{4:2:full_lagrangian}
{\cal L}_{\rm eff}
={\cal L}_{\pi}+{\cal L}_{\pi N}
={\cal L}_\pi^{(2)}+{\cal L}_\pi^{(4)}+\cdots
+{\cal L}_{\pi N}^{(1)}+{\cal L}_{\pi N}^{(2)}
+\cdots,
\end{equation}
which are organized in a derivative and quark-mass expansion.
   Tree-level calculations involving the sum ${\cal L}_\pi^{(2)}+{\cal L}_{\pi N}^{(1)}$
reproduce the current algebra results.
   The higher-order Lagrangians of the $\pi N$ sector can be found in \cite{Gasser:1987rb},
\cite{Ecker:1995rk}, \cite{Fettes:2000gb}.
   When studying higher orders in perturbation theory in terms of loop corrections one encounters
ultraviolet divergences.
   As a preliminary step, the loop integrals are regularized,
typically by means of dimensional regularization.
   In the process of renormalization the
counter terms are adjusted such that they absorb all the ultraviolet
divergences occurring in the calculation of loop diagrams.
   This will be possible, because we include in the Lagrangian all
of the infinite number of interactions allowed by symmetries
\cite{Weinberg:1995mt}.
   At the end the regularization is removed by taking the limit
$n\to 4$.
   Moreover, when renormalizing, we still have the freedom of choosing
a renormalization condition.
   As we will see, the power counting is intimately connected with choosing a suitable
renormalization condition.

\subsubsection{\label{genct}The generation of counter terms}
   Before discussing the power-counting problem and its solution,
let us briefly recall the principles of the renormalization procedure which
will then allow us to set up a consistent power counting.
   At the beginning, the Lagrangian is written down in terms of bare, i.e.,
unrenormalized parameters and fields.
   In order to illustrate the procedure let us discuss
${\cal L}_{\pi N}^{(1)}$ of Eq.~(\ref{4:1:2:l1pin})
and consider the free part in combination with the $\pi N$
interaction term with the smallest number of pion fields,
\begin{equation}
{\cal L}_{\pi N}^{(1)}=\bar \Psi_B \left( i\gamma^\mu
\partial_\mu -m_B -\frac{1}{2}\frac{{\texttt{g}_A}_B}{F_B}
\gamma^\mu
\gamma_5 \partial_\mu {\phi_i}_B  \tau_i \right) \Psi_B +\cdots,
\label{4:2:1:pieceoflolagr}
\end{equation}
where the subscript $B$ denotes bare quantities.
   The renormalization is performed by expressing all the bare parameters and
bare fields of the effective Lagrangian in terms of renormalized quantities
(see, e.g., Refs.\ \cite{Collins:xc}, \cite{Weinberg:1995mt} for details).
   Introducing the renormalized fields through
\begin{equation}
\label{4:2:1:renf}
\Psi=\frac{\Psi_B}{\sqrt{Z_\Psi}},\quad
\phi_i=\frac{{\phi_i}_B}{\sqrt{Z_\phi}},
\end{equation}
we express the field redefinition constants $\sqrt{Z_\Psi}$ and
$\sqrt{Z_\phi}$
and the bare quantities in terms of renormalized parameters:
\begin{eqnarray}
Z_\Psi&=& 1+\delta Z_\Psi\left(m,\texttt{g}_A, g_i, \nu \right),\nonumber\\
Z_{\phi}&=&1+\delta Z_\phi\left(m,\texttt{g}_A, g_i, \nu \right),\nonumber \\
m_B &=&m(\nu)+\delta m\left(m,\texttt{g}_A, g_i, \nu \right),\nonumber \\
{\texttt{g}_A}_B&=&\texttt{g}_A(\nu)+\delta \texttt{g}_A\left(m,\texttt{g}_A, g_i, \nu \right),
\label{4:2:1:bare}
\end{eqnarray}
where $g_i$, $i=1,\cdots, \infty$, collectively denote all the
renormalized parameters which correspond to bare parameters
${g_i}_B$ of the full effective Lagrangian of Eq.~(\ref{4:2:full_lagrangian}).
   The parameter $\nu$ indicates the dependence on the choice of the
renormalization condition.
   We emphasize that the usual choice $m(\nu)=m$, where $m$
is the nucleon pole mass in the chiral limit, is only one among an infinite
number of possibilities.
   Substituting Eqs.\ (\ref{4:2:1:renf}) and (\ref{4:2:1:bare}) into
Eq.\ (\ref{4:2:1:pieceoflolagr}), we obtain
\begin{equation}
\label{4:2:1:bct}
{\cal L}_{\pi N}^{(1)}={\cal L}_{\rm basic}+{\cal L}_{\rm ct}+\cdots
\end{equation}
with the so-called basic and counter-term Lagrangians,
respectively,
\begin{eqnarray}
\label{4:2:1:lbasic}
{\cal L}_{\rm basic}&=&
\bar \Psi \left( i\gamma^\mu\partial_\mu - m
-\frac{1}{2} \frac{\texttt{g}_{A}}{F}\gamma^\mu\gamma_5\partial_\mu \phi_i\tau_i\right) \Psi,\\
\label{4:2:1:lcounterterm}
{\cal L}_{\rm ct}&=&\delta Z_\Psi \bar \Psi i\gamma^\mu\partial_\mu
\Psi
-\delta\{m\}\bar{\Psi}\Psi
-\frac{1}{2}\delta\left\{\frac{\texttt{g}_{A}}{F}\right\}
\bar{\Psi}\gamma^\mu \gamma_5  \partial_\mu \phi_i  \tau_i\Psi,
\end{eqnarray}
where we introduced the abbreviations
\begin{eqnarray*}
\delta\{m\}&\equiv&\delta Z_\Psi m+Z_\Psi\delta m,\\
\delta\left\{\frac{\texttt{g}_{A}}{F}\right\}&\equiv&
\delta Z_\Psi \frac{\texttt{g}_{A}}{F}\sqrt{Z_\pi}
+Z_\Psi\left(\frac{{\texttt{g}_{A}}_B}{F_B}-
\frac{\texttt{g}_{A}}{F}
\right)\sqrt{Z_\phi}
+\frac{\texttt{g}_{A}}{F}(\sqrt{Z_\phi}-1).
\end{eqnarray*}
   In Eq.\ (\ref{4:2:1:lbasic}), $m$, $\texttt{g}_A$,
and $F$ denote the chiral limit of the physical nucleon mass, the axial-vector coupling
constant, and the pion-decay constant, respectively.
   Expanding the counter-term Lagrangian of Eq.\ (\ref{4:2:1:lcounterterm})
in powers of the renormalized coupling constants generates an infinite
series.
   By adjusting the expansion coefficients suitably, the individual terms are responsible for
the subtractions of loop diagrams.
   For example, the divergences occurring in dimensionally
regularized one-loop calculations involving vertices of ${\cal L}_2$ are absorbed in the
renormalization of the bare coefficients $L_i$ [see Eq.~(\ref{3:3:1:lihi})].

\subsubsection{Power counting for renormalized diagrams}
\label{subsection_pcrd}
   In the following, whenever we speak of renormalized diagrams, we refer to
diagrams which have been calculated with a basic Lagrangian and to which the
contribution of the counter-term Lagrangian has been added.
   Counter-term contributions are typically denoted by a cross.
   One also says that the diagram has been subtracted, i.e., the unwanted
contribution has been removed with the understanding that this
can be achieved by a suitable choice for the coefficient of the
counter-term Lagrangian.
   In this context we will adjust the {\em finite} pieces of the renormalized
couplings such that renormalized diagrams satisfy the following power counting:
   a loop integration in $n$ dimensions counts as $q^n$,
pion and fermion propagators count as $q^{-2}$ and $q^{-1}$,
respectively, vertices derived from ${\cal L}_{\pi}^{(2k)}$ and
${\cal L}_{\pi N}^{(k)}$ count as $q^{2k}$ and $q^k$, respectively.
   Here, $q$ collectively stands for a small quantity such as the pion
   mass, small external four-momenta of the pion, and small external
three-momenta of the nucleon.
   The power counting does not uniquely fix the renormalization scheme,
i.e., there are different renormalization schemes leading to the
above specified power counting.

\subsubsection{The power-counting problem}
\begin{figure}[t]
\begin{center}
\epsfig{file=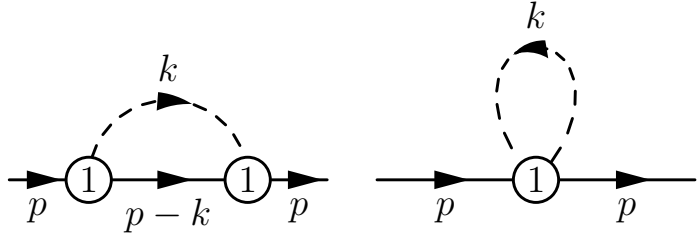,width=0.5\textwidth}
\caption{\label{4:2:3:SEDia}
One-loop contributions to the nucleon self energy. The
number 1 in the interaction blobs refers to ${\cal L}_{\pi
N}^{(1)}$.}
\end{center}\end{figure}

   In the mesonic sector, the combination of dimensional regularization and
the modified minimal subtraction scheme $\widetilde{\mbox{MS}}$
[see Eq.~(\ref{3:3:1:lihi})] leads to a straightforward correspondence between
the chiral and loop expansions.
   By studying the one-loop contributions of Fig.\ \ref{4:2:3:SEDia} to the
nucleon self energy, we will see that this correspondence, at first sight,
seems to be lost in the baryonic sector.

   In the following we will calculate the mass $m_N$ of the nucleon up to and
including ${\cal O}(q^3)$.
   As in the case of the Goldstone bosons, the physical mass is defined through
the pole of the full propagator, but here at $\slashed{p}=m_N$.
   In terms of the nucleon self energy $\Sigma(\slashed{p})$ we will solve the equation
\begin{equation}
\label{4:2:3:MassDef}
m_N-m-\Sigma(m_N)=0,
\end{equation}
where $m$ denotes the nucleon mass in the chiral limit.

   According to the power counting specified above, we need to calculate the two types of
one-loop contributions shown in Fig.\ \ref{4:2:3:SEDia} together with the
corresponding counter-term contribution and a tree-level contribution.
   After renormalization, we would like
to have the orders $D=n\cdot 1-2\cdot 1-1\cdot 1+2\cdot 1=n-1$ for the first loop diagram and
$n\cdot 1-2\cdot 1+1\cdot 1=n-1$ for the second loop diagram.

   The basic interaction Lagrangian obtained from expanding $\mathcal{L}_{\pi N}^{(1)}$ up
to and including two pion fields reads
\begin{displaymath}
\label{5:3:lpin}
{\cal L}_{\rm int}^{(1)}=-\frac{1}{2}\frac{{\texttt
g}_{A}}{F} \bar{\Psi}\gamma^\mu\gamma_5
\partial_\mu\phi_i\tau_i\Psi
-\frac{1}{4F^2}\bar{\Psi}\gamma^\mu\vec{\phi}
\times\partial_\mu\vec{\phi}\cdot\vec{\tau}\,\Psi.
\end{displaymath}
   The corresponding Feynman rules are given by
\begin{center}
\begin{tabular}{cc}
\parbox{0.15\textwidth}{\epsfig{file=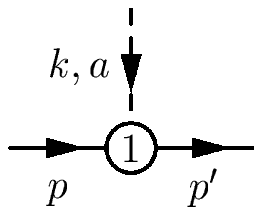,width=0.15\textwidth}}&
\parbox{0.2\textwidth}{\begin{equation}
\label{4:2:3:feynmanrule1}
-\frac{{\texttt g}_{A}}{2F}\,\slashed{k}\gamma_5\tau_a,\end{equation}}\\
\\
\parbox{0.2\textwidth}{\epsfig{file=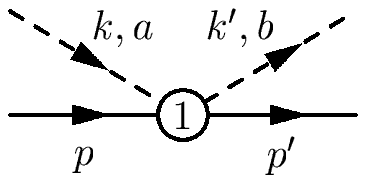,width=0.2\textwidth}}&
\parbox{0.3\textwidth}{\begin{equation}
\label{4:2:3:feynmanrule2}
\frac{1}{4F^2}\,(\slashed{k}+\slashed{k}')\epsilon_{abc}\tau_c.
\end{equation}}\\
\\
\end{tabular}
\end{center}
   From the next-to-leading-order $\pi N$ Lagrangian $\mathcal{L}_{\pi N}^{(2)}$ we only need
one term, namely,
\begin{equation}
\label{4:2:3:LpiN2}
\mathcal{L}_{\pi N}^{(2)}=c_1\,\mbox{Tr}(\chi U^\dagger +U\chi^\dagger)\bar{\Psi}\Psi+\cdots,
\end{equation}
resulting in the constant tree-level contribution
\begin{equation}
\Sigma^{\rm tree}_2=-4 c_1 M^2.
\end{equation}
   Moreover, there are no tree-level contributions from the
Lagrangian $\mathcal{L}_{\pi N}^{(3)}$.
   The second diagram of Fig.~\ref{4:2:3:SEDia} is zero, because the
contraction $\epsilon_{aac}=0$ in the Feynman rule of Eq.~(\ref{4:2:3:feynmanrule2}) vanishes.
   In dimensional regularization, the first diagram of
Fig.~\ref{4:2:3:SEDia} generates the contribution
\begin{equation}
\label{4:2:3:Loop1}
-i\Sigma^{\rm loop}(\slashed{p})
=-i\frac{3\texttt{g}_{A}^2}{4F^2}\,
    i\mu^{4-n}\int\frac{\mbox{d}^nk}{(2\pi)^n}\frac{\slashed{k}
    (\slashed{p}-m-\slashed{k})\slashed{k}}
    {[(p-k)^2-m^2+i0^+](k^2-M^2+i0^+)}.
\end{equation}
Using $ \{\gamma_\mu, \gamma_\nu\}=2g_{\mu\nu}$, the numerator of the integrand
is written as
\begin{displaymath}
\label{Num}
    -(\slashed{p}+m)k^2+(p^2-m^2)\slashed{k}
    -\left[(p-k)^2-m^2\right]\slashed{k},
\end{displaymath}
yielding the intermediate result
\begin{eqnarray}
\label{4:2:3:LoopRes}
\Sigma^{\rm loop}(\slashed{p})&=&
  \frac{3\texttt{g}_{A}^2}{4F^2}\,
  \left\{-(\slashed{p}+m)\mu^{4-n}i\int\frac{\mbox{d}^nk}{(2\pi)^n}
  \frac{1}{(p-k)^2-m^2+i0^+}\right.\nonumber\\
&&-(\slashed{p}+m)M^2\mu^{4-n}i\int\frac{\mbox{d}^nk}{(2\pi)^n}
  \frac{1}{[(p-k)^2-m^2+i0^+](k^2-M^2+i0^+)}\nonumber\\
&&
+(p^2-m^2)\mu^{4-n}i\int\frac{\mbox{d}^nk}{(2\pi)^n}
  \frac{\slashed{k}}{[(p-k)^2-m^2+i0^+](k^2-M^2+i0^+)}\nonumber\\
&& \left.-\mu^{4-n}i\int\frac{\mbox{d}^nk}{(2\pi)^n}
\frac{\slashed{k}}{k^2-M^2+i0^+}\right\}.
\end{eqnarray}
   The last term in Eq.~(\ref{4:2:3:LoopRes}) vanishes since the
integrand is odd in $k$.
   We use the following convention for scalar loop integrals,
\begin{equation}
\label{scalarInt}
I_{N\cdots\pi\cdots}(p_1,\cdots,q_1,\cdots)
=\mu^{4-n}i\int\frac{\mbox{d}^nk}{(2\pi)^n}
    \frac{1}{[(k+p_1)^2-m^2+i0^+]\cdots[(k+q_1)^2-M^2+i0^+]\cdots}\,.
\end{equation}
   The vector integral in the third line of Eq.~(\ref{4:2:3:LoopRes}) is
determined using the ansatz
\begin{equation}\label{vecInt}
    \mu^{4-n}i\int\frac{\mbox{d}^nk}{(2\pi)^n}
  \frac{k_\mu}{[(p-k)^2-m^2+i0^+](k^2-M^2+i0^+)}=
  p_\mu \, C.
\end{equation}
Multiplying Eq.~(\ref{vecInt}) by $p^\mu$, one obtains for $C$,
\begin{equation}\label{C}
    C=\frac{1}{2p^2}\left[I_N-I_\pi+(p^2-m^2+M^2)I_{N\pi}(-p,0)\right].
\end{equation}
   In terms of the above convention for the scalar loop integrals the loop contribution
to the nucleon self energy reads
\begin{eqnarray}
\label{4:2:3:LoopResInt}
\Sigma^{\rm loop}(\slashed{p})
&=&-\frac{3\texttt{g}_{A}^2}{4F^2}\,\Bigg\{
(\slashed{p}+m)I_N+(\slashed{p}+m)M^2 I_{N\pi}(-p,0)\nonumber\\
&&-(p^2-m^2)\frac{\slashed{p}}{2p^2}
\left[I_N-I_\pi+(p^2-m^2+M^2)I_{N\pi}(-p,0)\right]\Bigg\}.
\end{eqnarray}
   The explicit expressions for the integrals are given by
\begin{eqnarray}\label{Int}
  I_{\pi}&=&\frac{M^2}{16\pi^2}\left[R+\ln{\left(\frac{M^2}{\mu^2}\right)}\right],\quad
  I_{N}=\frac{m^2}{16\pi^2}
\left[R+\ln{\left(\frac{m^2}{\mu^2}\right)}\right],\nonumber\\
  I_{N\pi}(p,0)&=&\frac{1}{16\pi^2}
\left[R+\ln{\left(\frac{m^2}{\mu^2}\right)}-1
+
\frac{p^2-m^2-M^2}{p^2}\ln{\left(\frac{M}{m}\right)}
+\frac{2m M}{p^2}F(\Omega)\right],
\end{eqnarray}
where $R$ is given in Eq.~(\ref{3:2:5:R}), $\Omega$ is defined as
\begin{displaymath}
\Omega=\frac{p^2-m^2-M^2}{2m M},
\end{displaymath}
and
\begin{displaymath}
F(\Omega)=\left\{\begin{array}{l@{\quad}l}
\sqrt{\Omega^2-1}\ln{\left(-\Omega-\sqrt{\Omega^2-1}\right)}, &
\Omega\le -1,\\
\sqrt{1-\Omega^2}\;\arccos(-\Omega),&-1 \le\Omega\le 1,\\
\sqrt{\Omega^2-1}\ln{\left(\Omega+\sqrt{\Omega^2-1}\right)}
-i\pi\sqrt{\Omega^2-1}, & 1 \le \Omega\,.
\end{array}\right.
\end{displaymath}
   Because of the terms proportional to $R$, the result for the self
energy contains divergences as $n\rightarrow 4$, so it has to be
renormalized.
   The counter-term Lagrangian must produce structures
which precisely cancel the divergences, because otherwise the result
for the nucleon mass will not be finite.
   For convenience, we choose the renormalization parameter $\mu=m$.

\begin{figure}[t]
\begin{center}
\epsfig{file=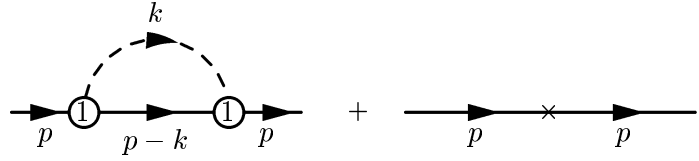,width=0.5\textwidth}
\caption{\label{4:2:3:ren_diag}
Renormalized one-loop self-energy diagram.}
\end{center}
\end{figure}

   In the modified minimal subtraction scheme $\widetilde{\mbox{MS}}$ all the
contributions proportional to $R$ are canceled by corresponding contributions
generated by the counter-term Lagrangian of Eq.~(\ref{4:2:1:lcounterterm}), but also
by counter-term Lagrangians resulting from higher-order terms of
Eq.~(\ref{4:2:full_lagrangian}).
   Operationally this means that we simply drop all terms proportional to
$R$ and indicate the renormalized coupling constants by a subscript $r$.
   Again, this is possible, because we include in the Lagrangian all
of the infinite number of interactions allowed by symmetries
\cite{Weinberg:1995mt}.
   The renormalized diagram is depicted in Fig.~\ref{4:2:3:ren_diag}, where
the cross generically denotes counter-term contributions.
   The $\widetilde{\rm MS}$-renormalized self-energy contribution then reads
\begin{equation}
\label{4:2:3:SErenorm}
\Sigma^{\rm
loop}_r(\slashed{p})=-\frac{3\texttt{g}_{Ar}^2}{4F^2}\,\Bigg\{
(\slashed{p}+m)M^2
I_{N\pi}^r(-p,0)
-(p^2-m^2)\frac{\slashed{p}}{2p^2}
\left[(p^2-m^2+M^2)I_{N\pi}^r(-p,0)-I_\pi^r\right]\Bigg\},
\end{equation}
where the superscript $r$ on the integrals means that the terms
proportional to $R$ have been dropped.
   Writing $\slashed{p}+m=2m+(\slashed p-m)$ and comparing the first term
of Eq.~(\ref{4:2:3:LoopResInt}) with Eq.~(\ref{4:2:3:SErenorm}), we
note that among other terms, the $\widetilde{\rm MS}$ renormalization
involves (even in the chiral limit) an infinite renormalization
yielding the relation between the bare and the renormalized mass
\cite{Gasser:1987rb}
\begin{displaymath}
m_B=m+\frac{3\texttt{g}_{Ar}^2}{32\pi^2 F^2}m^3 R+\cdots.
\end{displaymath}
Using
\begin{displaymath}
I_{N\pi}^r(-p,0)=-\frac{1}{16\pi^2}+\cdots,
\end{displaymath}
we see that the $\widetilde{\rm MS}$-renormalized self energy produces a contribution
of ${\cal O}(q^2)$ which is in conflict with the power counting assigned above.
   For a long time this was interpreted as the absence of a systematic power counting in the
relativistic formulation of ChPT.

   We can now solve Eq.~(\ref{4:2:3:MassDef}) for the nucleon mass,
\begin{equation}
m_N=m+\Sigma^{\rm tree}_{2r}(m_N)+\Sigma^{\rm loop}_r(m_N)=m-4c_{1r}M^2+\Sigma^{\rm loop}_r(m_N).
\end{equation}
   We have for the difference $m_N-m={\cal O}(q^2)$.
Since our calculation is only valid up to ${\cal O}(q^3)$, it is sufficient
to determine $\Sigma^{\rm loop}_r(m_N)$ to that order.
   In fact, using $\arccos{(-\Omega)}=\frac{\pi}{2}+\cdots,$ the expansion of $I_{N\pi}^r$
is given by
\begin{equation}
\label{4:2:3:INpExp}
    I_{N\pi}^r=\frac{1}{16\pi^2}\left(-1+\frac{\pi M}{m}+\cdots\right),
\end{equation}
from which we obtain for the nucleon mass in the $\widetilde{\rm MS}$ scheme
\cite{Gasser:1987rb},
\begin{equation}
\label{4:2:3:MassMStilde}
    m_N=m-4c_{1r}M^2+
    \frac{3\texttt{g}_{Ar}^2M^2}{32\pi^2F^2}m
    -\frac{3\texttt{g}_{Ar}^2M^3}{32\pi^2F^2}.
\end{equation}
   The solution to the power-counting problem is the observation
that the term violating the power counting, namely, the third on the right-hand side
of Eq.~(\ref{4:2:3:MassMStilde}), is \emph{analytic} in the quark mass
and can thus be absorbed in counter terms.
   In addition to the $\widetilde{\rm MS}$ scheme we have to perform an additional
{\em finite} renormalization.
   For that purpose we rewrite
\begin{equation}
\label{4:2:3:cRenorm}
    c_{1r}=c_1+\delta c_1,\quad \delta c_1 =\frac{3 m {\texttt g}_A^2}{128 \pi^2 F^2}+\cdots
\end{equation}
in Eq.~(\ref{4:2:3:MassMStilde}) which then gives the
final result for the nucleon mass at ${\cal O}(q^3)$:
\begin{equation}
\label{4:2:3:MassFinal}
    m_N=m-4c_{1}M^2
    -\frac{3\texttt{g}_{A}^2M^3}{32\pi^2F^2}.
\end{equation}
   To summarize, we have shown that the validity of a power-counting scheme is intimately
connected with a suitable renormalization condition.
   In the case of the nucleon mass, the $\widetilde{\rm MS}$ scheme alone does not
suffice to bring about a consistent power counting.
   We will shortly outline two methods, the infrared renormalization \cite{Becher:1999he}
and the extended on-mass-shell renormalization \cite{Fuchs:2003qc}, which both produce
a systematic power counting in a manifestly Lorentz-invariant framework.

\subsection{Solutions to the power-counting problem}

\subsubsection{Heavy-baryon approach}
\label{hb}
   The first solution to the power-counting problem
was provided by the heavy-baryon formulation of ChPT \cite{Jenkins:1990jv},
\cite{Bernard:1992qa}.
   The basic idea consists in dividing an external nucleon four-momentum
into a large piece close to on-shell kinematics and a soft residual contribution:
$p = m v +k_p$, $v^2=1$, $v^0\ge 1$
[often $v^\mu = (1,0,0,0)$].
   The relativistic nucleon field is expressed in terms of
velocity-dependent fields,
\begin{displaymath}
\Psi(x)=e^{-im v \cdot x} ({\cal N}_v +{\cal H}_v),
\end{displaymath}
with
\begin{displaymath}
{\cal N}_v=e^{+im v\cdot x}\frac{1}{2}(1+v\hspace{-.5em}/)\Psi,\quad
{\cal H}_v=e^{+im v\cdot x}\frac{1}{2}(1-v\hspace{-.5em}/)\Psi.
\end{displaymath}
   Using the equation of motion for ${\cal H}_v$, one can
eliminate ${\cal H}_v$ and obtain a Lagrangian for ${\cal N}_v$
which, to lowest order, reads \cite{Bernard:1992qa}
\begin{displaymath}
\widehat{\cal L}^{(1)}_{\pi N}=\bar{\cal N}_v(iv\cdot D + \texttt{g}_A S_v\cdot u)
{\cal N}_v+{\cal O}(1/m),\quad S^\mu_v=\frac{i}{2}\gamma_5\sigma^{\mu\nu}v_\nu.
\end{displaymath}
   The result of the heavy-baryon reduction is a $1/m$ expansion of the
Lagrangian similar to a Foldy-Wouthuysen expansion \cite{Foldy:1949wa}.
   In higher orders in the chiral expansion, the expressions due to
$1/m$ corrections of the Lagrangian become increasingly complicated
\cite{Fettes:2000gb}.

\begin{figure}[t]
\begin{center}
\epsfig{file=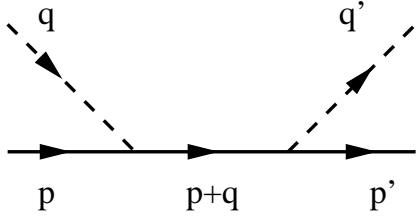,width=0.3\textwidth}
\caption{$s$-channel pole diagram of $\pi$N scattering.}
\label{4:3:1:fig:schannel}
\end{center}
\end{figure}
   Moreover---and what is more important---the approach sometimes generates problems
regarding analyticity which can be illustrated by considering the example of
pion-nucleon scattering \cite{Becher:2000mb}.
   The invariant amplitudes describing the scattering amplitude develop
poles for $s=m_N^2$ and $u=m_N^2$.
   For example, the singularity due to the nucleon pole in the $s$ channel
(see Fig.\ \ref{4:3:1:fig:schannel}) is understood in terms of the relativistic
propagator
\begin{equation}
\label{relprop}
\frac{1}{(p+q)^2-m_N^2}=\frac{1}{2p\cdot q+M_\pi^2},
\end{equation}
which, of course, has a pole at $2p\cdot q=-M_\pi^2$ or, equivalently,
$s=m_N^2$.
  Analogously, a second pole results from the $u$ channel at $u=m_N^2$.
  Although both poles are not in the physical region of pion-nucleon
scattering, analyticity of the invariant amplitudes requires these poles
to be present in the amplitudes.
   Let us compare the situation with a heavy-baryon type of expansion,
where, for simplicity, we choose as the four-velocity $p^\mu=m_N v^\mu$,
\begin{equation}
\label{4:3:1:relpropv}
\frac{1}{2p\cdot q+M_\pi^2}=
\frac{1}{2m_N}\frac{1}{v\cdot q+ \frac{M_\pi^2}{2m_N}}
=\frac{1}{2 m_N}\frac{1}{v\cdot q}\left(1-\frac{M_\pi^2}{2 m_N v\cdot q}
+\cdots\right).
\end{equation}
   Clearly, to any finite order the heavy-baryon expansion produces poles
at $v\cdot q=0$ instead of a simple pole at $v\cdot q=-M_\pi^2/(2 m_N)$ and
will thus not generate the (nucleon) pole structures of the invariant
amplitudes unless an infinite number of diagrams is summed.
   For a comprehensive overview of calculations performed in the heavy-baryon
framework the reader is referred to Ref.\ \cite{Bernard:1995dp}.

\subsubsection{Master integral}
\label{subsubsection_master_integral}
   We have seen that the modified minimal subtraction scheme $\widetilde{\rm MS}$
does not produce the desired power counting.
   We will discuss the power-counting problem in terms of the dimensionally regularized
one-loop integral
\begin{eqnarray}
\label{4:3:2:Hdef}
H(p^2,m^2,M^2;n)
&\equiv& -i\int \frac{\mbox{d}^n
k}{(2\pi)^n} \frac{1}{[(k-p)^2-m^2+i0^+](k^2-M^2+i0^+)}\nonumber\\
&=&-i\int \frac{\mbox{d}^n k}{(2\pi)^n} \frac{1}{[k^2-2p\cdot k
+(p^2-m^2)+i0^+](k^2-M^2+i0^+)}.
\end{eqnarray}
   We are interested in nucleon four-momenta close to the mass-shell condition,
$p^2\approx m^2$, counting $p^2-m^2$ as ${\cal O}(q)$ and $M^2$ as ${\cal O}(q^2)$.
   In order to conform with Ref.~\cite{Becher:1999he}, we have omitted
the factor $\mu^{4-n}$ and have reversed the overall sign in comparison with
our previous definition of $I_{N\pi}$.
   Let us turn to the discussion of $H(p^2,m^2,M^2;n)$.
   To that end, we make use of the Feynman parametrization
\begin{equation}
\label{4:3:2:feynmanparh}
{1\over ab}=\int_0^1 \mbox{d}z  {1\over [az+b(1-z)]^2}
\end{equation}
with $a=(k-p)^2-m^2+i0^+$ and $b=k^2-M^2+i0^+$, interchange the
order of integrations, and perform the shift $k\to k+zp$, to obtain
\begin{displaymath}
H(p^2,m^2,M^2;n)=-i \int_0^1 \mbox{d}z\int\frac{\mbox{d}^n
k}{(2\pi)^n}\frac{1}{[k^2-A(z) +i0^+]^2},
\end{displaymath}
where
\begin{displaymath}
A(z)=z^2 p^2-z(p^2-m^2+M^2)+M^2.
\end{displaymath}
Making use of
\begin{displaymath}
\int \frac{\mbox{d}^n k}{(2\pi)^n} \frac{(k^2)^p}{(k^2-A)^q}
=\frac{i(-)^{p-q}}{(4\pi)^{\frac{n}{2}}}
\frac{\Gamma\left(p+\frac{n}{2}\right)\Gamma\left(q-p-\frac{n}{2}\right)}{
\Gamma\left(\frac{n}{2}\right)\Gamma(q)} A^{p+\frac{n}{2}-q}
\end{displaymath}
with $p=0$ and $q=2$, we find
\begin{equation}
\label{4:3:2:HPmMn}
H(p^2,m^2,M^2;n)=\frac{1}{(4\pi)^{\frac{n}{2}}}
\Gamma\left(2-\frac{n}{2}\right) \int_0^1 \mbox{d}z
[A(z)-i0^+]^{\frac{n}{2}-2}.
\end{equation}
   The relevant properties can nicely be displayed at the threshold
$p^2_{\rm thr}=(m+M)^2$, where $A(z)=[z(m+M)-M]^2$
is particularly simple.
   The small imaginary part can be dropped in this case, because $A(z)$
is never negative.
    Splitting the integration interval into $[0,z_0]$ and $[z_0,1]$ with
$z_0=M/(m+M)$, we have, for $n>3$,
\begin{eqnarray*}
\int_0^1 \mbox{d}z [A(z)]^{\frac{n}{2}-2}&=&\int_0^{z_0}\mbox{d}z [M-z(m+M)]^{n-4}
+\int_{z_0}^1\mbox{d}z [z(m+M)-M]^{n-4}\\
&=&\frac{1}{(n-3)(m+M)}(M^{n-3}+m^{n-3}),
\end{eqnarray*}
yielding, through analytic continuation, for arbitrary $n$
\begin{equation}
\label{4:3:2:defhthr}
 H((m+M)^2,m^2,M^2;n)=
\frac{\Gamma\left(2-\frac{n}{2}\right)}{(4\pi)^{\frac{n}{2}}(n-3)}
\left(\frac{M^{n-3}}{m+M}+\frac{m^{n-3}}{m+M}\right).
\end{equation}
   The first term, proportional to $M^{n-3}$, is defined as the so-called
infrared singular part $I$.
   Since $M\to 0$ implies $p^2_{\rm thr} \to m^2$ this term is
singular for $n\leq 3$.
   The second term, proportional to $m^{n-3}$, is defined as the
infrared regular part $R$.
   Note that for non-integer $n$ the infrared singular
part contains non-integer powers of $M$, while an
expansion of the regular part always contains non-negative integer powers
of $M$ only.

\subsubsection{Infrared regularization}
   Let us now turn to a {\em formal} definition of the infrared singular
and regular parts for arbitrary $p^2$ \cite{Becher:1999he} which makes use of the Feynman
parametrization of Eq.\ (\ref{4:3:2:HPmMn}).
   Introducing the dimensionless variables
\begin{equation}
\label{4:3:3:alphaomegadef}
\alpha=\frac{M}{m}={\cal O}(q),\quad \Omega=\frac{p^2-m^2-M^2}{2m M}={\cal O}(q^0),
\end{equation}
we rewrite $A(z)$ as
\begin{displaymath}
A(z)=m^2[z^2-2\alpha\Omega z(1-z)+\alpha^2(1-z)^2]\equiv m^2 C(z),
\end{displaymath}
so that $H$ is now given by
\begin{equation}
\label{4:3:3:defh2}
H(p^2,m^2,M^2;n)=\kappa(m;n) \int_0^1 \mbox{d}z
[C(z)-i0^+]^{\frac{n}{2}-2},
\end{equation}
where
\begin{equation}
\label{4:3:3:kappan}
\kappa(m;n)
=\frac{\Gamma\left(2-\frac{n}{2}\right)}{(4\pi)^{\frac{n}{2}}}m^{n-4}.
\end{equation}
   The infrared singularity originates from small values of $z$, where
the function $C(z)$ goes to zero as $M\to 0$.
   In order to isolate the divergent part one scales the integration variable
$z\equiv \alpha x$ so that the upper limit $z=1$ in Eq.\ (\ref{4:3:3:defh2})
corresponds to $x=1/\alpha\to
\infty$ as $M\to 0$.
   An integral $I$ having the same infrared singularity as $H$ is then
defined which is identical to $H$ except that the upper limit is
replaced by $\infty$:
\begin{equation}
\label{4:3:3:Idef}
I\equiv\kappa(m;n)\int_0^\infty \mbox{d}z[C(z)-i0^+]^{\frac{n}{2}-2}
=\kappa(m;n) \alpha^{n-3}\int_0^\infty \mbox{d}x[D(x)-i0^+]^{\frac{n}{2}-2},
\end{equation}
where
\begin{displaymath}
D(x)=1-2\Omega x+x^2+2\alpha x(\Omega x-1)+\alpha^2 x^2.
\end{displaymath}
   (The pion mass $M$ is not sent to zero.)
   Accordingly, the regular part of $H$ is defined as
\begin{equation}
\label{4:3:3:Rdef}
R\equiv-\kappa (m;n)\int_1^\infty \mbox{d}z [C(z)-i0^+]^{\frac{n}{2}-2},
\end{equation}
so that
\begin{equation}
\label{4:3:3:hir}
H=I+R.
\end{equation}
   Let us verify that the definitions of Eqs.\ (\ref{4:3:3:Idef}) and
(\ref{4:3:3:Rdef}) indeed reproduce the behavior of Eq.\ (\ref{4:3:2:defhthr}).
   To that end we make use of $\Omega_{\rm thr}=1$, yielding
\begin{equation}
\label{4:3:3:ithr1}
I_{\rm thr}=\kappa(m;n)\alpha^{n-3}\int_0^\infty \mbox{d}x
\left\{[(1+\alpha)x-1]^2-i0^+
\right\}^{\frac{n}{2}-2},
\end{equation}
which converges for $n<3$.
   In order to continue the integral to $n>3$, write
\cite{Becher:1999he}
\begin{eqnarray*}
\left\{[(1+\alpha)x-1]^2-i0^+\right\}^{\frac{n}{2}-2}=
\frac{(1+\alpha)x-1}{(1+\alpha)(n-4)}\frac{\mbox{d}}{\mbox{d}x}
\left\{[(1+\alpha)x-1]^2-i0^+\right\}^{\frac{n}{2}-2},
\end{eqnarray*}
and make use of a partial integration
\begin{eqnarray*}
\lefteqn{\int_0^\infty \mbox{d}x
\left\{[(1+\alpha)x-1]^2-i0^+\right\}^{\frac{n}{2}-2}
=}\\
&&\left[\frac{(1+\alpha)x-1}{(1+\alpha)(n-4)}
\left\{[(1+\alpha)x-1]^2-i0^+\right\}^{\frac{n}{2}-2}
\right]_0^\infty-\frac{1}{n-4}
\int_0^\infty \mbox{d}x \left\{[(1+\alpha)x-1]^2-i0^+\right\}^{\frac{n}{2}-2}.
\end{eqnarray*}
   For $n<3$, the first expression vanishes at the upper limit and,
at the lower limit, yields $1/[(1+\alpha)(n-4)]$.
   Bringing the second expression to the left-hand side, we may then
continue the integral analytically as
\begin{equation}
\label{5:6:intanc}
\int_0^\infty \mbox{d}x\left\{[(1+\alpha)x-1]^2-i0^+
\right\}^{\frac{n}{2}-2}=\frac{1}{(n-3)(1+\alpha)},
\end{equation}
so that we obtain for $I_{\rm thr}$
\begin{equation}
\label{4:3:3:ithr}
I_{\rm thr}=\kappa(m;n)\alpha^{n-3}\frac{1}{(n-3)(1+\alpha)}
=
\frac{\Gamma\left(2-\frac{n}{2}\right)}{(4\pi)^{\frac{n}{2}}(n-3)}
\frac{M^{n-3}}{m+M},
\end{equation}
which agrees with the infrared singular part $I$ of Eq.\ (\ref{4:3:2:defhthr}).

  The threshold value of the regular part of
Eq.\ (\ref{4:3:3:Rdef}) is obtained by analytic continuation from $n<3$
to $n>3$:
\begin{eqnarray}
\label{4:3:3:Rthr}
 R_{\rm thr}&=&-\frac{\Gamma\left(2-\frac{n}{2}\right)}{(4\pi)^{\frac{n}{2}}}
\int_1^\infty \mbox{d}z [z(m+M)-M_\pi]^{n-4}\nonumber\\
&=& -\frac{\Gamma\left(2-\frac{n}{2}\right)}{(4\pi)^{\frac{n}{2}}}
\frac{1}{(n-3)(m+M)}(\infty^{n-3}-m^{n-3})\nonumber\\
&\stackrel{\mbox{$n<3$}}{=}&\frac{\Gamma\left(2-\frac{n}{2}\right)}{(4\pi)^{\frac{n}{2}}(n-3)}
\frac{m^{n-3}}{m+M},
\end{eqnarray}
which is indeed the regular part $R$ of Eq.\ (\ref{4:3:2:defhthr}).

   What distinguishes $I$ from $R$ is that, for non-integer values of
$n$, the chiral expansion of $I$ gives rise to non-integer powers of
${\cal O}(q)$, whereas the regular part $R$ may be
expanded in an ordinary Taylor series.
   For the threshold integral, this can nicely be seen by expanding
$I_{\rm thr}$ and $R_{\rm thr}$ in the pion mass counting as
${\cal O}(q)$.
   On the other hand, it is the regular part which does not satisfy
the counting rules.
    The basic idea of the infrared renormalization consists of replacing
the general integral $H$ of Eq.\ (\ref{4:3:2:HPmMn})
by its infrared singular part $I$, defined in Eq.\ (\ref{4:3:3:Idef}),
and dropping the regular part $R$, defined in Eq.\ (\ref{4:3:3:Rdef}).
   In the low-energy region $H$ and $I$ have the same analytic properties
whereas the contribution of $R$, which is of the type of an infinite
series in the momenta, can be included by adjusting the coefficients of
the most general effective Lagrangian.
   This is the infrared renormalization condition.

   As discussed in detail in Ref.\ \cite{Becher:1999he}, the method can
be generalized to an arbitrary one-loop graph (see also Ref.\ \cite{Semke:2005sn}).
   It is first argued that tensor integrals involving
an expression of the type $k^{\mu_1}\cdots k^{\mu_2}$ in the numerator
may always be reduced to scalar loop integrals of the form
\begin{displaymath}
-i\int\frac{\mbox{d}^n k}{(2\pi)^n}\frac{1}{a_1\cdots a_m}\frac{1}{b_1\cdots b_n},
\end{displaymath}
where $a_i=(q_i+k)^2-M^2+i0^+$ and $b_i=(p_i-k)^2-m^2+i0^+$ are
inverse meson and nucleon propagators, respectively.
   Here, the $q_i$ refer to four-momenta of ${\cal O}(q)$ and the $p_i$ are
four-momenta which are not far off the nucleon mass shell, i.e.,
$p_i^2=m^2+{\cal O}(q)$.
    Using the Feynman parametrization, all pion propagators and all nucleon
propagators are separately combined, and the result is written in such
a way that it is obtained by applying $(m-1)$ and $(n-1)$ partial derivatives
with respect to $M^2$ and $m^2$, respectively, to a master formula.
   A simple illustration is given by
\begin{displaymath}
\frac{1}{a_1 a_2}=\int_0^1 \mbox{d}z \frac{1}{[a_1 z+a_2(1-z)]^2}
=\frac{\partial}{\partial M^2}\int_0^1 \mbox{d}z \frac{1}{a_1 z+a_2 (1-z)},
\end{displaymath}
where $a_i=(q_i+k)^2-M^2+i0^+$.
   Of course, the expressions become more complicated for larger numbers
of propagators.
   The relevant property of the above procedure is that the result of
combining the meson propagators is of the type $1/A$ with
$A=(k+q)^2-M^2+i0^+$, where $q$ is a linear combination of the $m$ momenta
$q_i$, with an analogous expression $1/B$ for the nucleon propagators.
   Finally, the expression
\begin{displaymath}
-i\int\frac{\mbox{d}^n k}{(2\pi)^n}\frac{1}{AB}
\end{displaymath}
may then be treated in complete analogy to $H$ of Eq.\ (\ref{4:3:2:Hdef}),
i.e., the denominators are combined as in Eq.\ (\ref{4:3:2:feynmanparh}),
and the infrared singular and regular pieces are identified by
writing $\int_0^1 \mbox{d}z\cdots
= \int_0^\infty \mbox{d}z \cdots-\int_1^\infty \mbox{d}z \cdots$.

\subsubsection{Extended on-mass-shell scheme}
   In the following, we will concentrate on yet another solution which
has been motivated in Ref.\ \linebreak\cite{Gegelia:1999gf} and has been worked out in
detail in Ref.\ \cite{Fuchs:2003qc}.
   The central idea of the
extended on-mass-shell (EOMS) scheme
consists of performing additional subtractions beyond
the $\widetilde{\rm MS}$ scheme such that
renormalized diagrams satisfy the power counting.
   Terms violating the power counting are analytic in small
quantities and can thus be absorbed in a renormalization of
counter terms.

   In order to illustrate the approach, let us consider a simplified version
of the integral $H$, namely its value in the chiral limit,
\begin{displaymath}
H(p^2,m^2,0;n)= -i\int \frac{\mbox{d}^n k}{(2\pi)^n}
\frac{1}{[(k-p)^2-m^2+i0^+](k^2+i0^+)},
\end{displaymath}
where
\begin{displaymath}
\Delta=\frac{p^2-m^2}{m^2}={\cal O}(q)
\end{displaymath}
is a small quantity.
   Applying the power-counting rules of Sec.~\ref{subsection_pcrd},
we want the renormalized integral to be of order $D=n-1-2=n-3$.
Introducing
$C(z,\Delta)=z^2-\Delta\, z(1-z)-i0^+$,
we obtain
\begin{equation}
\label{Hcl}
H(p^2,m^2,0;n)=\kappa(m;n)\int_0^1 \mbox{d}z\, [C(z,\Delta)]^{\frac{n}{2}-2},
\end{equation}
where $\kappa(m;n)$ is given in Eq.~(\ref{4:3:3:kappan}).
   For the purpose of evaluating the integral of Eq.\ (\ref{Hcl})
we write
\begin{displaymath}
\int_0^1 \mbox{d}z\, [C(z,\Delta)]^{\frac{n}{2}-2}=
(-\Delta)^{\frac{n}{2}-2}\int_0^1 \mbox{d}z\, z^{\frac{n}{2}-2}
\left(1-\frac{1+\Delta}{\Delta}z\right)^{\frac{n}{2}-2}
\end{displaymath}
and apply Eqs.\ 15.3.1 and 15.3.4 of Ref.\ \cite{Abramowitz} to obtain
\begin{equation}
\label{Hclresult}
H(p^2,m^2,0;n)=\kappa(m;n)\frac{\Gamma\left(\frac{n}{2}-1\right)}{\Gamma
\left(\frac{n}{2}\right)}F\left(1,2-\frac{n}{2};\frac{n}{2};\frac{p^2}{m^2}
\right),
\end{equation}
where $F(a,b;c;z)$ is the hypergeometric function \cite{Abramowitz}.
   In order to discuss the power-counting properties of $H$ (in the chiral
limit) in terms of $\Delta$, we make use of Eq.\ 15.3.6 of Ref.\ \cite{Abramowitz} to re-write
Eq.\ (\ref{Hclresult}) as
\begin{eqnarray}
\label{4:3:4:j0dircalcrewr}
H(p^2,m^2,0;n)&=&\frac{m^{n-4}}{(4\pi)^\frac{n}{2}}\left[
\frac{\Gamma\left(2-\frac{n}{2}\right)}{n-3}
F\left(1,2-\frac{n}{2};4-n;-\Delta\right)\right.\nonumber\\
&&\left.+(-\Delta)^{n-3}\,\Gamma\left(\frac{n}{2}-1\right)
\Gamma(3-n)F\left(\frac{n}{2}-1,n-2;n-2;-\Delta\right)\right].
\end{eqnarray}
   Making use of
\begin{equation}
\label{fexpansion}
F(a,b;c;z)=1+\frac{ab}{c}z+\frac{a(a+1)b(b+1)}{c(c+1)}\frac{z^2}{2}+\cdots
\end{equation}
for $|z|<1$ and the fact that $\Delta$ counts as a small quantity of
${\cal O}(q)$, we immediately see that the first term of
Eq.\ (\ref{4:3:4:j0dircalcrewr}) contains a contribution which does not
satisfy the above power counting, i.e., which is not proportional
to ${\cal O}(q)$ as $n\to 4$.
   Using the expansion of Eq.\ (\ref{fexpansion}) together with
$\Gamma(1+x)=x\Gamma(x)$ we obtain, as $n\to 4$,
\begin{equation}
H=\frac{m^{n-4}}{(4\pi )^{\frac{n}{2}}}
\left[\frac{\Gamma \left(2-\frac{n}{2}\right)}{n-3}
+\left( 1-{p^2\over m^2}\right)\ln \left( 1-{p^2\over m^2}
\right)+
\left( 1-{p^2\over m^2}\right)^2\ln \left( 1-{p^2\over m^2}\right)
+\cdots\right],
\label{j0dircalcisolated}
\end{equation}
where $\cdots$ refers to terms which are at least of ${\cal O}(q^3)$
or $O(n-4)$.
   Note that we count a term of the type $-\Delta \ln(-\Delta)$ as ${\cal O}(q)$.
   If we subtract
\begin{equation}
\label{subtraction}
\frac{m^{n-4}}{(4\pi )^{\frac{n}{2}}}
\frac{\Gamma \left( 2-\frac{n}{2}\right)}{n-3}
\label{subtrterm}
\end{equation}
from Eq.\ (\ref{j0dircalcisolated}) we obtain as the renormalized integral
\begin{equation}
H_R(p^2,m^2,0;n)={m^{n-4}\over (4\pi )^{n/2}}\left[ \left( 1-{p^2\over m^2}
\right)
\ln \left( 1-{p^2\over m^2}\right)+
\left( 1-{p^2\over m^2}\right)^2\ln \left( 1-{p^2\over m^2}\right)
+\cdots\right].
\label{j0ren}
\end{equation}
   The subtracted term of Eq.\ (\ref{subtrterm}) is local in the external
momentum $p$, i.e., it is a {\em polynomial}\, in $p^2$
and can thus be obtained by a finite number of counter terms in the
most general effective Lagrangian.

   We have seen in Eq.\ (\ref{4:3:4:j0dircalcrewr}) that the one-loop integral
is of the type
\begin{displaymath}
H\sim F(n,\Delta)+\Delta^{n-3}G(n,\Delta),
\end{displaymath}
where $F$ and $G$ are hypergeometric functions and are analytic in $\Delta$ for
any $n$.
   The observation central for the setting up of a systematic method is the fact
that the part proportional to $F$ can be obtained by first expanding
the integrand in small quantities and {\em then} performing the integration
for each term \cite{Gegelia:1994zz} (see Sec.\ \ref{section_dimensional_counting_analysis}
for an illustration of the general method).
   We now apply a conventional renormalization prescription which allows us to identify
those terms which we subtract from a given integral without {\em explicitly}
calculating the integral beforehand.
   In essence we work with a modified integrand which is obtained from
the original integrand by subtracting a suitable number of
counter terms.
   The meaning of suitable in the present context will be explained in
a moment.
   To that end we consider the series
\begin{eqnarray}
\label{4:3:4:series}
\lefteqn{
\sum_{l=0}^\infty \frac{(p^2-m^2)^l}{l!}\left[
\left(\frac{1}{2p^2}p_\mu\frac{\partial}{\partial p_\mu}\right)^l
\frac{1}{[k^2-2k\cdot p+(p^2-m^2)+i0^+](k^2+i0^+)}\right]_{p^2=m^2}}\nonumber\\
&=&\left.\frac{1}{(k^2-2k\cdot p+i0^+)(k^2+i0^+)}\right|_{p^2=m^2}\nonumber\\
&&+(p^2-m^2)\left[\frac{1}{2m^2}\frac{1}{(k^2-2k\cdot p+i0^+)^2}
-\frac{1}{2m^2}\frac{1}{(k^2-2k\cdot p+i0^+)(k^2+i0^+)}\right.\nonumber\\
&&\left. -\frac{1}{(k^2-2k\cdot p+i0^+)^2(k^2+i0^+)}
\right]_{p^2=m^2}+\cdots,
\end{eqnarray}
   where $[\cdots ]_{p^2=m^2}$ means that we consider the {\em coefficients}
of $(p^2-m^2)^l$ only for four-momenta $p^\mu$ satisfying the on-mass-shell
condition.
   Although the coefficients still depend on the direction of $p^\mu$,
after integration of this series with respect to the loop momentum $k$ and
evaluation of the resulting coefficients for $p^2=m^2$, the integrated series
is a function of $p^2$ only.
   In fact, as was shown in Ref.\ \cite{Gegelia:1994zz}, the integrated
series exactly reproduces the first term of  Eq.\ (\ref{4:3:4:j0dircalcrewr}).
   At this point we stress that
\begin{displaymath}
\left.
-i\int \frac{\mbox{d}^n k}{(2\pi)^n}
\frac{1}{(k^2-2k\cdot p+i0^+)(k^2+i0^+)}\right|_{p^2=m^2}
\end{displaymath}
and
\begin{displaymath}
\left[-i\int \frac{\mbox{d}^n k}{(2\pi)^n}
\frac{1}{(k^2-2k\cdot p+p^2-m^2+ i0^+)(k^2+i0^+)}\right]_{p^2=m^2}
\end{displaymath}
are not the same for $n\leq 3$.

   The formal definition of the EOMS renormalization scheme is then as follows:
we subtract from the integrand of $H(p^2,m^2,0;n)$ those terms of the
series of Eq.\ (\ref{4:3:4:series}) which violate the power counting.
    These terms are always analytic in the small parameter and do not
contain infrared singularities.
   In the above example we only need to subtract the first term.
   All the higher-order terms contain infrared singularities.
   For example, the last term of the second coefficient would generate a
behavior $k^3/k^4$ of the integrand for $n=4$.
   The integral of the first term of Eq.\ (\ref{4:3:4:series}) is given
by Eq.\ (\ref{subtraction}), and we end up with Eq.\ (\ref{j0ren}) for
the renormalized integral:
\begin{displaymath}
H_R=H-H_{\rm subtr}={\cal O}(q^{n-3}).
\end{displaymath}
   Since the subtraction point is $p^2=m^2$, the renormalization condition
is denoted ``extended on-mass-shell'' (EOMS) scheme
in analogy with the on-mass-shell renormalization scheme in renormalizable
theories.
   In the general case including the pion mass, one would consider the series
\begin{eqnarray*}
&& \left.\frac{1}{\left( k^2-2 k\cdot p+i 0^+ \right)\left( k^2+i
0^+\right)}\right|_{p^2=m^2}+(p^2-m^2)\left[ \frac{1}{2 m^2} \frac{1}{\left( k^2-2 k\cdot p+i
0^+
\right)^2}+\cdots \right]_{p^2=m^2}\\
&&+M^2 \left. \frac{1}{\left( k^2-2 k\cdot p+i 0^+
\right)\left( k^2+i 0^+\right)^2 }\right|_{p^2=m^2}+\cdots
\end{eqnarray*}
instead of Eq.~(\ref{4:3:4:series}).
   However, it would still be only the contribution resulting from the first
term that were to be subtracted.

   Within the EOMS framework it is straightforward to obtain
a consistent power counting in manifestly Lorentz-invariant
baryon ChPT including, e.g., vector mesons \cite{Fuchs:2003sh} or
the $\Delta(1232)$ resonance \cite{Hacker:2005fh} as explicit
degrees of freedom.
   Moreover, the infrared regularization of Becher and Leutwyler
can be reformulated in a form analogous to the EOMS renormalization
scheme and can thus be applied straightforwardly to multi-loop diagrams
with an arbitrary number of particles with arbitrary masses
\cite{Schindler:2003xv} (see also Refs.\ \cite{Lehmann:2001xm}, \cite{Bruns:2004tj},
\cite{Bruns:2008ub}).
    The application of both infrared and extended on-mass-shell renormalization
schemes to multi-loop diagrams was explicitly demonstrated by means of a
two-loop self-energy diagram \cite{Schindler:2003je}.

\subsubsection{Dimensional counting analysis}
\label{section_dimensional_counting_analysis}
   In this section we provide an illustration of the dimensional counting
analysis \cite{Gegelia:1994zz} in terms of a specific example.
   To that end let us consider the one-loop integral of Eq.~(\ref{4:3:2:Hdef}),
\begin{equation}
H(p^2,m^2,M^2;n)=
-i\int \frac{\mbox{d}^n k}{(2\pi)^n} \frac{1}{k^2-2p\cdot k
+p^2-m^2+i0^+}\frac{1}{k^2-M^2+i0^+}.
\end{equation}
   One would like to know how the integral behaves for small values of
$M$ and/or $p^2-m^2$ as a function of $n$.
   If we consider, for fixed $p^2\neq m^2$, the limit $M\to 0$, the
integral $H$ can be represented as
\begin{equation}
H(p^2,m^2,M^2;n)=\sum_i M^{\beta_i} F_i(p^2,m^2,M^2;n),
\label{ippsiexp}
\end{equation}
   where the functions $F_i$ are analytic in $M^2$ and are obtained
as follows.
   First, one re-writes the integration variable as
$k=M^{\alpha_i}\tilde k$, where $\alpha_i$ is an arbitrary
non-negative real number.
   Next, one isolates the overall factor of $M^{\beta_i}$ so
that the remaining integrand can be expanded in positive powers of
$M^2$ and interchanges the integration and summation.
   The resulting series represents the expansion of $F_i(p^2,m^2,M^2;n)$ in
powers of $M^2$.
   The sum of all possible re-scalings with subsequent expansions with
non-trivial coefficients then reproduces the expansion of the result
of the original integral.

   To be specific, let us apply this program to $H$:
\begin{equation}
H(p^2,m^2,M^2;n) = -i\int\frac{M^{n \alpha_i} \mbox{d}^n\tilde k}{(2\pi)^n}
\frac{1}{\tilde k^2 M^{2 \alpha_i}-2 p\cdot\tilde k M^{\alpha_i}+ p^2-m^2+i0^+}
\frac{1}{\tilde k^2 M^{2 \alpha_i} -M^2+i0^+}. \label{ippsichv}
\end{equation}
   From Eq.~(\ref{ippsichv}) we see that the first fraction does
not contribute to the overall factor $M^{\beta_i}$ for any $\alpha_i$.
   It will be expanded in (positive) powers of $(\tilde k^2 M^{2\alpha_i}
-2 p\cdot \tilde k M^{\alpha_i})$ except for $\alpha_i=0$.
   For $0<\alpha_i<1$, we re-write the second fraction as
\begin{equation}
\frac{1}{M^{2 \alpha_i}} \ \frac{1}{(\tilde k^2-M^{2-2 \alpha_i}+i 0^+)}
=\frac{1}{M^{2 \alpha_i}}
\frac{1}{\tilde k^2+i0^+}\left(1+\frac{M^{2-2\alpha_i}}{\tilde k^2+i0^+}+\cdots
\right).
\label{1pr}
\end{equation}
   On the other hand, if $1<\alpha_i$ we re-write the second fraction as
\begin{equation}
\frac{1}{M^{2}} \ \frac{1}{(\tilde k^2 M^{2 \alpha_i-2}-1+i 0^+) }
=-\frac{1}{M^{2}}\left(1+\tilde k^2 M^{2 \alpha_i-2}+\cdots\right).
\label{1pr2}
\end{equation}
    In both cases one obtains integrals of the type
$\int \mbox{d}^n\tilde k \ \tilde k^{\mu_1}\cdots \tilde k^{\mu_m}$ as the coefficients of the
expansion.
   However, such integrals vanish in dimensional regularization.
   Therefore, the only non-trivial terms in the sum of Eq.~(\ref{ippsiexp})
correspond to either $\alpha_i=0$ or $\alpha_i=1$.
   Thus we obtain
\begin{equation}
H(p^2,m^2,M^2;n)=H^{(0)}(p^2,m^2,M^2;n)+H^{(1)}(p^2,m^2,M^2;n),
\label{ippsidecomp}
\end{equation}
where
\begin{equation}
H^{(0)}(p^2,m^2,M^2;n)=-i\sum_{j=0}^{\infty}
\left( M^2\right)^j \int \frac{\mbox{d}^n k}{(2\pi)^n}
\frac{1}{k^2 -2 p\cdot k+ p^2-m^2+i0^+}
\frac{1}{(k^2 +i0^+)^{j+1}},
\label{ippsi2}
\end{equation}
and
\begin{equation}
H^{(1)}(p^2,m^2,M^2;n)=-i \frac{M^{n-2}}{p^2-m^2+i0^+}
\sum_{j=0}^{\infty}
\frac{(-1)^j M^j}{(p^2-m^2+i0^+)^j}
\int\frac{\mbox{d}^n\tilde k}{(2\pi)^n} \frac{\left(\tilde k^2 M-2 p\cdot\tilde
k\right)^j}{\tilde k^2-1+i0^+}. \label{ippsi1}
\end{equation}
   A comparison with the direct calculation of $H$ shows that
the dimensional counting method indeed leads to the correct
expressions \cite{Gegelia:1994zz}.
   While the loop integrals of Eq.~(\ref{ippsi1}) have a simple
analytic structure in $p^2-m^2$, the same technique can
be repeated for the loop integrals of Eq.~(\ref{ippsi2}) when
$p^2-m^2\to 0$, now using the change of variable $k=(p^2-m^2)^{\gamma_i}
\tilde k$ with arbitrary non-negative real numbers $\gamma_i$.

\section{Applications}
\label{section_applications}
   In the following we will illustrate a few selected applications
of the manifestly Lorentz-invariant framework to the one-nucleon sector.

\subsection{Nucleon mass and sigma term at ${\cal O}(q^4)$}
\begin{figure}[t]
\begin{center}
\epsfig{file=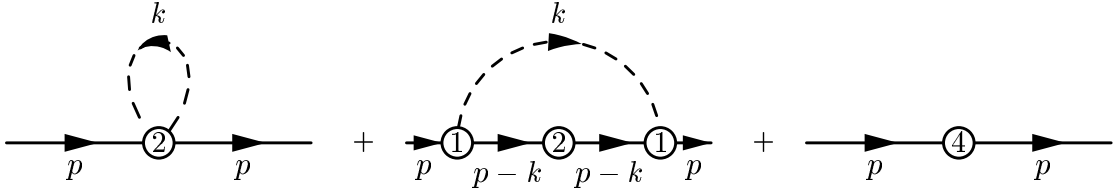,width=0.8\textwidth}
\caption{\label{5:1:1:SEDiagrams}
Contributions to the nucleon self energy at ${\cal O}(q^4)$.
The number $n$ in the interaction blobs refers to ${\cal L}_{\pi N}^{(n)}$.
The Lagrangian ${\cal L}_{\pi N}^{(2)}$ does not produce a contribution to the $\pi NN$ vertex.}
\end{center}
\end{figure}

   A full one-loop calculation of the nucleon mass also includes
${\cal O}(q^4)$ terms (see Fig.~\ref{5:1:1:SEDiagrams}).
   The quark-mass expansion up to and including ${\cal O}(q^4)$ is given by
\begin{equation}
\label{mnoq4}
m_N=m+k_1 M^2+k_2 M^3+k_3 M^4\ln\left(\frac{M}{m}\right)+k_4 M^4
+{\cal O}(M^5),
\end{equation}
where the coefficients $k_i$ in the EOMS scheme read \cite{Fuchs:2003qc}
\begin{eqnarray}
\label{parki}
k_1&=&-4 c_1,\quad
k_2=-\frac{3 {\texttt{g}_A}^2}{32\pi F^2},\quad
k_3=-\frac{3}{32\pi^2 F^2 m}\left(\texttt{g}_A^2-8c_1m+c_2m+4 c_3m\right),
\nonumber\\
k_4&=&\frac{3 {\texttt{g}_A}^2}{32 \pi^2 F^2 m}(1+4 c_1 m)
+\frac{3}{128\pi^2F^2}c_2-\hat{e}_1.
\end{eqnarray}
   Here, $\hat{e}_1= 16 e_{38}+2e_{115}+2e_{116}$ is a linear combination
of ${\cal O}(q^4)$ coefficients \cite{Fettes:2000gb}.
   A comparison with the results using the infrared regularization \cite{Becher:1999he}
shows that the lowest-order correction ($k_1$ term) and those terms which are non-analytic
in the quark mass $\hat{m}$ ($k_2$ and $k_3$ terms) coincide.
   On the other hand, the analytic $k_4$ term ($\sim M^4$) is different.
   This is not surprising; although both renormalization schemes satisfy
the power counting specified in Sec.~\ref{subsection_pcrd},
the use of different renormalization conditions is compensated by different
values of the renormalized parameters.

   For an estimate of the various contributions of Eq.\ (\ref{mnoq4}) to the nucleon mass,
we make use of the parameter set
\begin{equation}
\label{parametersci}
c_1=-0.9\,m_N^{-1},\quad
c_2=2.5\, m_N^{-1},\quad
c_3=-4.2\, m_N^{-1},\quad
c_4=2.3\, m_N^{-1},
\end{equation}
which was obtained in Ref.\ \cite{Becher:2001hv} from a (tree-level) fit to
the $\pi N$ scattering threshold parameters.
   Using the numerical values
\begin{equation}
\label{numericalvalues}
g_A=1.267,\quad F_\pi=92.4\,\mbox{MeV},\quad
m_N=m_p=938.3\,\mbox{MeV},\quad M_\pi=M_{\pi^+}=139.6\,\mbox{MeV},
\end{equation}
one obtains for the mass of nucleon in the chiral limit (at
fixed $m_s\neq 0$):
\begin{equation}
m=m_N-\Delta m
=[938.3-74.8+15.3+4.7+1.6-2.3\pm 4]\,\mbox{MeV}
=(883\pm 4)\, \mbox{MeV}
\end{equation}
with $\Delta m=(55.5\pm 4)\,\mbox{MeV}$.
   Here, we have made use of an estimate for $\hat{e}_1M^4=(2.3\pm 4)$ MeV obtained from
the $\sigma$ term.
   (Note that errors due to higher-order corrections are not taken into account.)
   In terms of the SU(2)$_L\times$SU(2)$_R$-chiral-symmetry-breaking mass term of the
QCD Hamiltonian,
\begin{equation}
{\cal H_{\rm sb}}=\hat{m}(\bar u u+\bar d d),
\end{equation}
the pion-nucleon $\sigma$ term is defined as the proton matrix element
\begin{equation}
\sigma=\frac{1}{2m_p}\langle p(p,s)|{\cal H_{\rm sb}}(0)|p(p,s)\rangle
\end{equation}
at zero momentum transfer.
   The $\sigma$ term provides a sensitive measure of explicit chiral symmetry
breaking in QCD, because it is a correction to a null result in the chiral limit rather
than a small correction to a non-trivial result \cite{Pagels:1974se}.
   The quark-mass expansion of the $\sigma$ term reads
\begin{equation}
\sigma=\sigma_1 M^2+\sigma_2 M^3 +\sigma_3 M^4 \ln\left(\frac{M}{m}\right)
+\sigma_4 M^4+{\cal O}(M^5),
\end{equation}
with
\begin{eqnarray}
\label{parsigmai}
\sigma_1&=&-4 c_1,\quad
\sigma_2=-\frac{9{\texttt{g}_{A}}^2}{64\pi F^2},\quad
\sigma_3=-\frac{3}{16\pi^2 F^2 m}\left(\texttt{g}_A^2-8c_1m+c_2m+4 c_3m\right),
\nonumber\\
\sigma_4&=&\frac{3}{8\pi^2 F^2m}\left[\frac{3 {\texttt{g}_{A}}^2}{8}
+c_1m(1+2 {\texttt{g}_A}^2)-\frac{c_3m}{2}\right]-2\hat{e}_1.
\end{eqnarray}
   We obtain [with $\hat{e}_1=0$ in Eq.\ (\ref {parsigmai})]
\begin{equation}
\label{sigmanum}
\sigma=(74.8-22.9-9.4-2.0)\,\mbox{MeV}=40.5\, \mbox{MeV}.
\end{equation}
   The result of Eq.\ (\ref{sigmanum}) has to be compared with, e.g., the
dispersive analysis $\sigma=(45\pm 8)$ MeV of Ref.\ \cite{Gasser:1990ce}
which would imply, neglecting higher-order terms, $-2\hat{e}_1 M^4 \approx (4.5\pm 8)$
MeV.
   As has been discussed, e.g., in Ref.\ \cite{Becher:1999he},
a fully consistent description would also require to determine
the low-energy coupling constant $c_1$ from a complete ${\cal O}(q^4)$
calculation of, say, $\pi N$ scattering.
   The results of Eqs.\ (\ref{parki}) and (\ref{parsigmai}) satisfy the
constraints as implied by the application of the Hellmann-Feynman theorem
to the nucleon mass \cite{Gasser:1987rb},
\begin{equation}
\sigma =M^2 \ {\partial m_N\over \partial M^2}.
\label{fhrelation}
\end{equation}

\subsection{Chiral expansion of the nucleon mass to ${\cal O}(q^6)$}
\begin{figure}[t]
\begin{center}
\epsfig{file=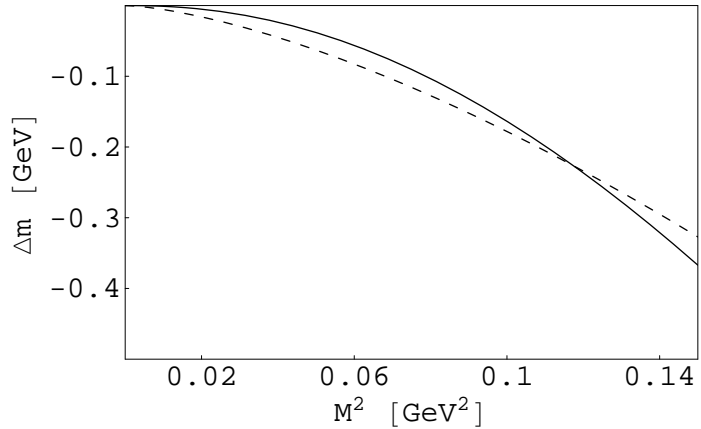,width=0.5\textwidth}
\end{center}
\caption{Pion mass dependence of the term $k_5 M^5 \ln(M/m_N)$
(solid line) for $M<400\,\mbox{MeV}$. For comparison also the term
$k_2 M^3$ (dashed line) is shown.\label{Mass:k2k5Low}}
\end{figure}

    So far, essentially all of the manifestly Lorentz-invariant calculations have been
restricted to the one-loop level.
   One of the exceptions is the chiral expansion of the nucleon mass which,
in the framework of the reformulated infrared regularization, has been calculated
up to and including ${\cal O}(q^6)$
\cite{Schindler:2006ha,Schindler:2007dr}:
\begin{equation}
\label{Mass:Exp}
    m_N = m +k_1 M^2 +k_2 \,M^3 +k_3 M^4 \ln\frac{M}{\mu}
+ k_4 M^4  + k_5 M^5\ln\frac{M}{\mu} + k_6 M^5 + k_7
M^6 \ln^2\frac{M}{\mu}+ k_8 M^6 \ln\frac{M}{\mu} + k_9 M^6.
\end{equation}
   We refrain from displaying the lengthy expressions for the coefficients
$k_i$ but rather want to discuss a few general implications \cite{Schindler:2007dr}.
   Chiral expansions like Eq.~(\ref{Mass:Exp}) currently play an important
role in the extrapolation of lattice QCD results to physical quark
masses.
   Unfortunately, the numerical contributions from higher-order terms cannot be
calculated so far since, starting with $k_4$, most expressions in
Eq.~(\ref{Mass:Exp}) contain unknown low-energy coupling
constants (LECs) from the Lagrangians of ${\cal O}(q^4)$ and
higher.
   The coefficient $k_5$ is free of higher-order LECs
and is given in terms of the axial-vector coupling constant
$\texttt{g}_A$ and the pion-decay constant $F$:
\begin{displaymath}
 k_5 = \frac{3 \texttt{g}_A^2}{1024\pi^3 F^4}\,\left(16\texttt{g}_A^2-3\right).
 \end{displaymath}
   While the values for both $\texttt{g}_A$ and $F$ should be taken in the chiral
limit, we evaluate $k_5$ using the physical values $g_A=1.2695(29)$
and $F_\pi=92.42(26)$ MeV.
   Setting $\mu=m_N$, $m_N=(m_p+m_n)/2=938.92$ MeV, and $M=M_{\pi^+}=139.57$ MeV
we obtain $k_5 M^5 \ln(M/m_N) = -4.8$ MeV.
   This amounts to approximately $31$\% of the leading non-analytic contribution
at one-loop order, $k_2 M^3$.
   Figure \ref{Mass:k2k5Low} shows the pion mass dependence of the term
$k_5 M^5 \ln(M/m_N)$ (solid line) in comparison with the term $k_2
M^3$ (dashed line) for pion masses below $400\,\mbox{MeV}$
which is considered a region where chiral extrapolations are valid
(see, e.g., Refs.~\cite{Meissner:2005ba}, \cite{Djukanovic:2006xc}).
   We see that already at $M \approx 360\,\mbox{MeV}$ the
term $k_5 M^5 \ln(M/m_N)$ becomes as large as the leading
non-analytic term at one-loop order, $k_2 M^3$, indicating the
importance of the fifth-order terms at unphysical pion masses.
   Our results for the renormalization-scheme-independent terms agree
with the heavy-baryon ChPT results of Ref.~\cite{McGovern:1998tm}.

\subsection{Form factors of the nucleon}

\subsubsection{Scalar form factor}
\begin{figure}[t]
\begin{minipage}[t]{0.45\textwidth}
\begin{center}
\epsfig{file=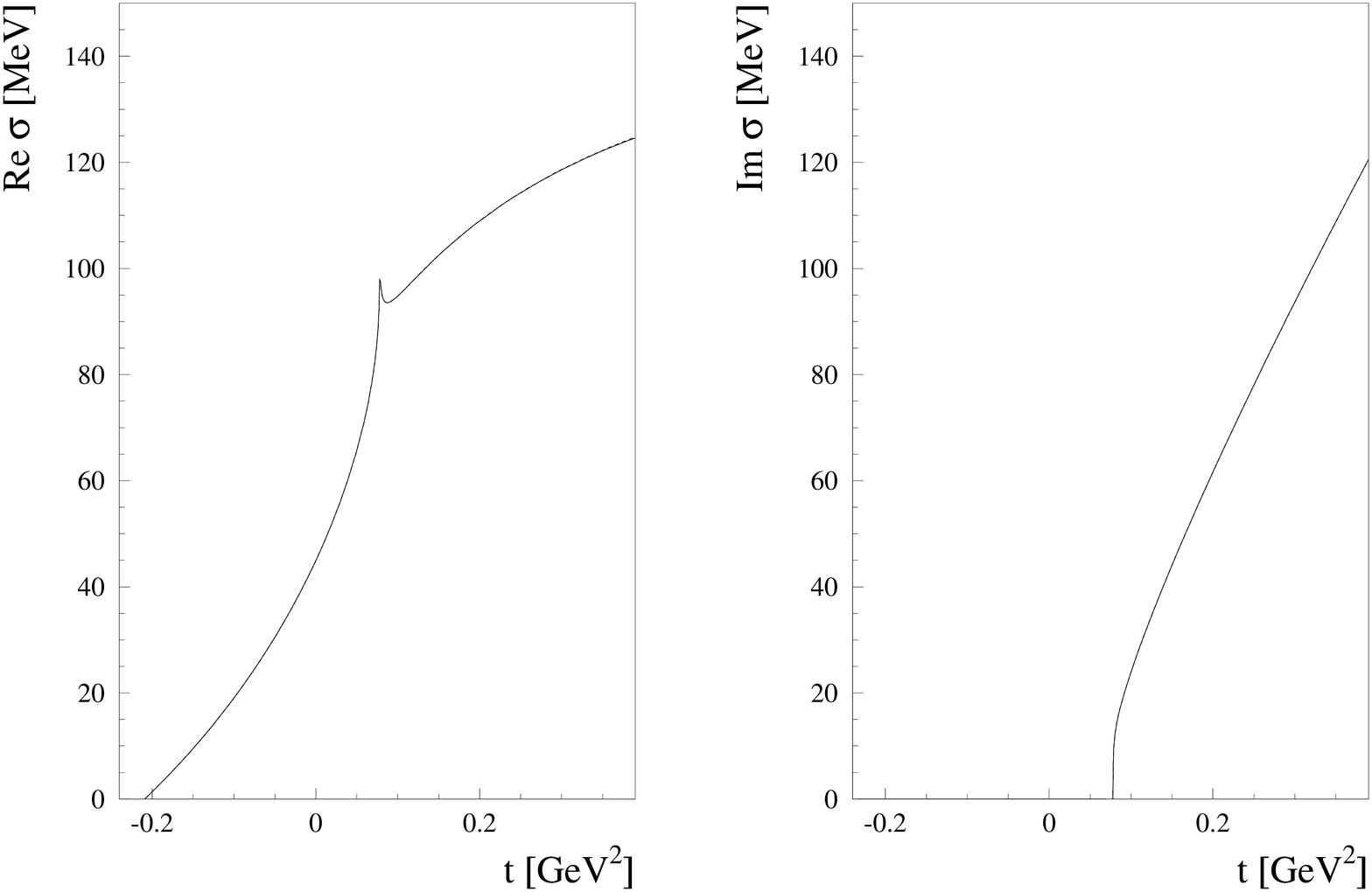,width=\textwidth}
\end{center}
\caption{Scalar form factor $\sigma(t)$ as a function of $t$ at
${\cal O}(q^4)$.}
\label{fig:scffq4new}
\end{minipage}
\hfill
\begin{minipage}[t]{0.45\textwidth}
\begin{center}
\epsfig{file=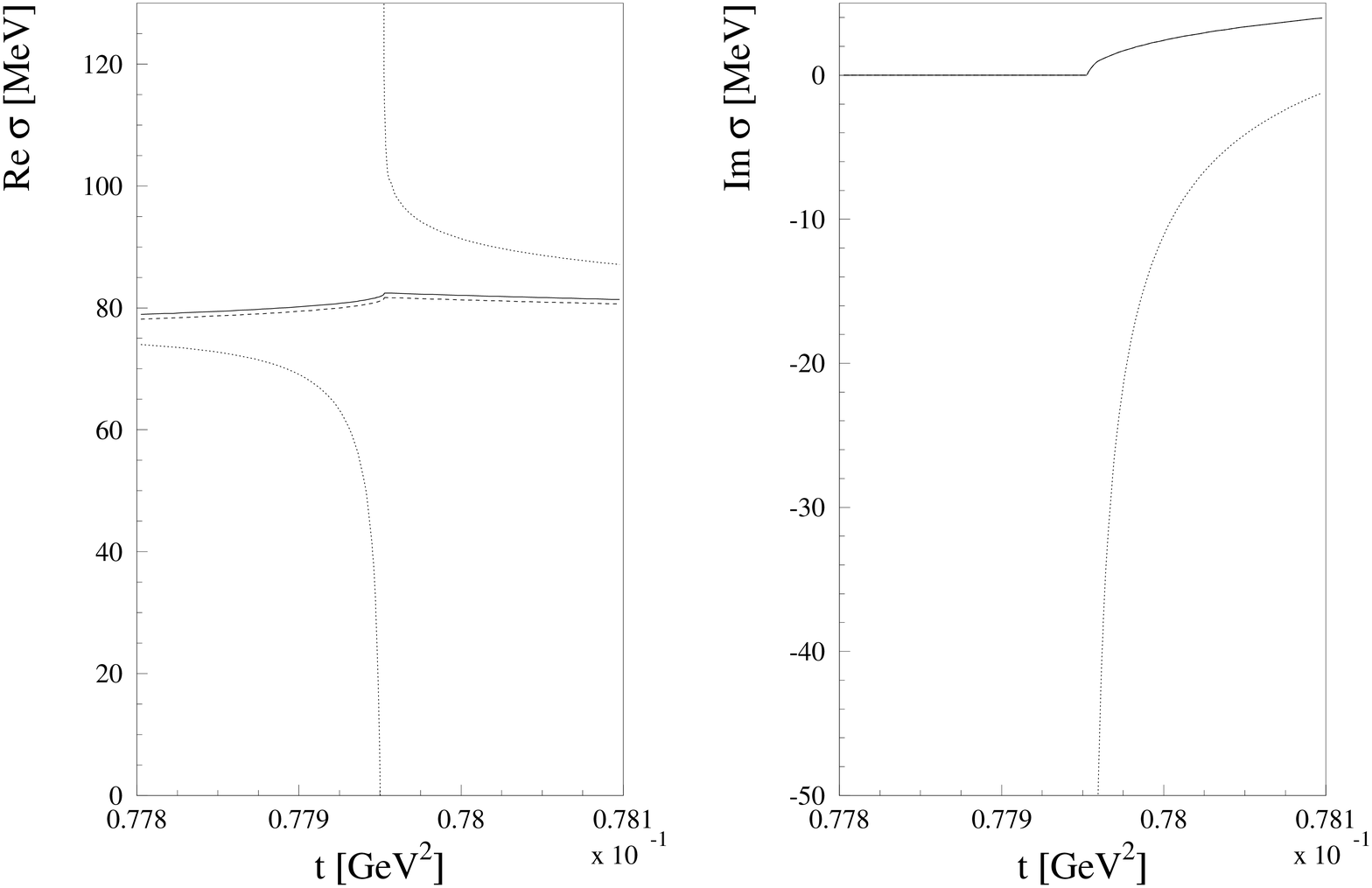,width=\textwidth}
\end{center}
\caption{Real and imaginary parts of the scalar form factor as a function of
$t$ at ${\cal O}(q^3)$ in the vicinity of $t=4 M_\pi^2$.
Solid lines: EOMS scheme;
dashed lines: infrared regularization (IR) of Ref.\ \cite{Becher:1999he};
dotted lines: HBChPT calculation of Ref.\ \cite{Bernard:1992qa}.
On this scale the (unphysical) divergence of both real and imaginary parts
of the heavy-baryon result becomes visible.
}
\label{fig:scffq3thrnew}
\end{minipage}
\end{figure}

   The pion-nucleon $\sigma$ term corresponds to the kinematical point
$t=0$ of the scalar form factor which is defined as
\begin{displaymath}
\langle p(p',s')|{\cal H_{\rm sb}}(0)|p(p,s)\rangle
=\bar{u}(p',s')u(p,s)\sigma(t),\quad t=(p'-p)^2.
\end{displaymath}
    The numerical results for the real and imaginary parts of the
scalar form factor at ${\cal O}(q^4)$ are shown in Fig.\
\ref{fig:scffq4new} for the extended on-mass-shell scheme
(solid lines) and the infrared regularization scheme (dashed lines).
   While the imaginary parts are identical in both schemes, the
differences in the real parts are practically indistinguishable.
   Note that for both calculations $\sigma(0)$ and $\Delta_\sigma\equiv\sigma(2 M_\pi^2) -\sigma(0)$ have
been adjusted to the dispersion results of Ref.\ \cite{Gasser:1990ce}, $\Delta_\sigma=(15.2\pm 0.4)$ MeV.

   Figure \ref{fig:scffq3thrnew} contains an enlargement near
$t\approx 4 M_\pi^2$ for the results at ${\cal O}(q^3)$ which
clearly displays how the heavy-baryon calculation
fails to produce the correct analytic behavior
not only at the tree level but also in higher-order loop diagrams.
   Both real and imaginary parts diverge as $t\to 4 M_\pi^2$.

\subsubsection{Electromagnetic form factors}
\label{subsectionemff}
   Imposing the relevant symmetries such as translational invariance,
Lorentz covariance, the discrete symmetries, and current
conservation, the nucleon matrix element of the electromagnetic
current operator ${\cal J}^\mu(x)$,
\begin{displaymath}
{\cal J}^{\mu}(x)
=\frac{2}{3}\,\bar{u}(x)\gamma^{\mu}u(x)
-\frac{1}{3}\,\bar{d}(x)\gamma^{\mu}d(x),
\end{displaymath}
can be parameterized in terms of two form factors,
\begin{equation}
\label{H1:emff:empar} \langle N(p',s')|{\cal J}^{\mu}(0)|N(p,s)\rangle=
\bar{u}(p',s')\left[F_1^N(Q^2)\gamma^\mu
+i\frac{\sigma^{\mu\nu}q_\nu}{2m_p}F_2^N(Q^2) \right]u(p,s),\quad
N=p,n,
\end{equation}
   where $q=p'-p$, $Q^2=-q^2$, and $m_p$ is the proton mass.
   At $Q^2=0$, the so-called Dirac and Pauli form factors $F_1$ and
$F_2$ reduce to the charge and anomalous magnetic moment in units of
the elementary charge $e$ and the nuclear magneton $e/(2m_p)$,
respectively,
\begin{displaymath}
F_1^{p}(0)=1,\quad F_1^{n}(0)=0,\quad F_2^{p}(0)=1.793,\quad
F_2^{n}(0)=-1.913.
\end{displaymath}
   The Sachs form factors $G_E$ and $G_M$ are linear combinations of
$F_1$ and $F_2$,
\begin{displaymath}
G_E^N(Q^2)=F_1^N(Q^2)-\frac{Q^2}{4m_p^2}F_2^N(Q^2),\quad
G_M^N(Q^2)=F_1^N(Q^2)+F_2^N(Q^2), \quad N=p,n,
\end{displaymath}
and, in the non-relativistic limit, their Fourier transforms are
commonly interpreted as the distribution of charge and magnetization
inside the nucleon.
   For a covariant interpretation in terms of the transverse charge
density see Refs.\ \cite{Miller:2007uy}, \cite{Carlson:2007xd}.
   The description of the electromagnetic form factors of the nucleon
presents a stringent test for any theory or model of the strong
interactions (see, e.g., Ref.~\cite{Perdrisat:2006hj} for a recent review).

\begin{figure}[t]
\begin{center}
\epsfig{file=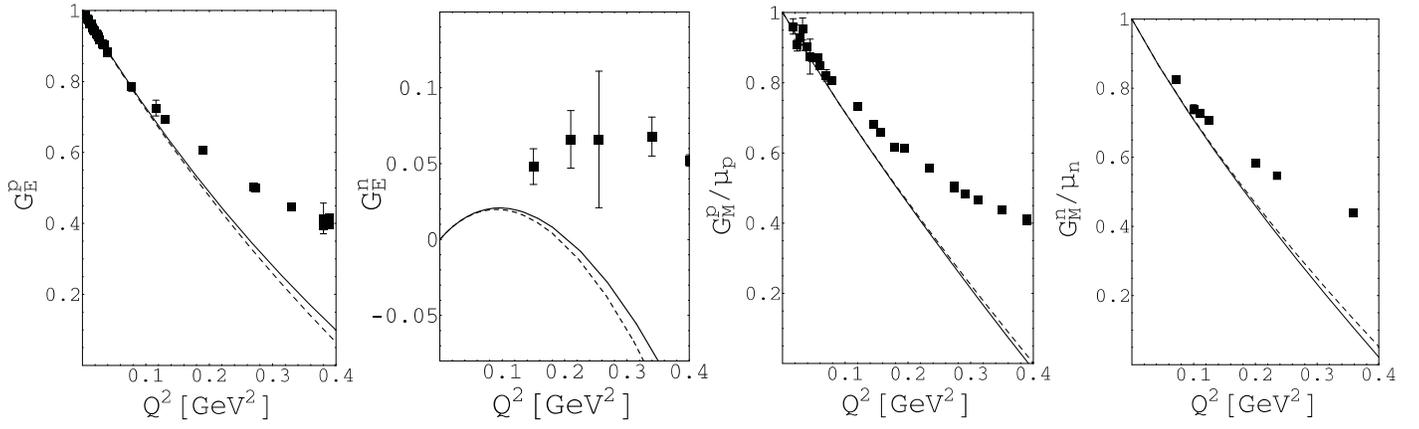,width=\textwidth}
\end{center}
\caption{\label{G_ohne} The Sachs form factors of the nucleon in manifestly
Lorentz-invariant chiral perturbation theory at ${\cal O}(q^4)$ without vector
mesons. Full lines: results in the extended on-mass-shell scheme; dashed lines:
results in infrared regularization. The experimental data are taken from
Ref.\ \cite{Friedrich:2003iz}.}
\end{figure}

    In the framework of chiral perturbation theory, the electromagnetic form
factors were calculated in the early relativistic approach \cite{Gasser:1987rb},
the heavy-baryon approach \cite{Bernard:1992qa}, \cite{Fearing:1997dp},
the small-scale expansion \cite{Bernard:1998gv}, the infrared regularization
\linebreak\cite{Kubis:2000zd}, and the EOMS scheme \cite{Fuchs:2003ir}.
   All these calculations have in common that they fail to describe the proton and nucleon
form factors for momentum transfers beyond $Q^2\sim 0.1\, \mbox{GeV}^2$.
   Moreover, up to and including ${\cal O}(q^4)$, the most general effective Lagrangian
provides sufficiently many independent parameters such that the empirical values of the
anomalous magnetic moments and the charge and magnetic radii are fitted rather than predicted.
   Figure \ref{G_ohne} shows the Sachs form factors in the momentum transfer region
$0\,{\rm GeV^2}\le Q^2\le 0.4\,{\rm GeV^2}$ in the EOMS scheme and the reformulated
infrared regularization \cite{Schindler:2005ke}.

   In Ref.\ \cite{Kubis:2000zd} it was shown that the
inclusion of vector mesons can result in the re-summation of
important higher-order contributions.
   In standard ChPT, such vector meson contributions manifest themselves
in terms of the values of the low-energy coupling constants.
   Symbolically, the contributions to certain LECs originate from
the expansion of the vector-meson propagator,
\begin{displaymath}
\frac{1}{q^2-M_{V}^2}=-\frac{1}{M_{V}^2}
\left[1+\frac{q^2}{M_{V}^2}+\left(\frac{q^2}{M_{V}^2}\right)^2+\mathcal{O}(q^6)\right]
\end{displaymath}
combined with the relevant vector-meson vertices.
   However, diagrams with internal vector-meson lines {\em inside} loops
were not considered, because a generalization
of ChPT which fully includes the effects of vector mesons as intermediate
states in loops was not yet available \cite{Kubis:2000zd}.
   On the other hand, the EOMS renormalization scheme of Ref.~\cite{Fuchs:2003qc} and
the reformulated version of infrared regularization of
Ref.~\cite{Schindler:2003xv} both allow to include virtual vector mesons {\em
systematically} in the region of the applicability of baryon chiral perturbation
theory \cite{Fuchs:2003sh} (see also Ref.\ \cite{Bruns:2008ub}).
   The standard power counting determines which diagrams
(including diagrams with vector mesons appearing in loops) should be taken into
account to a given order in the chiral expansion.

\begin{figure}[t]
\begin{center}
\epsfig{file=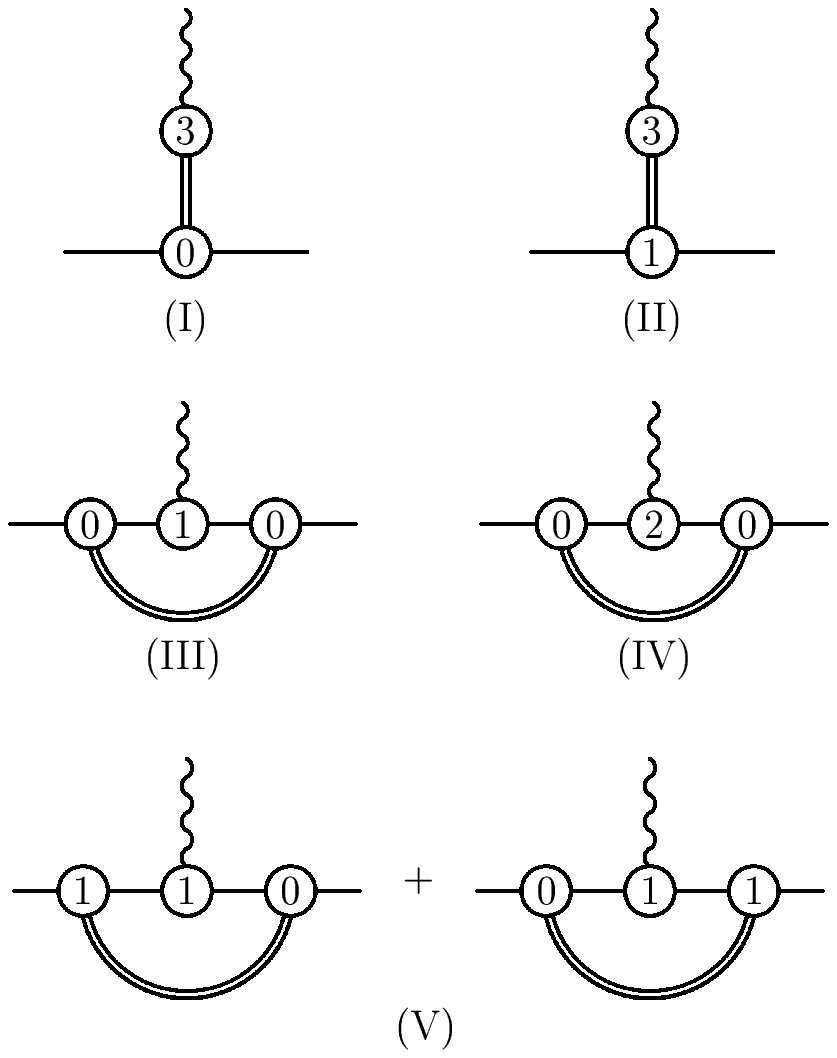,width=0.4\textwidth}
\caption{\label{Dia_mit} Feynman diagrams
including vector mesons that contribute to the electromagnetic form factors of
the nucleon up to and including ${\cal O}(q^4)$. External leg corrections are not
shown. Solid, wiggly, and double lines refer to nucleons, photons, and vector
mesons, respectively. The numbers in the interaction blobs denote the order of
the Lagrangian from which they are obtained. The direct coupling of the photon to
the nucleon is obtained from ${\cal L}_{\pi N}^{(1)}$ and ${\cal L}_{\pi
N}^{(2)}$.}
\end{center}
\end{figure}

   In Ref.\ \cite{Schindler:2005ke} the electromagnetic form factors
were calculated with the $\rho$, $\omega$, and $\phi$
mesons as explicit degrees of freedom.
   In the vector-field representation of Ref.\ \cite{Ecker:1989yg} the $\rho$ meson is
represented by $\rho^{\mu}=\rho^{\mu}_i\tau_i$ and the $\omega$ and $\phi$ mesons by
$\omega^\mu$ and $\phi^\mu$, respectively.
   The coupling of the vector mesons to pions and external fields is at least of
${\cal O}(q^3)$,
\begin{equation}\label{PionLagrange}
   \mathcal{L}_{{\pi}V}^{(3)}=-f_\rho\mbox{Tr}(\rho^{\mu\nu}f^+_{\mu\nu})
-f_\omega\omega^{\mu\nu}f^{(s)}_{\mu\nu}-f_\phi\phi^{\mu\nu}f^{(s)}_{\mu\nu}+\cdots,
\end{equation}
where the field strength tensors are given by
\begin{displaymath}
f^{(s)}_{\mu\nu}=\partial_\mu v^{(s)}_\nu-\partial_\nu
v^{(s)}_\mu,\quad
f^+_{\mu\nu}=u^+f^R_{\mu\nu}u+u f^L_{\mu\nu} u^+,
\end{displaymath}
with $f^R_{\mu\nu}$ and $f^L_{\mu\nu}$ defined in Eqs.~(\ref{3:2:3:fr}) and (\ref{3:2:3:fl}),
respectively.
   For the case of a coupling to an external electromagnetic potential ${\cal A}_\mu$,
the external fields are given by Eq.\ (\ref{2:1:6:rlasu2}).
   Furthermore, in terms of the connection $\Gamma^\mu$ of Eq.~(\ref{4:1:2:gamma}),
we define
\begin{displaymath}
    \rho^{\mu\nu}=\nabla^{\mu}\rho^{\nu}-\nabla^{\nu}\rho^{\mu},\quad
    \nabla^{\mu}\rho^{\nu}=\partial^{\mu}\rho^{\nu}+\left[\Gamma^{\mu},\rho^{\nu}\right],
\end{displaymath}
and, finally,
\begin{displaymath}
\omega^{\mu\nu}=\partial^\mu\omega^\nu-\partial^\nu\omega^\mu,\quad
\phi^{\mu\nu}=\partial^\mu\phi^\nu-\partial^\nu\phi^\mu.
\end{displaymath}
   The lowest-order Lagrangian for the coupling to the nucleon is
given by
\begin{equation}\label{LagNV0}
   \mathcal{L}_{VN}^{(0)}=\frac{1}{2}\sum_{V=\rho,\omega,\phi}g_V\,\bar{\Psi}\gamma^{\mu}V_{\mu}\Psi,
\end{equation}
and the ${\cal O}(q)$ Lagrangian reads
\begin{equation}\label{LagNV1}
   \mathcal{L}_{VN}^{(1)}=\frac{1}{4}\sum_{V=\rho,\omega,\phi}G_V
   \bar{\Psi}\sigma^{\mu\nu}V_{\mu\nu}\Psi.
\end{equation}
   The additional power-counting rules state that vertices from $\mathcal{L}_{{\pi}V}^{(3)}$ count as
$\mathcal{O}(q^3)$ and vertices from $\mathcal{L}_{VN}^{(i)}$ as
$\mathcal{O}(q^i)$, respectively, while the vector-meson propagators count as
$\mathcal{O}(q^0)$.
   The additional diagrams involving vector mesons that contribute in the calculation of
the form factors up to and including  ${\cal O}(q^4)$ using the Lagrangians
of Eqs.\ (\ref{PionLagrange}), (\ref{LagNV0}), and (\ref{LagNV1}) are shown in
Fig.~\ref{Dia_mit}.
     The parameters of the vector-meson Lagrangian of Eq.\ (\ref{PionLagrange})
for the coupling to external fields have been taken from Ref.~\cite{Ecker:1989yg},
and those of Eqs.\ (\ref{LagNV0}) and (\ref{LagNV1}) for the coupling of vector
mesons to the nucleon from the dispersion relations of
Refs.~\cite{Mergell:1995bf}, \cite{Hammer:2003ai}.

\begin{figure}[t]
\begin{center}
\epsfig{file=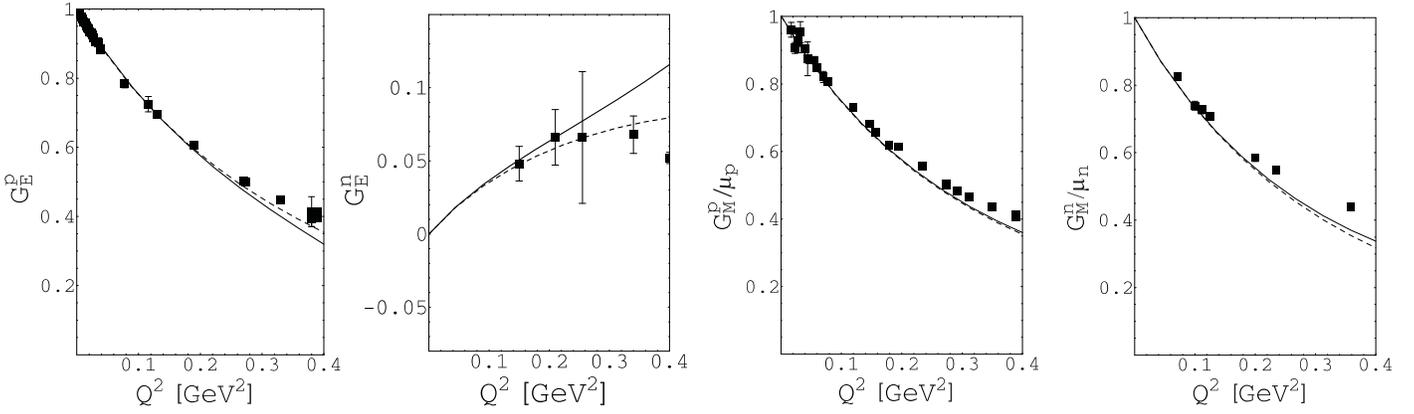,width=\textwidth}
\caption{\label{H1:emff:G:neu} The Sachs form factors of the nucleon
in manifestly Lorentz-invariant chiral perturbation theory at ${\cal
O}(q^4)$ including vector mesons as explicit degrees of freedom.
Full lines: results in the extended on-mass-shell scheme; dashed
lines: results in infrared regularization. The experimental data are
taken from Ref.\ \cite{Friedrich:2003iz}.}
\end{center}
\end{figure}

   As expected on phenomenological grounds, the quantitative description of the
data has improved considerably for $Q^2\geq 0.1$ GeV$^2$   (see
Fig.~\ref{H1:emff:G:neu}).
   The small difference between the two renormalization schemes is due to the
way how the regular higher-order terms of loop integrals are treated.
   Numerically, the results are similar to those of Ref.\ \cite{Kubis:2000zd}.
   Due to the renormalization condition, the contribution of the vector-meson
loop diagrams either vanishes (IR) or turns out to be small (EOMS).
   Thus, in hindsight our approach puts the traditional phenomenological
vector-meson-dominance model on a more solid theoretical basis.
   In the sense of a strict chiral expansion in terms of small external momenta $q$ and
quark masses $m_q$ at a fixed ratio $m_q/q^2$ \cite{Gasser:1983yg}, up to and including
${\cal O}(q^4)$ the results with and without explicit vector mesons are completely equivalent.
   The additional vector-meson contributions up to this order are
compensated by a readjustment of the low-energy constants pertaining to the
theory including vector mesons as dynamical degrees of freedom.
   On the other hand, the inclusion of vector-meson degrees of freedom in the present
framework results in a reordering of terms which, in an ordinary chiral
expansion, would show up at higher orders beyond ${\cal O}(q^4)$.
   It is these terms which change the form factor results favorably for larger
values of $Q^2$.
   It should be noted, however, that this re-organization proceeds according to
well-defined rules so that a controlled, order-by-order, calculation of
corrections is made possible.
   In contrast to the calculation without vector mesons, the Sachs form factors
$G_E^p$, $G_M^p$, and $G_M^n$ now show sufficient curvature to generate a more
accurate  phenomenology for values of $Q^2$, where the ordinary chiral expansion
to the same order is no longer reliable.

\subsubsection{Axial and induced pseudoscalar form factors}
   Assuming isospin symmetry, the most general
parametrization of the isovector axial-vector current evaluated
between one-nucleon states is given by
\begin{equation}\label{H1_axff_FFDef}
\langle N(p',s')| A^\mu_i(0) |N(p,s) \rangle = \bar{u}(p',s')
\left[\gamma^\mu\gamma_5 G_A(Q^2) +\frac{q^\mu}{2m_N}\gamma_5
G_P(Q^2) \right] \frac{\tau_i}{2}u(p,s),
\end{equation}
where $q=p'-p$, $Q^2=-q^2$, and $m_N$ denotes the nucleon mass.
   $G_A(Q^2)$ is called the axial form factor and
$G_P(Q^2)$ is the induced pseudoscalar form factor.
     The value of the axial form factor at zero momentum transfer is defined as
the axial-vector coupling constant, $g_A=G_A(Q^2=0) =1.2695(29)$ \cite{PDG_2008},
and is quite precisely determined from neutron beta decay.
   The $Q^2$ dependence of the axial form factor can be obtained
either through neutrino scattering or pion electroproduction
(see \cite{Bernard:2001rs} and references therein).
   The second method makes use of the so-called Adler-Gilman relation
\cite{Adler:1966gd} which provides a chiral Ward identity
establishing a connection between charged pion electroproduction at
threshold and the isovector axial-vector current evaluated between
single-nucleon states (see, e.g., Ref.\ \cite{Scherer:1991cy} for more details).
   The induced pseudoscalar form factor $G_P(Q^2)$ has been investigated in
ordinary and radiative muon capture as well as pion
electroproduction (see Ref.\ \cite{Gorringe:2002xx} for a review).

   For the analysis of experimental data, $G_A(Q^2)$ is conventionally
parameterized using a dipole form as
\begin{equation}
\label{GAPara}
    G_A(Q^2)=\frac{g_A}{(1+\frac{Q^2}{M^2_A})^2},
\end{equation}
where the axial mass $M_A$ is related to the axial root-mean-square
radius by $\langle r^2_A\rangle^\frac{1}{2}=2\sqrt{3}/M_A$.
    The global average for the axial mass extracted from neutrino scattering
experiments given in Refs.\ \linebreak \cite{Liesenfeld:1999mv}, \cite{Bernard:2001rs} is
\begin{equation}\label{MAv1}
    M_A = (1.026 \pm 0.021)\,\mbox{GeV}.
\end{equation}
  The extraction of the axial mean-square radius from
charged pion electroproduction at threshold is motivated by
the current algebra results and the PCAC hypothesis.
   At threshold (the spatial components of) the center-of-mass transition current
for pion electroproduction can be written in terms of two s-wave amplitudes
$E_{0+}$ and $L_{0+}$,
\begin{displaymath}
\left.e\vec{M}\right|_{\rm thr}=\frac{4\pi W}{m_N}\left[i\vec\sigma_\perp
E_{0+}(k^2) +i\vec{\sigma}_\parallel L_{0+}(k^2)\right],
\end{displaymath}
where $W$ is the total center-of-mass energy, $k^2$ is the four momentum transfer
squared of the virtual photon, and $\vec{\sigma}_\parallel=\vec \sigma\cdot\hat
k\hat k$ and $\vec\sigma_\perp= \vec \sigma-\vec{\sigma}_\parallel$.
   The reaction $p(e,e'\pi^+)n$ has been measured at MAMI at an invariant mass
of $W=1125$ MeV (corresponding to a pion center-of-mass momentum of
$|\vec{q}^\ast|=112$ MeV) and photon four-momentum transfers of $-k^2=0.117,
0.195$ and 0.273 GeV$^2$ \cite{Liesenfeld:1999mv}.
   Using an effective-Lagrangian model
an axial mass of
\begin{displaymath}
\bar{M}_A=(1.077\pm 0.039)\,\mbox{GeV}
\end{displaymath}
was extracted, where the bar is used to distinguish the result from the
neutrino scattering value.
    In the meantime, the experiment has been repeated including an
additional value of $-k^2=0.058$ GeV$^2$ and is currently
being analyzed.
   The global average from several pion electroproduction experiments
is given by \cite{Bernard:2001rs}
\begin{equation}\label{MApi}
    \bar{M}_A=(1.069\pm 0.016)\,\mbox{GeV}.
\end{equation}
   It can be seen that the values of Eqs.~(\ref{MAv1})
for the neutrino scattering experiments are smaller than Eq.~(\ref{MApi}) for
the pion electroproduction experiments.
   The discrepancy was explained in heavy-baryon chiral
perturbation theory \cite{Bernard:1992ys}.
   It was shown that at ${\cal O}(q^3)$ pion
loop contributions modify the $k^2$ dependence of the electric dipole amplitude
from which $\bar{M}_A$ is extracted.
   These contributions result in a change of
\begin{equation}\label{deltaMA}
    \Delta M_A = 0.056 \,\mbox{GeV},
\end{equation}
bringing the neutrino scattering and pion electroproduction
results for the axial mass into agreement.
     In a recent analysis \cite{Bodek:2007vi} updated expressions for the vector form
factors have been taken into account together with the hadronic correction
of Eq.~(\ref{deltaMA}) to produce an average from both neutrino and electroproduction
experiments,
\begin{equation}\label{MAv2}
    M_A = (1.014 \pm 0.014)\,\mbox{GeV}.
\end{equation}

\begin{figure}[t]
\begin{center}
\epsfig{file=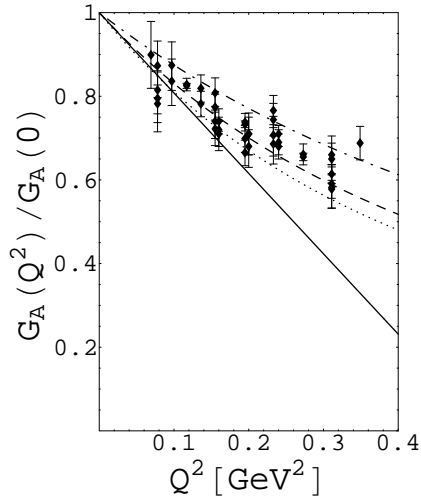,width=0.3\textwidth}
\caption{\label{GAwithout}
The axial form factor $G_A$ in manifestly Lorentz-invariant ChPT
at ${\cal O}(q^4)$. Full line: result in infrared renormalization with
parameters fitted to reproduce the axial mean-square radius corresponding
to the dipole parametrization with $M_A=1.026$ GeV (dashed line).
  The dotted and dashed-dotted lines refer to dipole parameterizations
with $M_A=0.95$ GeV and $M_A=1.20$ GeV, respectively.
   The experimental values are
taken from \cite{Bernard:2001rs}.}
\end{center}
\end{figure}

  Earlier calculations of the axial form factor were performed in the framework of
heavy-baryon ChPT \cite{Bernard:1992ys}, \cite{Fearing:1997dp} and the small-scale
expansion \cite{Bernard:1998gv}.
   In Ref.\ \cite{Schindler:2006it} the form factors $G_A$ and
$G_P$ have been calculated in manifestly Lorentz-invariant baryon
ChPT up to and including ${\cal O}(q^4)$.
   The axial form factor can be written as
\begin{equation}\label{GAexpansion}
    G_A(Q^2)=g_A-\frac{1}{6}\,g_A\,\langle r^2_A\rangle\, Q^2 +
    \frac{\texttt{g}_A^3}{4F^2}L(Q^2),
\end{equation}
where $\langle r^2_A\rangle$ is the axial mean-square radius
and $L$ contains loop contributions and satisfies $L(0)=L'(0)=0$.
   The result for $G_A$ in the momentum-transfer region
$0\,\mbox{GeV}^2\leq Q^2 \leq 0.4\,\mbox{GeV}^2$
is shown in Fig.~\ref{GAwithout}.
   The parameters have been determined such as to reproduce the axial
mean-square radius corresponding to the dipole parametrization with
$M_A=1.026$ GeV (dashed line).
  The dotted and dashed-dotted lines refer to dipole parameterizations
with $M_A=0.95$ GeV and $M_A=1.20$ GeV, respectively.
   The loop contributions from $L(Q^2)$ are small
and the result does not produce enough curvature to describe the data
for momentum transfers $Q^2 \ge 0.1\, \mbox{GeV}^2$.
   The situation is similar to the electromagnetic case of Fig.~\ref{G_ohne},
where ChPT at ${\cal O}(q^4)$ also fails to
describe the form factors beyond $Q^2 \ge 0.1\, \mbox{GeV}^2$.

\begin{figure}[t]
\begin{center}
\epsfig{file=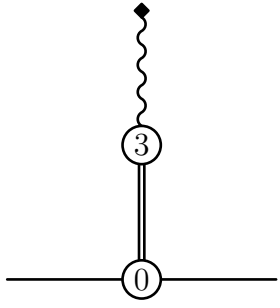,width=0.2\textwidth}\caption{\label{AVMDia}
Diagram containing the axial-vector meson (double line) contributing
to the form factors at ${\cal O}(q^4)$.}
\end{center}
\end{figure}

\begin{figure}[t]
\begin{minipage}[b]{0.3\textwidth}
\includegraphics[width=\textwidth]{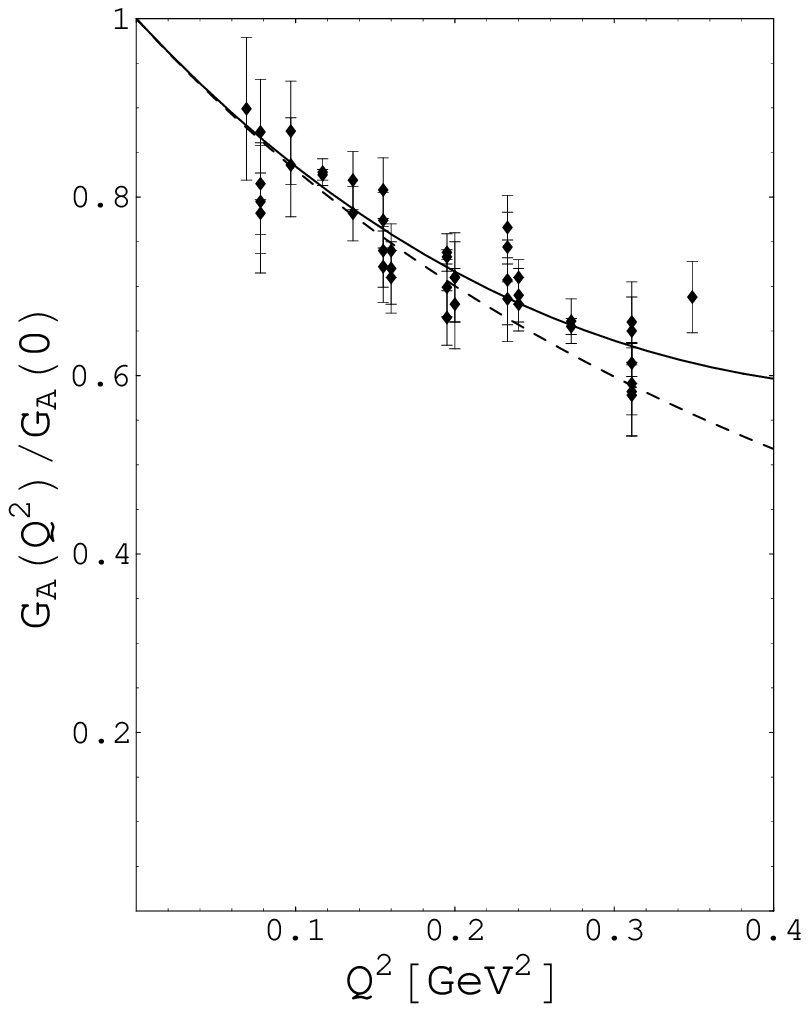}
\end{minipage}
\hspace{2em}
\begin{minipage}[b]{0.5\textwidth}
\includegraphics[width=\textwidth]{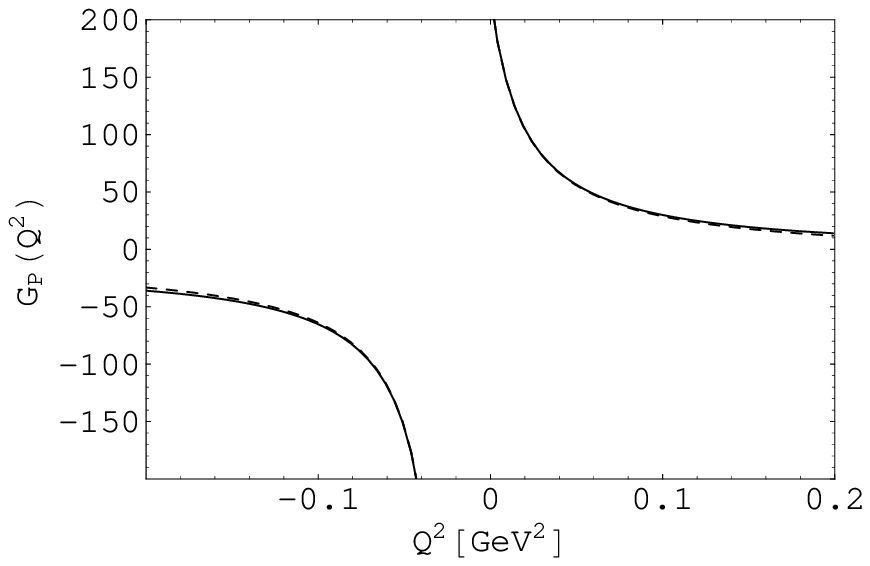}
\end{minipage}
\caption{\label{H1_axff_GAwith}Left panel: Axial form factor $G_A$
in manifestly Lorentz-invariant ChPT at ${\cal O}(q^4)$ including
the axial-vector meson $a_1(1260)$ explicitly. Full line: result in
infrared renormalization, dashed line: dipole parametrization. The
experimental data are taken from Ref.\ \cite{Bernard:2001rs}.
 Right
panel:
   The induced pseudoscalar form factor $G_P$ in manifestly Lorentz-invariant ChPT
at ${\cal O}(q^4)$ including the axial-vector meson $a_1(1260)$
explicitly. Full line: result with axial-vector meson; dashed line:
result without axial-vector meson.
   One can clearly see the dominant pion pole contribution at $Q^2\approx
   -M_\pi^2$.}
\end{figure}

   In addition to the standard treatment including the
nucleon and pions, the axial-vector meson $a_1(1260)$ has also been
considered as an explicit degree of freedom \cite{Schindler:2006it}.
   In the vector-field formulation of \cite{Ecker:1989yg} the
$a_1(1260)$ meson is represented by $A^\mu=A^\mu_i \tau_i$.
   The advantage of this formulation is that the coupling of the axial-vector
mesons to pions and external sources is at least of ${\cal O}(q^3)$.
   The calculation of the contributions to the isovector axial-vector
form factors only requires the term
\begin{equation}
\label{LagAVMmeson}
{\cal L}_{{\pi}A}^{(3)} = \frac{f_A}{4}\mbox{Tr}(A_{\mu\nu}f^{\mu\nu}_{-}),
\end{equation}
where the field strength tensor is defined as
\begin{displaymath}
f_-^{\mu\nu}=u^+f_R^{\mu\nu}u-u f_L^{\mu\nu} u^+,
\end{displaymath}
and
\begin{displaymath}
A_{\mu\nu}=\nabla_\mu A_\nu-\nabla_\nu A_\mu,
\quad\nabla_\mu A_\nu=\partial_\mu A_\nu+[\Gamma_\mu,A_\nu],
\end{displaymath}
with the connection of Eq.~(\ref{4:1:2:gamma}).
   The coupling of the axial-vector meson to the nucleon starts at
${\cal O}(q^0)$.
   The corresponding Lagrangian reads
\begin{equation}
\label{LagAVMNuc}
{\cal L}_{AN}^{(0)} = \frac{g_{a_1}}{2} \bar{\Psi} \gamma^{\mu}\gamma_5 A_\mu \Psi.
\end{equation}
   A calculation up to and including ${\cal O}(q^4)$ would in principle
also require the Lagrangian of ${\cal O}(q)$.
   However, there is no term at this order that is allowed by the symmetries.
   The additional power-counting rules are as in Section \ref{subsectionemff}
for the vector mesons.
   We count the axial-vector meson propagator as ${\cal O}(q^0)$, vertices from
${\cal L}_{{\pi}A}^{(3)}$ as ${\cal O}(q^3)$, and vertices from
${\cal L}_{AN}^{(0)}$ as ${\cal O}(q^0)$, respectively.
     The contributions of the axial-vector meson to the form
factors $G_A$ and $G_P$ at ${\cal O}(q^4)$ originate from the diagram in
Fig.~\ref{AVMDia}.
   The inclusion of the axial-vector meson effectively results in one additional
low-energy coupling constant which has been determined by a fit to
the data for $G_A(Q^2)$.
   The inclusion of the axial-vector meson results in an improved
description of the experimental data for $G_A$ (see
Fig.~\ref{H1_axff_GAwith}), while the contribution to $G_P$ is
small.

\subsubsection{Pion-nucleon form factor}
   The pion-nucleon form factor $G_{\pi N}(Q^2)$ may be defined in terms of
the pseudoscalar quark density $P_i=i\bar{q}\gamma_5 \tau_i q$ and
the average light-quark mass $\hat m$ as \cite{Gasser:1987rb}
\begin{equation}
\label{GpiN} \hat m \langle N(p',s')| P_i (0) | N(p,s) \rangle =
         \frac{M_\pi^2 F_\pi}{M_\pi^2 + Q^2}
         G_{\pi N}(Q^2)i\bar{u}(p',s') \gamma_5 \tau_i u(p,s),
\end{equation}
where $q=p'-p$, $Q^2=-q^2$, and $\Phi_i(x)\equiv \frac{\hat m P_i
(x)}{M_\pi^2 F_\pi}$ is the corresponding interpolating pion field.
   The pion-nucleon coupling constant is given by
$g_{\pi N}=G_{\pi N}(-M_\pi^2)$.
   Using the (QCD-) partially conserved axial-vector current (PCAC) relation,
$\partial_\mu A^\mu_i= \hat m P_i$, the pion-nucleon form factor
is completely given in terms of the axial and the induced
pseudoscalar form factors,
\begin{displaymath}
   2m_N G_A(Q^2) - \frac{Q^2}{2m_N} G_P(Q^2) =
       2\frac{M_\pi^2 F_\pi}{M_\pi^2 + Q^2} G_{\pi N}(Q^2).
\end{displaymath}
   This is an exact relation which holds true for any value of
   $Q^2$.
   The result at ${\cal O}(q^4)$ is given by \cite{Schindler:2006it}
\begin{displaymath}
\label{GpiNresult} G_{\pi N}(Q^2)=\frac{m_N g_A}{F_\pi}-g_{\pi
N}\Delta \frac{Q^2}{M_\pi^2} + \cdots
\end{displaymath}
where $\Delta=1-\frac{m_N g_A}{F_\pi g_{\pi N}}$ denotes the
Goldberger-Treiman discrepancy.
   The chiral expansion of the pion-nucleon coupling constant can
   be found in Ref.\ \cite{Schindler:2006it}.

\section{Conclusion}

   Effective field theory has become a very important tool for investigating
the dynamics of the strong interactions.
   In particular, mesonic chiral perturbation theory is a full-grown and mature
area of low-energy particle physics which has successfully been
applied at the two-loop level.
   Whether the predictions for the electromagnetic polarizabilities of the
charged pion are really in conflict with empirical data remains to be seen.
   In the baryonic sector new renormalization conditions have reconciled
the manifestly Lorentz-invariant approach with the standard power counting.
   Phenomenological extensions allowing for the rigorous inclusion of (axial-)
vector-meson degrees of freedom (and also of the $\Delta(1232)$ resonance)
have opened the door to an extended kinematic region.
   Unfortunately, the question of convergence in the three-flavor
sector remains a controversial issue \cite{Lehnhart:2004vi}, even
though the manifestly Lorentz-invariant approach might yield better
phenomenological results \cite{Geng:2008mf}.
   Finally, beyond the one-nucleon sector the covariant framework has been used
in the discussion of relativistic corrections to the nucleon-nucleon potential
(see, e.g., Refs.\ \cite{Higa:2003jk}, \cite{Robilotta:2006xp}) or may be applied
to the nuclear many-body problem (see, e.g., Refs.\ \cite{Furnstahl:2003cd}, \cite{Serot:2003ed}).

\medskip
   It is a pleasure to thank D.~Djukanovic, H.W.~Fearing, T.~Fuchs, J.~Gegelia,
G.~Japaridze, and M.R.~Schindler for the fruitful collaboration on the topics
of this article.
   This work was made possible by the financial support from the Deutsche
Forschungsgemeinschaft (SFB 443 and SCHE 459/2-1).

\end{document}